\pdfoutput=1
\documentclass[11pt]{article}

\usepackage[a4paper,margin=25mm]{geometry}
\usepackage[T1]{fontenc}
\usepackage[utf8]{inputenc}
\usepackage{times}
\usepackage{amsmath,amssymb}
\usepackage{graphicx}
\usepackage{booktabs}
\usepackage{array}
\usepackage{enumitem}
\usepackage{caption}
\usepackage{natbib}
\usepackage{microtype}
\usepackage[colorlinks=true,allcolors=black]{hyperref}

\renewcommand{\baselinestretch}{1.28}\normalsize
\setcitestyle{authoryear,round,comma,aysep={,},yysep={,}}

\newcommand{\code}[1]{\texttt{\small #1}}
\newcommand{\vect}[1]{\mathbf{#1}}
\newcommand{\mat}[1]{\mathbf{#1}}
\newcommand{\snote}[1]{\subsection*{Supplementary Note #1}}
\newcolumntype{L}[1]{>{\raggedright\arraybackslash}p{#1}}

\newcommand{\SnBackwardBound}{1.8\times10^{-5}}
\newcommand{\SnCohCross}{275}
\newcommand{\SnCohGtThresh}{42}

\newcommand{\SnCohMedianAll}{0.825840}
\newcommand{\SnCohMedianShift}{6.7\times10^{-4}}
\newcommand{\SnCohPairs}{1,770}

\newcommand{\SnCohSeconds}{29}
\newcommand{\SnCohWithinMedian}{0.871}
\newcommand{\SnCohZoneCands}{5}

\newcommand{\SnCrossNets}{9}
\newcommand{\SnCrossSpread}{10}

\newcommand{\SnDwCoords}{24}
\newcommand{\SnDwLam}{2.85\times10^{-6}}
\newcommand{\SnDwTurb}{1.15\times10^{-8}}

\newcommand{\SnExtResid}{4.263\times10^{-14}}
\newcommand{\SnExtTol}{1.00\times10^{-4}}
\newcommand{\SnExtWorst}{5.56\times10^{-5}}

\newcommand{\SnHealthBad}{\texttt{ky4}, \texttt{Net3}}
\newcommand{\SnHealthBadN}{2}
\newcommand{\SnHealthCond}{\texttt{Pescara}}
\newcommand{\SnHealthNets}{7}
\newcommand{\SnInvCandidates}{60}

\newcommand{\SnKappaLtown}{1.64\times10^{11}}
\newcommand{\SnKyFourAB}{12.1--12.5}
\newcommand{\SnKyFourAC}{50.4--54.0}
\newcommand{\SnKyFourBC}{4.11--4.36}

\newcommand{\SnKyFourFwd}{6.6--6.9}

\newcommand{\SnLtownMemDense}{29\,096}
\newcommand{\SnLtownMemSparse}{1\,286}

\newcommand{\SnModenaAB}{10.6--10.9}
\newcommand{\SnModenaBC}{0.21--0.22}

\newcommand{\SnModenaFwdEight}{1.21--1.24}
\newcommand{\SnModenaFwdSixtyFour}{0.71--0.73}
\newcommand{\SnNetOneAB}{1.97--2.11}
\newcommand{\SnNetOneWorst}{0.06}
\newcommand{\SnNetThreeFbEight}{0.97--1.04}

\newcommand{\SnPlanKyFour}{3.99--4.45}
\newcommand{\SnPlanKyFourPlans}{16}
\newcommand{\SnPlanModena}{2.33--2.43}
\newcommand{\SnPlanSlotsCost}{2438}
\newcommand{\SnPlanSlotsPen}{24.3--25.8}

\newcommand{\SnSweepMaxNet}{\texttt{pub\_fossolo\_poly1}}

\newcommand{\SnSweepMinNet}{\texttt{pub\_net3}}

\newcommand{\SnSweepSlopeDll}{0.99}
\newcommand{\SnSweepSlopeOurs}{1.01}

\newcommand{\WrAuditItems}{22}
\newcommand{\WrAuditOper}{6}
\newcommand{\WrAuditWorstFrac}{0.57}
\newcommand{\WrAuditZero}{7}
\newcommand{\WrAugCoverKAll}{56}
\newcommand{\WrAugCoverKEighty}{32}
\newcommand{\WrAugCoverSat}{57}
\newcommand{\WrAugCoverTwentyMed}{15.5}
\newcommand{\WrAugCoverTwentyPrior}{19.1}

\newcommand{\WrAugDoptTwentyCrlbSub}{222}
\newcommand{\WrAugDoptTwentyKSub}{36}
\newcommand{\WrAugDoptTwentyMed}{11.5}
\newcommand{\WrAugDoptTwentyPrior}{21.9}
\newcommand{\WrAugHanoiCoverK}{1}
\newcommand{\WrAugHanoiDoptK}{4}
\newcommand{\WrAugHanoiSZeroRmse}{18.7}
\newcommand{\WrAugHanoiTwentyRmse}{7.0}
\newcommand{\WrAugInfoRmseMax}{24.3}
\newcommand{\WrAugInfoRmseMin}{23.2}
\newcommand{\WrAugLeakGtMax}{34}
\newcommand{\WrAugLeakGtMin}{18}
\newcommand{\WrAugLeakMedMax}{0.90}
\newcommand{\WrAugLeakMedMin}{0.79}
\newcommand{\WrAugLeakRandTThree}{3}
\newcommand{\WrAugLeakRivalMax}{0.9997}
\newcommand{\WrAugLeakRivalMin}{0.9994}
\newcommand{\WrAugLeakSZeroGt}{48}
\newcommand{\WrAugLeakSZeroMedian}{0.928}
\newcommand{\WrAugLeakTOneFootprint}{0.09}
\newcommand{\WrAugLtownKAll}{36}
\newcommand{\WrAugLtownKEighty}{24}
\newcommand{\WrAugLtownLostRankA}{19}
\newcommand{\WrAugLtownLostRankB}{32}
\newcommand{\WrAugLtownReselLost}{99}
\newcommand{\WrAugSZeroCrlbSub}{742}
\newcommand{\WrAugSZeroInfoRmse}{25.6}
\newcommand{\WrAugSZeroInfoRmseThree}{26.3}
\newcommand{\WrAugTwentyInfoRmseThreeMax}{30.3}
\newcommand{\WrAugTwentyInfoRmseThreeMin}{29.5}

\newcommand{\WrCityDClampMax}{6.45\times10^{-7}}
\newcommand{\WrCityDExtWorst}{4.78\times10^{-5}}

\newcommand{\WrClusTauCityD}{0.999}
\newcommand{\WrClusTauLtown}{0.995}
\newcommand{\WrClusTopThreeKm}{7.42}
\newcommand{\WrClusTopThreeRad}{750}
\newcommand{\WrClusTopThreeShare}{8.8}

\newcommand{\WrCohKFortyCIHi}{0.379}
\newcommand{\WrCohKFortyCILo}{0.032}
\newcommand{\WrCohKFortyP}{$1.90\times10^{-1}$}
\newcommand{\WrCohKTwentyP}{$4.76\times10^{-2}$}
\newcommand{\WrCohLOneStageOneP}{$2.39\times10^{-3}$}
\newcommand{\WrCohLThreeStageOneP}{$8.92\times10^{-1}$}
\newcommand{\WrCohLTwoFisherP}{$1.05\times10^{-7}$}
\newcommand{\WrCohLTwoStageOneP}{$2.10\times10^{-6}$}
\newcommand{\WrCohPooledCIHi}{0.204}
\newcommand{\WrCohPooledCILo}{0.016}
\newcommand{\WrCtlAnytownWas}{6.08\times10^{-12}}
\newcommand{\WrCtlBigWas}{6.836}
\newcommand{\WrCtlExactZero}{25}
\newcommand{\WrCtlExactZeroWas}{20}
\newcommand{\WrCtlMaxdQFixed}{7.11\times10^{-15}}
\newcommand{\WrCtlNetSixFrames}{609}
\newcommand{\WrCtlNetSixWas}{2.30\times10^{-5}}

\newcommand{\WrCtlNotExact}{27}
\newcommand{\WrCtlNullNets}{eight}

\newcommand{\WrCtlWorstdH}{1.14\times10^{-13}}
\newcommand{\WrCtlWorstdQ}{1.42\times10^{-14}}
\newcommand{\WrDW}{1}
\newcommand{\WrDisagreeN}{22}
\newcommand{\WrDisagreeOutOf}{48}
\newcommand{\WrDriftInfoHi}{412}
\newcommand{\WrDriftInfoLo}{279}
\newcommand{\WrDriftRankHi}{56}
\newcommand{\WrDriftRankLo}{20}
\newcommand{\WrEmitVarWorst}{2.39\times10^{-5}}
\newcommand{\WrExactZero}{20}
\newcommand{\WrExampleNets}{3}

\newcommand{\WrFrames}{8,140}

\newcommand{\WrGdWall}{631}
\newcommand{\WrHW}{51}
\newcommand{\WrHeadNoiseShare}{96}
\newcommand{\WrHybridSeeds}{one}
\newcommand{\WrHybridWall}{4,756}
\newcommand{\WrHybridWallEight}{595}

\newcommand{\WrIdentKRef}{98}
\newcommand{\WrIdentNull}{334}
\newcommand{\WrIdentPool}{541}
\newcommand{\WrIdentPriorRmse}{28.92}
\newcommand{\WrLeTwelve}{47}

\newcommand{\WrMsIdent}{25}

\newcommand{\WrMsTrain}{0.4}
\newcommand{\WrMsVal}{1.4}
\newcommand{\WrNets}{52}

\newcommand{\WrOperationalNets}{3}
\newcommand{\WrPublicNets}{23}
\newcommand{\WrRegOper}{7}

\newcommand{\WrRegTotal}{54}
\newcommand{\WrSweepCityD}{21}
\newcommand{\WrSweepCityH}{47}
\newcommand{\WrSweepN}{16}
\newcommand{\WrSweepPubMax}{73}
\newcommand{\WrSweepPubMin}{26}
\newcommand{\WrSyntheticNets}{23}
\newcommand{\WrThreeWayCoords}{61}
\newcommand{\WrThreeWayWorst}{5.85\times10^{-8}}
\newcommand{\WrUnitsCMH}{2}
\newcommand{\WrUnitsGPM}{13}
\newcommand{\WrUnitsLPS}{37}

\renewcommand{\SnHealthBad}{ky4, Net3}
\renewcommand{\SnHealthCond}{Pescara}
\renewcommand{\SnSweepMaxNet}{Fossolo}
\renewcommand{\SnSweepMinNet}{Net3}

\title{\bfseries Water-network decisions share one hydraulic gradient,
and it can now be computed exactly}

\author{%
Tianwei Mu$^{1,2,3}$, Yue Wang$^{3}$, Mingzhe Yuan$^{1,4,*}$, Wenhong Wang$^{1}$,\\
Qing Luo$^{2}$, Min Xiao$^{2}$, Jun Li$^{3}$, Hui Yang$^{3}$, Manhong Huang$^{5}$}
\date{}

\begin{document}

\maketitle
\vspace{-8mm}

\begin{sloppypar}\noindent{\small
$^{1}$Guangzhou Institute of Industrial Intelligence, Guangzhou 510000, China.
$^{2}$Key Laboratory of Ecological Restoration of Regional Contaminated
Environment, Ministry of Education, College of Environment, Shenyang
University, Shenyang 110044, China.
$^{3}$School of Municipal Engineering and Environment, Shenyang Jianzhu
University, Shenyang 110168, China.
$^{4}$Shenyang Institute of Automation, Chinese Academy of Sciences,
Shenyang 110169, China.
$^{5}$College of Environmental Science and Engineering, Donghua University,
Shanghai 201620, China.
$^{*}$Corresponding author: \href{mailto:mzyuan@sia.cn}{mzyuan@sia.cn}.}
\end{sloppypar}

\vspace{4mm}
\noindent\textbf{\large Abstract}\\[2pt]
\begin{sloppypar}\noindent
Calibration, leak localisation and sensor placement on water distribution
networks (WDNs) are decisions about continuous parameters, yet the hydraulic engine
that defines the physics returns a solution and no derivatives, so practice
falls back on derivative-free search or on surrogates whose error the answer
inherits. We make the global gradient algorithm itself exactly differentiable: the
forward pass reproduces the reference engine's discrete devices, status
switching and low-flow linearisation included, and the backward pass solves
the implicit adjoint by reusing the forward pass's terminal factorisation, so
one extra sparse solve returns every parameter's gradient at once, batched
over scenarios on one graphics processor. Across 52 public,
synthetic and operational networks and 8,140 simulation frames, every network
meets the acceptance criterion, the largest head deviation from EPANET~2.2 is
$1.137\times10^{-13}$~ft and 25 agree exactly. One adjoint solve
replaces the 906 simulations a finite-difference roughness Jacobian costs on
the 905-pipe L-TOWN benchmark, and a leak-inversion training loop runs at
463--470~ms per optimiser step for 256 scenarios, 191 times the prior
pipeline. Gradient calibration reaches
its endpoint within a median 595 model calls, where the strongest of five
tuned metaheuristics needs 8,060 to match it on the training loss and two
never do within 20,000. On a 554-link operating network, one adjoint
pass audits, pipe by pipe, which roughness parameters the installed sensors
can constrain and which sensors to add, on the model the utility already
operates.
\end{sloppypar}

\vspace{2mm}

\vspace{4mm}

\section{Introduction}\label{sec:intro}

Most quantitative work on a water distribution network (WDN) is a search over
continuous parameters: which roughness values reproduce the observed
heads, which junction is leaking and by how much, where the next pressure
sensor should go. Each runs against the same physics, the global
gradient algorithm (GGA) of \citet{todini1988gradient,todini2013unified},
whose reference implementation EPANET~2.2 \citep{rossman2020epanet22} the
field treats as ground truth; each is gradient-shaped, yet each is solved
without gradients, because the engine returns a solution and nothing
else. The workaround is derivative-free search
\citep{maier2014evolutionary}, whose cost grows with parameter count and,
with leakage management under regulatory and economic pressure worldwide
\citep{sousa2026leakage}, is paid again at network scale whenever the model
changes.

Two lines of work point at the missing ingredient. One makes models
differentiable by replacing them: learned surrogates support state estimation
and calibration \citep{truong2024graph,kerimov2023assessing}, and
gradient-based control of urban drainage became tractable because a neural
internal model supplied the derivatives the physics engine could not
\citep{zhang2026differentiable}. A surrogate is fast, but its gradient is the
surrogate's and its approximation error enters every downstream decision
unbounded. The other keeps the physics: \citet{ulusoy2022biobjective} showed
on an operational network that gradient-based optimisation outperforms
evolutionary search on large continuous design-for-control problems. What
that argument needs, and what no published tool supplies, is the exact
gradient of the deployed hydraulic engine.

A separate acceleration literature makes the forward solve faster, by
topological reduction, domain decomposition and parallel computing
\citep{guo2024decomposed} or by replacing the linear solver. All of it
optimises one network for one demand frame; this work optimises over a batch
of operating conditions and requires the result to be differentiable, so the
two are complementary (Section~\ref{sec:discussion}).

We therefore take the reference numerical scheme exactly as it ships and make
it differentiable. Analytic sensitivities of the idealised steady state are
classical \citep{piller2017local} and we claim no priority over them, but a
solver in production switches link status, clamps options at parse time and
linearises the head-loss law below a flow threshold, and the object
differentiated here includes those devices. Three contributions follow.
(i)~\emph{The gradient object exists and can be trusted.} The backward pass is
the implicit adjoint of the very iteration the utility runs, discrete devices
included, not of a smoothed stand-in, and the forward pass it hangs on is
verified against the reference binary across 52 networks and 8,140 frames,
largest head deviation $1.137\times10^{-13}$~ft, none excluded
(Section~\ref{sec:res-fidelity}), so transfer to the utility's own model
carries no surrogate residual. (ii)~\emph{The object is affordable at decision
scale.} Batching the whole solver, status machine included, on one graphics
processor and reusing each scenario's terminal factorisation for the adjoint
brings one forward-plus-backward L-TOWN scenario from 333~ms to 1.90~ms at
batch size 256; per single solve it is not a faster simulator, and
Section~\ref{sec:res-batch} keeps the qualifications in view.
(iii)~\emph{Three decisions, one object.} Calibration is least squares on the
adjoint Jacobian, the success or failure of leak search is the coherence of
its columns, and sensor placement is its Fisher information, so coherence can
be read before a search and the Cram\'er--Rao lower bound (CRLB) prices the
identifiability added sensors buy. The chain runs end to end on public
benchmarks and on a
utility's operating network, with leak candidates drawn from its own work
orders; placement keeps the sensors the utility has and prescribes, in
simulation, additions that restore parameter identifiability but do not
recover the leaks noise defeated, while a second objective, against coherence
itself, recovers one of the three as coherence predicts
(Section~\ref{sec:res-place}).

\section{Materials and methods}\label{sec:methods}

\subsection{Networks and data}\label{sec:networks}

The verification suite has 52 networks: 23 public benchmark models, 21
fetched from their upstream sources and checked against recorded SHA-256
digests plus two from the Kentucky dataset release; 3 EPANET distribution
examples; 23 randomly generated networks released with the code; and 3
operational models: City~D (542 nodes, 554 links, 79 throttle-control
valves), its emitter variant, and City~H (921 nodes, 1,038 links, 6 pumps).
Figure~\ref{fig:networks} draws nine of them.
The mainline public network is L-TOWN, the BattLeDIM benchmark
\citep[CC~BY~4.0, Zenodo record 4017659]{vrachimis2020dataset,
vrachimis2022battledim}, a cleaned copy differing from the published file by
one removed default-pattern line and a line-ending change, both digests
recorded. Public calibration studies use a synthetic 25-frame diurnal
profile; City~D the utility's own 24-hour pattern.

Both operational models are released with the permission of the operating
utility, in anonymised form: coordinates carry a rigid transform that removes
georeferencing, and identifiers are replaced. In the text they are reported by
counts and positional labels, and the three injected leaks are labelled
$L_1$--$L_3$.

\subsection{The global gradient algorithm, and a bit-faithful replica}
\label{sec:gga}

For a network with unknown junction heads $\vect{H}$ and link flows
$\vect{Q}$, one GGA iteration solves the symmetric positive definite Schur
system $\mat{A}\vect{H}^{(k+1)}=\vect{F}^{(k)}$ with
$\mat{A}=\mat{A}_{21}\mat{D}^{-1}\mat{A}_{12}$, where
$\mat{D}=\mathrm{diag}(\mathrm{d}h_{L,k}/\mathrm{d}Q_k)$ collects the
head-loss derivatives, then corrects every flow by an independent scalar
update satisfying nodal continuity exactly at every iterate. EPANET~2.2 adds
the engineering devices a replica must reproduce: guard constants ($10^{8}$
for closed links and active pressure-reducing-valve (PRV) penalties, a
$10^{-7}$ floor linearising the head-loss law at vanishing flow), a discrete
status machine for check valves, pumps, PRVs and tanks with hysteresis and
throttled checking, and convergence requiring status-consistency, not only a
small flow change \citep{elhay2011dealing}.

\begin{figure}[htbp]
\centering
\includegraphics[width=\linewidth]{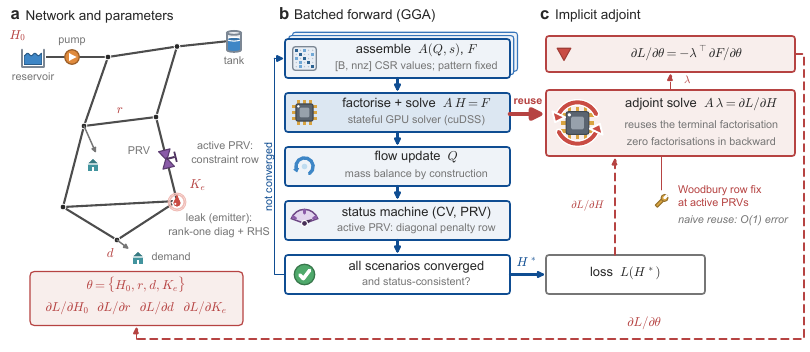}
\caption{The differentiable solver. \textbf{a},~Parameter classes: reservoir
head, pipe resistance, demand, emitter coefficient; a leak is a rank-one
diagonal update. \textbf{b},~Batched forward pass with the full status
machine. \textbf{c},~The implicit adjoint, reusing the terminal
factorisation, returning all gradient classes at once.}
\label{fig:method}
\end{figure}

The implementation is built on an automatic-differentiation tensor library
\citep{paszke2019pytorch} over standard numerical Python, and carries two paths: a
replica path for bit-level comparison against the reference engine, and a
batched differentiable path \citep{mu2026hydrograd}
(Fig.~\ref{fig:method}). Transcribing formulas is not what makes the replica agree to $10^{-14}$~ft;
four things invisible in the mathematics are. The symbolic factorisation's
elimination order and left-looking Cholesky loop were ported verbatim
\citep{george1981computer}, since any other ordering changes the sequence of
floating-point additions; the power and logarithm routines are taken from the reference
binary's own C runtime, which is why bit-level claims are platform-specific;
literal constants are reproduced, down to that build's truncated fallback
$\pi$; and so are parse-time clamps, down to its silent clamping of the engine's
convergence-tolerance option. Backward error analysis
says nothing finer than the arithmetic order is verifiable: with
$\kappa_2(\mat{A})=1.64\times10^{11}$ on L-TOWN with all three PRVs active,
two implementations differing only in reduction order may disagree at the
$10^{-4}$~ft level, and the same batched solve on CPU and GPU differs by
$1.023\times10^{-5}$~ft. A re-implementation agreeing at $10^{-6}$~ft is
therefore at its noise floor (Supplementary Note~1).

\subsection{Exact gradients: the implicit adjoint}\label{sec:adjoint}

Two reverse-mode routes are provided. A truncated $K$-step unrolling
differentiates the solver itself, the route for learning inside the
iteration; it is checked against a looser threshold and ships with a health
check, because its truncated gradient does not converge in $K$ on every
network. The workhorse is the implicit-function adjoint: writing the
converged state as the root of the fixed-point residual assembled from
EPANET's own update rules, clamped head-loss branches included, block
elimination returns exactly the Schur matrix $\mat{A}$ of
Section~\ref{sec:gga} augmented by the emitter diagonal, so the adjoint
system has the same sparsity, conditioning and factorisation as the forward
solve. One backward solve and one contraction return the gradient of a scalar
loss for every parameter of four closed-form classes (demand, emitter
coefficient, fixed head, pipe resistance) at memory cost independent of the
iteration count, the standard implicit construction of differentiable
optimisation layers \citep{amos2017optnet} applied to a new object.
Finite-differencing one loss against all 905 L-TOWN pipe roughnesses costs
906 simulations; one forward-plus-adjoint pair returns the same gradient, and
a full sensor-by-pipe Jacobian one adjoint solve per sensor
(Section~\ref{sec:res-batch}). Where the forward pass factorised a PRV
penalty matrix, naive reuse corrupts the demand gradient at the valve's
downstream node by order one while nothing visibly fails; a Woodbury row
replacement over the $p$ replaced rows restores relative error to
$1.4\times10^{-10}$ (Supplementary Notes~2 and~3). On clamped branches the
derivative is exactly zero because the engine's own guard is the definition;
status switching is combinatorial; the backward pass freezes the
configuration the forward pass converged to.

\subsection{The leak perturbation operator and the signature dictionary}
\label{sec:leakop}

Model-based leak localisation was posed on the steady-state equations by
\citet{pudar1992leaks} and carried into practice by sensitivity-matrix
methods \citep{perez2011methodology}; what follows changes its price and its
diagnosability, not its formulation.
EPANET represents a leak three ways: an emitter $q_{E,i}=C_ip_i^{\gamma}$,
pressure-driven demand \citep{wagner1988water}, or a known discharge added to
nodal demand; none touches an off-diagonal entry of $\mat{A}$. Inserting
a leak at junction $i$ is
$\mat{A}\mapsto\mat{A}+\beta_i\vect{e}_i\vect{e}_i^{\!\top}$ plus a
right-hand-side component: a rank-one diagonal update leaving the ordering,
fill-in and symbolic factorisation invariant, so a candidate sweep over $M$
nodes needs $M$ solves, not $M$ factorisations
\citep{sherman1950adjustment}, and the adjoint returns
$\partial L/\partial C_i$ for all of them simultaneously. Stacking the
normalised sensor responses of each candidate gives the signature dictionary
any sparse localisation method implicitly works with; its mutual coherence
\citep{tropp2007signal} decides whether sparse recovery can succeed;
Section~\ref{sec:res-limit} measures it, and grouping candidates into
coherent, network-adjacent clusters converts it into a group-sparse district
search (Supplementary Note~11). The exponent $\gamma$ is held fixed and only
the coefficient estimated (Supplementary Note~4).

\subsection{Batched scenarios on one GPU}\label{sec:batch}

The batched port runs EPANET's two cadences of discrete logic for the whole
batch by masks: status branches are all evaluated and combined by one-hot
selection, an active PRV becomes a penalty on the diagonal, and the loop ends
when every scenario is simultaneously converged and status-consistent. On the
networks carried in this paper's mainline and operating-network results the
assembled matrix is bit-wise symmetric and one factorisation serves both
passes; on two large benchmarks outside the batched path, parallel links in
mixed directions perturb $\mat{A}-\mat{A}^{\!\top}$ at the last bit, so the
reuse there is symmetric to rounding rather than bit-exactly. The sparse route builds the compressed-row pattern once (a leak,
a closed link or an active PRV changes only values) and feeds a stateful GPU
sparse direct solver whose symbolic analysis is planned once per batch shape
and cached; an undersized cache silently re-plans every step at
$4.0$--$4.5\times$ the cost. The backward pass adds no factorisation,
verified by independent counters. Unsupported features raise at construction
rather than degrade. Timings come from two RTX~5090 nodes measured
independently in double precision, ranges being minimum--maximum; memory is
peak device residency in one fresh process per cell (Supplementary
Notes~5--7).

\subsection{Calibration problem, baselines and fairness protocol}
\label{sec:calibsetup}

Calibration estimates Hazen--Williams roughness coefficients from noisy
junction heads: on Hanoi, 34 free pipes from 25 synthetic-diurnal frames at
31 junctions (25 training sensors, 20 training frames, 5 validation frames),
with $\sigma=0.1$~ft Gaussian head noise per seed; on City~D, 432 free pipes
after freezing 43 structurally unidentifiable ones. The gradient calibrator
runs Adam, then L-BFGS, then a Levenberg--Marquardt polish whose Jacobian
comes from the adjoint. Five tuned derivative-free baselines, differential
evolution (DE), particle swarm (PSO), CMA-ES, simulated annealing (SA) and a
DE$\to$Levenberg--Marquardt hybrid, were each tuned over four configurations
and three dedicated seeds
($\approx$240,000 model calls each on Hanoi; SA 60,576, reduced because it is
serial, and declared; 3,168--3,600 on City~D), then evaluated on 30 (Hanoi)
or 5 (City~D) fresh seeds at matched model-call budgets, with backward passes
costed at their measured $\approx$1\,\% of a forward call. Baselines received
batched forward evaluation, a more converged forward solve, ground-truth-prior
initialisation and generous snapshots; effect sizes are Vargha--Delaney
A12 with two-sided Wilcoxon tests, cross-checked against brute-force
implementations (Supplementary Note~8). Enhanced gradient arms, two
preconditioners and batched multi-start, share the same pipeline
(Supplementary Note~13).

\subsection{Sensor placement objective}\label{sec:placeobj}

Placement selects $k$ pressure-sensor junctions to make roughness
identifiable. From the adjoint-built sensitivity matrix of every candidate
junction head to every pipe parameter, a Bayesian D-optimal objective
(log-determinant of the prior-regularised Fisher information,
$\sigma_{\mathrm{prior}}=15$, $\sigma_{\mathrm{noise}}=0.1$~ft) is maximised
by lazy greedy search \citep{krause2008sensor}, with an optimality
certificate and a submodularity check. An augmentation mode holds an installed set fixed
and adds $k$ sensors greedily, for the same objective or for coverage (pipes
crossing the census threshold), asserting at every step that no pipe loses
identifiability; every added sensor is virtual (Supplementary Note~10).

\subsection{Verification protocol}\label{sec:verify}

Forward verification compares every network frame by frame against the
double-precision EPANET~2.2 dynamic library obtained through WNTR
\citep{klise2017wntr}, requiring $\max|\Delta H|<10^{-6}$~ft,
$\max|\Delta Q|<10^{-6}$~cfs and per-frame equality of time steps, statuses,
settings and Newton iteration counts. Every input field class, reservoir
heads and valve settings included, is read from the model file's text rather
than through a unit round-trip (Section~\ref{sec:res-fidelity}). Gradient
verification is four-fold:
Richardson-extrapolated central differences, an unrolled-vs-implicit
cross-check, a 22-item audit of degenerate cases (six on the
operating networks; all pass, the worst reaching 0.57 of its
threshold), and external checks driving finite differences through the compiled
reference library. A 54-item regression suite (53 numerical checks, all passing, including seven
on the operational models, plus one packaging guard) covers alignment,
replay, extended-period simulation and the symmetry and status-schedule
guards. The full protocol is Supplementary Note~9; per-network and
per-coordinate results are Tables~\ref{tab:full}--\ref{tab:regression}.

\section{Results}\label{sec:results}

Every result below reads one object: the Jacobian of the deployed engine's
converged state with respect to its parameters, delivered whole by the
implicit adjoint of Section~\ref{sec:adjoint}.

\subsection{Exact gradients of the solver as deployed}\label{sec:res-fidelity}

\begin{figure}[htbp]
\centering
\includegraphics[width=\linewidth]{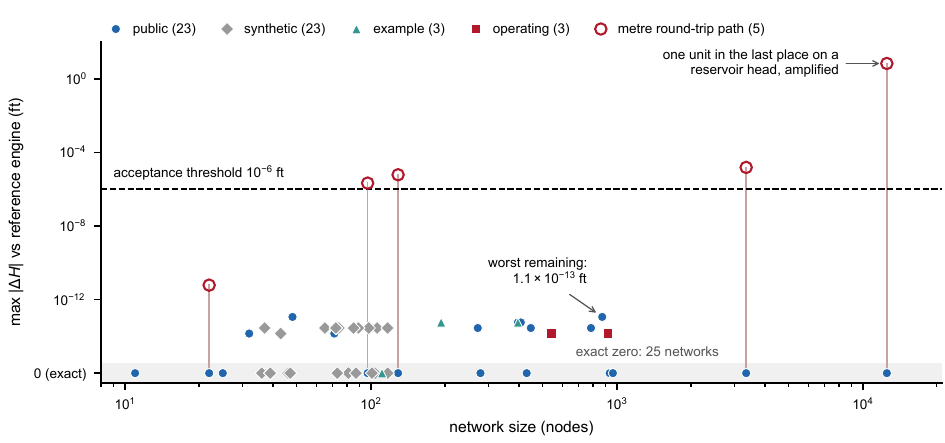}
\caption{Forward fidelity: per-network maximum head deviation from the
reference engine against network size, all 52 networks. Twenty-five agree
exactly (floor row) and the worst is $1.1\times10^{-13}$~ft; open circles
mark the five networks that reach the floor only when reservoir heads and
valve settings are read from the input text (Table~\ref{tab:exactinput}).}
\label{fig:fidelity}
\end{figure}

Fidelity is the warrant for transfer, and it is scoped: bit-level figures
belong to the serial replica path, the batched path that runs the decision
chain agrees within the conditioning envelope of Section~\ref{sec:gga}, and
``no surrogate residual'' means nothing beyond the reference engine's own
rounding, not GPU bit-identity. Over the 52-network suite and 8,140 frames,
all 52 meet the acceptance criterion; the largest head deviation is
$1.137\times10^{-13}$~ft (flow $1.421\times10^{-14}$~cfs), 25 agree
\emph{exactly}, and per-frame Newton iteration counts are identical on all 52
(Table~\ref{tab:fidelity}, Fig.~\ref{fig:fidelity}). One input path
deserves a caution. Every field class the solver reads can be rebuilt bit for
bit from the input text, but a parser that takes reservoir heads and valve
settings through a metre representation and back loses one unit in the last
place on models written in US customary units. A reservoir head is a boundary
condition, and the suite's largest network (12,527 nodes) amplifies one:
perturbing a single roughness value inside the reference library by one unit
in the last place moves its own answer by $1.3\times10^{-6}$ to 6.9~ft. Reading
those two field classes from the input text is what brings five networks to
zero over every frame, one of them a 609-frame model, and it changes neither
a field nor a digit on eight control networks (Table~\ref{tab:exactinput},
Supplementary Note~1). The $1.137\times10^{-13}$~ft that remains has another
cause.

\begin{table}[htbp]
\centering\footnotesize
\caption{Forward fidelity against EPANET~2.2 over the 52-network suite
by provenance. Deviation columns are maxima over every frame of
every network, none excluded, with reservoir heads and valve settings
read from the input text (Section~3.1); five rows reach these values
only that way (Table~\ref{tab:exactinput}).
Per-network rows: Table~\ref{tab:full}.}
\label{tab:fidelity}
\begin{tabular}{lrrrrr}
\toprule
Provenance & Networks & Frames & $\le10^{-12}$~ft & Exact zero & Worst $|\Delta H|$ (ft) \\
\midrule
Public benchmark models & 23 & 7,964 & 23 & 13 & $1.137\times10^{-13}$ \\
EPANET distribution examples & 3 & 78 & 3 & 1 & $5.684\times10^{-14}$ \\
Randomly generated & 23 & 23 & 23 & 11 & $2.842\times10^{-14}$ \\
Operational (City D, variant, City H) & 3 & 75 & 3 & 0 & $1.421\times10^{-14}$ \\
\midrule
All & 52 & 8,140 & 52 & 25 & $1.137\times10^{-13}$ \\
\bottomrule
\end{tabular}
\end{table}

The backward pass meets the same standard. Against Richardson-extrapolated
central differences the implicit adjoint reaches a worst relative error of
$5.9\times10^{-8}$ over 61 coordinates and four parameter classes per network,
the operating network included, against a $10^{-6}$ threshold. Externally,
with finite differences driven through the compiled EPANET library, eight
coordinates on the public Hanoi model with five emitters agree to a worst
$5.56\times10^{-5}$ against a $10^{-4}$ threshold, and 24 non-clamped City~D
pipes spanning 1.7 decades of sensitivity to $4.78\times10^{-5}$; a second
check on the emitter variant covers four parameter classes to
$2.39\times10^{-5}$. The six clamped pipes are proven signal-free: the
analytic gradient is exactly zero while the finite-difference residual, at
most $6.45\times10^{-7}$, matches the independently predicted noise floor
and sits a factor $\ge1.6\times10^{4}$ below the smallest non-clamped
gradient.

\subsection{What batching buys}\label{sec:res-batch}

\begin{figure}[htbp]
\centering
\includegraphics[width=\linewidth]{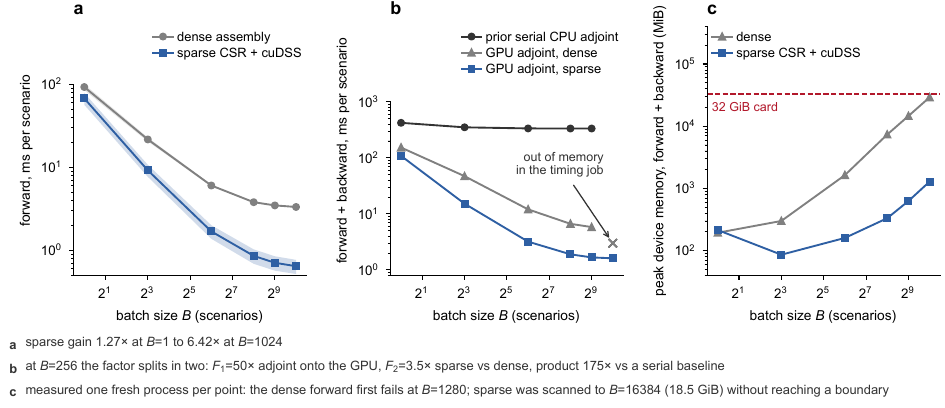}
\caption{Cost on the public 782-junction L-TOWN benchmark; bands are
minimum--maximum over two GPU nodes, double precision. \textbf{a},~Forward
wall time per scenario against batch size. \textbf{b},~Forward-plus-backward
time, the end-to-end factor a product of two measured factors against a
serial CPU-adjoint baseline. \textbf{c},~Peak device memory.}
\label{fig:cost}
\end{figure}

On L-TOWN (Fig.~\ref{fig:cost}, Table~\ref{tab:ltowncost}), one
forward-plus-backward scenario falls
from 333~ms to 1.90~ms at $B=256$ (175--176$\times$), peak memory at
$B=1{,}024$ is 22.6$\times$ below the dense path's, and the sparse-over-dense
crossover sits near 300 junctions. In a real inversion loop the steady state
costs 463--470~ms per optimiser step for 256 scenarios, 190.7--191.6$\times$
the prior pipeline. A whole City~D
sensitivity matrix is 20.0--21.4$\times$ faster than sequential backward
passes, 1.22--1.32$\times$ end to end.

Four qualifications travel with these factors. First, the end-to-end factor is
two things never to be quoted as one: the GPU adjoint contributes
$F_1=2.8\times$ at $B=1$ rising to $57\times$ at $B=512$, sparse-versus-dense
linear algebra $F_2=3.1$--$3.8\times$ for $B\ge8$. Second, the baseline is our
own prior serial CPU-adjoint pipeline, not EPANET, which supplies no gradient
at any price; a perfectly eight-way-parallel CPU adjoint would divide these
factors by up to eight, arithmetically rather than by measurement. Third, the win is
not universal: below the crossover the sparse route is \emph{slower} than the
dense one, down to $0.06\times$ at $B=1{,}024$ on a nine-junction network.
Fourth, per single solve the dense default path is slower than the reference
engine, by 26 to 73$\times$ across a 14-network public size sweep
(Table~\ref{tab:sweep}).

\subsection{Calibration: least squares on the gradient object, at matched
budget}\label{sec:res-calib}

\begin{figure}[htbp]
\centering
\includegraphics[width=\linewidth]{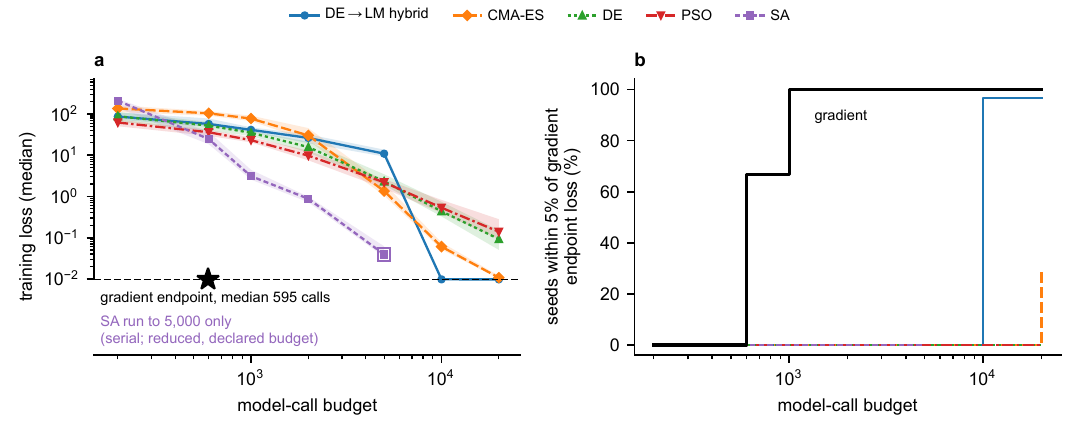}
\caption{Calibration at matched model-call budget (Hanoi, $\sigma=0.1$~ft,
30 seeds). \textbf{a},~Median training loss against budget for five tuned
baselines; the star marks the gradient endpoint (median 595 calls).
\textbf{b},~Fraction of seeds reaching within 5\,\% of the gradient
endpoint; SA is right-censored at its declared 5,000-call ceiling.}
\label{fig:calib}
\end{figure}

Calibration under uncertainty is a live problem here
\citep{kerimov2023assessing,du2026decoupling}. On Hanoi at $\sigma=0.1$~ft
over 30 seeds, the gradient calibrator reaches its endpoint in a median 595
model calls. At every budget from 200 to 5,000 calls all five tuned baselines
are worse on every metric, the training-loss separation total
(Vargha--Delaney A12~$=0.00$, Wilcoxon $p=1.86\times10^{-9}$;
Fig.~\ref{fig:calib}). Read as run length, the
DE$\to$LM hybrid reaches within 5\,\% of the gradient endpoint on 96.7\,\% of
seeds at a median 8,060 calls; CMA-ES on 30\,\%; DE and PSO never within
20,000; and SA (serial, with a reduced, declared budget) not within the 5,000
calls it was run to. At 10,000 and 20,000 calls the hybrid closes the gap: the
effect size collapses to negligible (A12~$=0.46$ and $0.47$) while the
difference stays detectable ($p=3.05\times10^{-5}$ and $0.031$), the gradient
endpoint still marginally ahead. Its refinement stage uses our own adjoint
Jacobian, so the strongest baseline is itself a consumer of the instrument
under test, and the equal-budget claim is bounded: it holds for 200--5,000
calls and should not be quoted beyond its range.

On City~D, on the utility's own diurnal pattern, 432 free pipes recover with
median final training error $8.4\times10^{-4}$, $9.4\times10^{-3}$ and
$8.5\times10^{-2}$ under head noise $\sigma\in\{0.03,0.1,0.3\}$~ft at five
seeds each (Table~\ref{tab:calib}), one full 432-pipe calibration costing
220--231~s of wall clock.
With truth demands perturbed by $\pm15\,\%$ and the inversion run on nominal
demands the median degrades only from $9.11\times10^{-3}$ to
$9.51\times10^{-3}$, though one of four repeats reaches $1.62\times10^{-2}$;
5\,\% sensor bias yields a median $9.86\times10^{-3}$. Eight Latin-hypercube
starts land within a $1.0087\times$ spread of final training error, so the
endpoint is not a local-minimum artefact. In the one configuration where a baseline wins,
the DE$\to$LM hybrid at a budget of 2,000 calls edges the gradient endpoint
($0.0094$ vs $0.00941$, A12~$=0.60$, $p=0.0625$, the attainable floor at
$n=5$), reported as a baseline win.

Two controls bound that comparison. Tuning budget: the City~D baselines are
tuned at 2,000 calls, equal to their evaluation budget, over four
configurations and three seeds, and each of the four algorithms selects the
same configuration as under the 3,168--3,600-call tuning of
Section~\ref{sec:calibsetup}; differential evolution at ten times the budget
still ends at $1.4\times10^{-2}$. Wall clock: 631~s for the gradient
configuration's 200 calls against 4,756~s for the hybrid's 1,997 (one seed);
divided by the eight-way parallelism the fairness protocol grants the
baseline, the hybrid's figure becomes 595~s, a tie, so the equal-budget claim
rests on the call count. Enhancements to the gradient search do not extend
it. A Schur-complement diagonal preconditioner improves the roughness error
on the design's own subspace, 17.71 to 12.40, on 10 of 10 seeds
($p=0.0020$), but on the layout-independent ruler of
Section~\ref{sec:res-place} the same comparison is 7 of 10 ($p=0.34$); an
exact Gauss--Newton diagonal is worse than the cheap approximation, and
batching $B$ starts buys at most $2.8\times$ per model call, never
$B\times$, at a cost in accuracy. No arm beats the 200-call gradient
configuration, which leads on both rulers (Supplementary Note~13,
Table~\ref{tab:optimiser}).

\subsection{Leak search with the same object's columns}\label{sec:res-leak}

\begin{figure}[htbp]
\centering
\includegraphics[width=\linewidth]{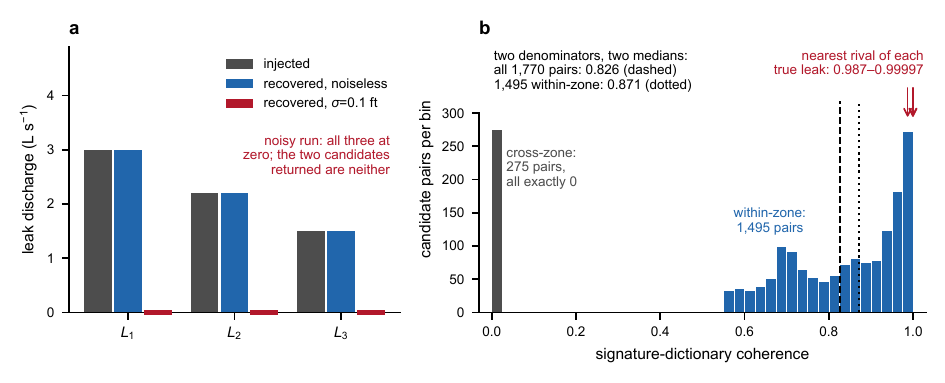}
\caption{Success and failure of leak search, and the mechanism.
\textbf{a},~Work-order search on the operating network: injected versus
recovered discharge for $L_1$--$L_3$, exact without noise and empty with it.
\textbf{b},~Signature-dictionary coherence on the L-TOWN configuration,
1,770 candidate pairs, both denominators marked; arrows mark each true
leak's nearest rival.}
\label{fig:leak}
\end{figure}

Because a leak is a rank-one update (Section~\ref{sec:leakop}), each
candidate contributes one column to the same Jacobian and one adjoint prices
them all at once. The test is candidates a crew would dig: 49 junctions
distilled from a year of the utility's work orders, 40 pressure sensors and
three injected leaks of 3.0, 1.5 and 2.2~L\,s$^{-1}$ on City~D.
The noiseless inversion recovers the exact three-leak support with a maximum
relative discharge error of $5.9\times10^{-7}$ and residual leakage exactly
zero on all 46 other candidates. That recovery is an inverse crime by construction, the leaks
being injected into the same model the inversion searches, and it does not
survive noise: under $\sigma=0.1$~ft the same inversion recovers none of the
three, returning two candidates that are neither of them
(Fig.~\ref{fig:leak}a).

The bracket is not specific to that network. On L-TOWN the test uses three
self-injected leaks rather than the BattLeDIM competition scenarios, so
published localisation results \citep{daniel2022lila} set context without
being numerically comparable; a 60-candidate, 33-sensor inversion returns a
true leak node as its largest coefficient but not the full support, the other
two finishing 19th and 32nd, beneath their coherent neighbours. The failure
is not an optimiser pathology but a measurable property of the columns.

\subsection{Where the search stops: the coherence of the columns}
\label{sec:res-limit}

The signature dictionary's mutual coherence on this L-TOWN configuration has
maximum $0.9999999981$ and median 0.8258 over all 1,770 candidate pairs,
including 275 made exactly orthogonal by a PRV, or 0.8713 over the 1,495
within-zone pairs: two denominators, both reported
(Fig.~\ref{fig:leak}b). Forty-two pairs, all within-zone, exceed 0.999, and
each true leak has an inseparable rival at 0.999965, 0.999231 and 0.987023;
the spurious recoveries are the high-coherence neighbours of the one node
found. Coherence this close to one violates every sufficient condition for
sparse recovery, so the inversion stops where the dictionary says it must,
and the whole check costs under a minute of single-process CPU time.

The same boundary in a crew's units: the candidates coherence above 0.99 makes
indistinguishable from L-TOWN's two lost leaks carry 984 and 512~m of
connected pipe, while the leak the inversion returns has no rival above that
threshold and 96~m at stake. The work-order dictionary (49 candidates,
40 sensors) reads the same way, with a median coherence of 0.934, 51 pairs
above 0.999 and a rival at 0.99999 for the 1.5~L\,s$^{-1}$ leak: both
failures are diagnosed post hoc by one quantity.

Ambiguity of that size is a unit of work. Grouping candidates mutually
coherent above a threshold $\tau$ and adjacent in the network's Voronoi
partition, then replacing the sparse penalty by its group form, searches
districts, not junctions (Supplementary Note~11 and Table~\ref{tab:cluster}).
The trade-off, not the hit rate, is the result. On City~D the single-point
inversion ranks the three injected leaks 49th, 15th and 16th of 49; at
$\tau=0.999$ two of the three fall inside the top three clusters, which hold
seven candidates and 7.42~km of main, 8.8\,\% of the network, the largest
750~m in radius. On
L-TOWN $\tau=0.995$ returns all three inside 22 clusters averaging 98~m.
Neither figure means anything without the radius: at $\tau=0$ City~D is one
9,043~m cluster with a hit rate of 3 of 3, the hit rate of naming the whole
network. Thresholds are chosen after seeing the tabulated sweep; against 20
size-matched random groupings that keep the cluster sizes and destroy topology
and coherence the design is never significant ($p\ge0.095$) and wins only on
radius, 179 against 1,700~m. The leak the amplitude limit defeats
stays defeated: signal-to-noise 0.50 against 166 and 151, a singleton at every
$\tau\ge0.5$, and 49th to 36th even when forced in with its twelve most
coherent neighbours. Coherence is a property of the sensor set as much as of
the network, which makes it a design variable rather than a fixed limit.

\subsection{Sensor placement: a prescription, and how far it closes the
loop}\label{sec:res-place}

\begin{figure}[htbp]
\centering
\includegraphics[width=\linewidth]{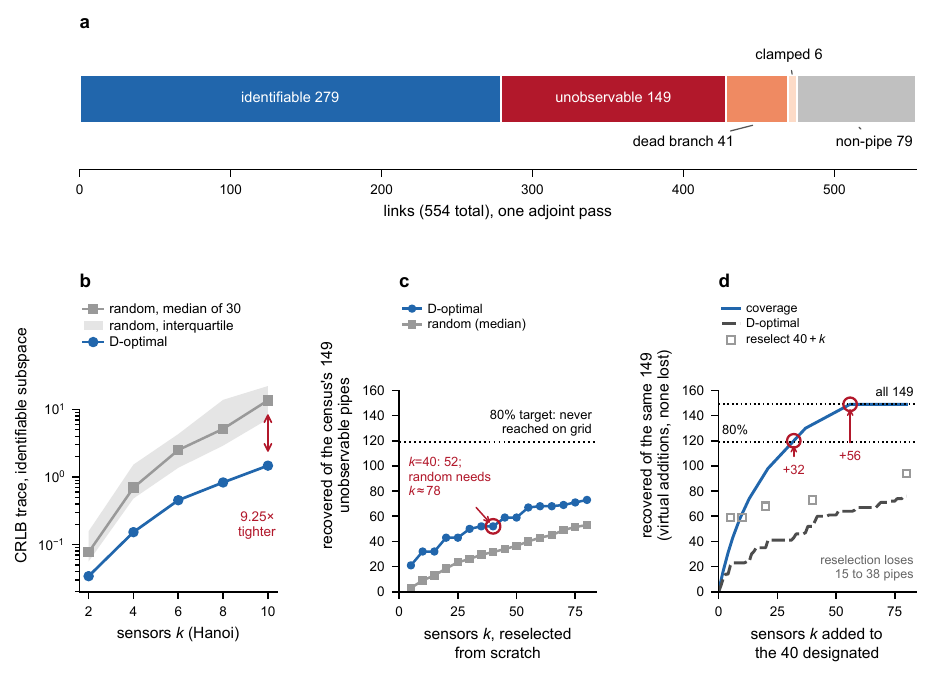}
\caption{Placement as a prescription, by counts only.
\textbf{a},~Gradient-attribution census of the operating network's 554
links. \textbf{b},~Identifiable-subspace CRLB trace against sensor count on
Hanoi, D-optimal versus 30 random layouts. \textbf{c},~Recovery of the 149
unobservable pipes by reselection from scratch. \textbf{d},~The same with the
40 designated sensors kept and $k$ added.}
\label{fig:place}
\end{figure}

Pressure-sensor placement is a standing topic here
\citep{zhou2024allpurpose,cheng2024graphlaplace} and information-based
sampling design is older still \citep{kapelan2005optimal}; what differs is
the input, exact adjoint sensitivities of the deployed engine rather than a
graph proxy, and the certificate. (No graph-spectral baseline is included, since only a
proxy of the published method would be available for comparison.)
On City~D, a gradient-attribution census (one adjoint pass over 25 frames)
partitions all 554 links into 279 whose roughness the designated 40-sensor
data can identify, 149 it cannot, 41 dead branches, 6 clamped at the low-flow
threshold and 79 non-pipe links (Fig.~\ref{fig:place}a); the zero classes sum
exactly to the 196 all-zero sensitivity columns, on each of which central
differences also return zero. The census precedes any calibration and
reveals, rather than causes, an unidentifiability the sensor configuration
already determines.

On Hanoi the discriminating number is the identifiability bound: at $k=10$
the D-optimal design attains an identifiable-subspace CRLB trace of 1.4746
against a random median of 13.642 (Table~\ref{tab:placement},
Fig.~\ref{fig:place}b), with a worst-case
optimality-certificate ratio of 0.763, zero submodularity violations in 200
samples and an 8-ms search. Hanoi has no structurally unobservable pipe, so
``zero unobservable'' claims would be vacuous there.

\begin{table}[htbp]
\centering\footnotesize\setlength{\tabcolsep}{4.5pt}
\caption{Sensor placement for roughness identifiability. Hanoi:
identifiable-subspace CRLB trace, Bayesian D-optimal greedy against
the median of 30 random layouts. Operating network: of the 149 pipes
unobservable under the 40 designated sensors, how many a layout
recovers and loses. Augmentation keeps the 40 and adds $k$;
every design is virtual.}
\label{tab:placement}
\begin{tabular}{lrrrrr}
\toprule
\multicolumn{6}{l}{\emph{Hanoi} (34 pipes, 31 candidate sites, 1 frame): CRLB trace}\\
$k$ & 2 & 4 & 6 & 8 & 10 \\
\midrule
D-optimal & 0.034 & 0.153 & 0.454 & 0.832 & 1.475 \\
Random (median) & 0.077 & 0.696 & 2.509 & 5.164 & 13.642 \\
\midrule
\multicolumn{6}{l}{\emph{Operating network} (475 pipes, 541 candidate sites, 25 frames): of the 149 unobservable}\\
$k$ sensors added to the 40 & 5 & 10 & 20 & 40 & 80 \\
\midrule
Augment, coverage objective: recovered & 37 & 62 & 95 & 133 & 149 \\
Augment, D-optimal: recovered & 21 & 23 & 35 & 52 & 76 \\
Augment, either objective: lost & 0 & 0 & 0 & 0 & 0 \\
Reselect $40+k$ from scratch: recovered / lost & 59/38 & 59/38 & 68/24 & 73/21 & 94/15 \\
Still unobservable (cover. / D-opt. / resel.) & 112/128/128 & 87/126/128 & 54/114/105 & 16/97/97 & 0/73/70 \\
\midrule
\multicolumn{6}{l}{Reselect 40 from scratch ($k=0$): 52 recovered, 38 lost, 135 still unobservable; random median 31.5 recovered}\\
\bottomrule
\end{tabular}
\end{table}

On City~D the question is which sensors to add to the 40 the utility has, not
where 40 would go if none existed. Reselecting from scratch answers the
latter: a $k=40$ D-optimal layout over 541 candidate junctions recovers 52 of
the 149 unobservable pipes against a random median of 31.5 but loses 38 the
designated sensors could see, so the unobservable total falls only to 135, and
no from-scratch layout on the $k\le80$ grid reaches 80\,\% recovery
(Fig.~\ref{fig:place}c). The prescription answers the former: the 40 stay and
$k$ are added greedily, for the D-optimal objective or for coverage under the
census's threshold, with the identifiable set asserted never to shrink (zero
violations in 80 steps). Coverage restores 37 of the 149 at $k=5$ and all 149
at $k=80$, D-optimal 21 and 76, neither losing a pipe at any budget
(Table~\ref{tab:placement}, Fig.~\ref{fig:place}d, Table~\ref{tab:augment}). On
L-TOWN, adding to the 33 sensors restores all 62 pipes they could not inform
at 36 coverage additions, where reselecting 33 from scratch loses 99
(Supplementary Note~10).

The downstream gain is graded on a ruler that does not move with the design.
Roughness error over a layout's own informative pipes is not comparable across
layouts, that set growing with the sensors added from 279 pipes to 412, so
designs are scored on the roughness error projected onto a reference subspace
fixed before any design was chosen: the 98 directions a fully instrumented
candidate pool would identify, against a prior of 28.92 (Supplementary
Note~12). Head misfit ranks nothing here: at 0.1~ft it is 96\,\% noise
variance, and over eight restarts on identical data it moves 0.4\,\% where
this roughness error moves 25\,\%. On that ruler coverage $+20$ scores 17.45
and D-optimal $+20$ 18.04 against a random median of 21.73, no draw among 100
random additions from the same pool at the same budget matching either
($p=0.0099$, the floor at $R=100$); all twelve budget-by-noise cells agree,
five only to the coarser floor $R=20$ allows (Tables~\ref{tab:augcalib}
and~\ref{tab:identmetric}). Two costs travel with it: in the 334 discarded directions the
designs do not beat random (60 of 100 at least as good), and in 22 of 48
cell-design-metric combinations the design leads every random layout on
roughness while trailing most on heads, so choosing a layout by head misfit
chooses backwards. Repeating the work-order
search with the augmented sets, same injected triple, noise and inverse-crime
setting, does not recover the leaks 40 sensors lost: on L-TOWN all ten
augmented sets return the same one of three, and on City~D every set returns
$L_2$ with 1--4\,\% discharge error, a hit a control attributes to an added
sensor on $L_2$'s own junction. $L_1$ moves no head by more than 0.09~ft,
below the noise; $L_3$ is recovered by three of five random $+20$ sets and
none of the six designed (Table~\ref{tab:leakrerun}). Placement for roughness identifiability does not close the
leak-search loop. A second objective, additions maximising the same
dictionary's coherence, does better: pooling every designed City~D set against
40 random draws, $L_2$ recovers 12/14 against 3/40 ($p=1.05\times10^{-7}$),
and what sorts recovery is each set's own rival-coherence gap, which separates
every recovering configuration from every losing one across all 55 sets
whatever objective chose it. $L_3$'s recovery is indistinguishable from
random ($p=0.51$) and $L_1$ by neither, its column norm an order of magnitude
below $L_2$'s: an amplitude limit, not a coherence one. Both
conclusions hold at the coarser stage the L-TOWN inversion shares, where the same
mechanism bounds the loop: of 746 positions only one and 14 separate its two
unrecovered leaks from their nearest rival (Table~\ref{tab:coherence}).

\section{Discussion}\label{sec:discussion}

Calibration, leak search and sensor siting were three separate black-box
searches because the engine defining them exposed no derivatives. With
derivatives exposed they are three readings of one object. The same coherence
that forbids a junction-level answer sizes the district-level one that
survives it: on the operating network the ambiguity a crew must walk is
7.42~km of main, not 84. The placement reading is a prescription, not a
redesign: the utility's sensors stay, the adjoint names the additions, and the
census's unidentifiable pipes become identifiable, though only additions
against coherence itself recover any leak noise defeated. Because the forward
pass reproduces the reference engine to $1.137\times10^{-13}$~ft, every link
transfers to the model a utility already operates with no surrogate residual,
and the census and coherence reading arrive as by-products of one pass, not
assumptions.

The forward-acceleration literature is orthogonal and the borrowing bounded.
The batch path is limited by per-matrix host dispatch, not arithmetic;
removing that floor is worth an estimated 4--9$\times$, an estimate rather
than a measurement. Three incompatibilities bound what the decomposition line
\citep{guo2024decomposed} can lend: it multiplies the per-matrix bottleneck
count, early stopping breaks the implicit adjoint's premise, and any
partition changes the elimination order the bit-faithful path reproduces.

Adoption asks a utility for a hydraulic model inside the documented feature
subset, the pressure records it logs and one CUDA GPU. The gradients are
those of steady-state frames: extended-period simulation is forward-verified,
but storage coupling between frames is not differentiated through, so
multi-day control synthesis is out of scope.

The binding limits bear repeating. Bit-level agreement is a claim about
one reference binary on one platform; elsewhere it degrades to
$\approx10^{-6}$~ft. Gradients are gradients of a frozen status
configuration, one-sided where a perturbation would flip a status. The
batched differentiable path covers a documented feature subset, refusing
what it does not implement. The speed factors carry the four
qualifications of Section~\ref{sec:res-batch}, including that the dense
default path is slower per solve than the reference engine. The
equal-budget calibration claim is bounded at 5,000 calls and rests on the
call count: on the wall clock the strongest baseline, given the eight-way
parallelism it is promised, draws level. The leak
demonstrations are one network and one injected triple each, with the
noiseless operating-network recovery an inverse crime whose noisy
counterpart fails; identifiability-driven additions did not change that
outcome, coherence-driven additions changed it for one of the three, for
the reason coherence predicts; what generalises is the diagnosis, not the
recovery. The cluster-level reading buys hits with inspection radius, at a
threshold chosen after the sweep, and never separates from size-matched
random grouping on hit rate alone. Placement gains are established on a
projected roughness error under a single noise realisation, and not in the
directions that projection discards. The sensor additions are
virtual, designed and verified in simulation, none installed. This is
research code, not an operational product.

\section{Conclusions}\label{sec:conclusions}

An exactly differentiable global gradient algorithm, verified against
EPANET~2.2 over 52 networks (to $1.137\times10^{-13}$~ft at worst, none
excluded) and batched on one GPU, hands three standing WDN
decisions one gradient object. Calibration reaches its endpoint in a median
595 model calls where the best of five tuned metaheuristics needs 8,060; leak
search runs over a utility's own work-order candidates, and where noise
defeats a junction-level answer the same object's column coherence says so in
advance and sizes the district-level answer that survives; sensor siting keeps
an operating network's 40 sensors and adds the few that restore, in
simulation, the identifiability of the pipes their data could not inform,
without recovering the leaks noise defeated. The same adjoint prices its own
limits, so a utility can decide where sensors, crews and trust should go
before committing them.

\section*{CRediT authorship contribution statement}

\textbf{T.\,Mu}: Conceptualization, Methodology, Software, Validation,
Investigation, Visualization, Writing -- original draft.
\textbf{Y.\,Wang}: Data curation, Investigation, Writing -- review \&
editing. \textbf{M.\,Yuan}: Conceptualization, Supervision, Funding
acquisition, Writing -- review \& editing. \textbf{W.\,Wang}: Software,
Formal analysis, Writing -- review \& editing. \textbf{Q.\,Luo}: Resources,
Funding acquisition, Writing -- review \& editing. \textbf{M.\,Xiao}:
Validation, Formal analysis. \textbf{J.\,Li}: Investigation, Resources.
\textbf{H.\,Yang}: Supervision, Project administration.
\textbf{M.\,Huang}: Supervision, Writing -- review \& editing.

\section*{Declaration of competing interest}

The authors declare that they have no known competing financial interests or
personal relationships that could have appeared to influence the work
reported in this paper.

\section*{Data availability}

The two operational network models, and the year of leak repair records
drawn from one of them, are released with the permission of the operating
utility, in anonymised form under CC~BY~4.0. The public networks are fetched from their upstream sources and
verified against recorded SHA-256 digests rather than redistributed; sources
and licences are listed with the code. The randomly generated networks, their generator, the
emitter-augmented Hanoi configuration and every measurement report cited here
are released with the code \citep{mu2026hydrograd}: development is public at
\url{https://github.com/mutianwei521/wdsgpu} and the submitted
version is archived with a DOI at Zenodo. L-TOWN is available under
CC~BY~4.0 as Zenodo record 4017659 \citep{vrachimis2020dataset}. HydroGrad is
the Python package \code{dgga}, MIT licence; the sparse route requires CUDA
and reference cross-checks the EPANET~2.2 library obtained through WNTR.

\section*{Acknowledgements}

This work was funded by the open fund of the Key Laboratory of Ecological
Restoration of Regional Contaminated Environment (Shenyang University),
Ministry of Education (grant KF-26-11), and by the Guangdong Province Natural
Science Foundation General Project (2026A1515011817). HydroGrad
re-implements the EPANET~2.2 hydraulic engine from its MIT-licensed C
sources; it is not endorsed by the US Environmental Protection Agency or the
OpenWaterAnalytics community. WNTR supplied input parsing and the reference
library.

\clearpage
\newgeometry{a4paper,margin=22mm}
\renewcommand{\baselinestretch}{1.15}\normalsize
\setcounter{table}{0}
\setcounter{figure}{0}
\renewcommand{\thetable}{S\arabic{table}}
\renewcommand{\thefigure}{S\arabic{figure}}
\captionsetup[table]{name=Table,labelsep=period}
\captionsetup[figure]{name=Figure,labelsep=period}

\begin{center}
{\Large\bfseries Supplementary Information}\\[6pt]
{\large Water-network decisions share one hydraulic gradient,\\
and it can now be computed exactly}\\[8pt]
Tianwei Mu, Yue Wang, Mingzhe Yuan, Wenhong Wang, Qing Luo, Min Xiao, Jun Li, Hui Yang, Manhong Huang
\end{center}

\vspace{-2mm}
\noindent\rule{\linewidth}{0.4pt}
\vspace{2mm}

\noindent\textbf{Contents}
\begin{itemize}[leftmargin=1.4em,itemsep=1pt,topsep=3pt]
\item \textbf{Supplementary Notes 1--13}: the implementation, derivation
      and protocol detail that Section~2 of the main text compresses:
      reproducing the reference arithmetic, including the input round-trip
      that costs one unit in the last place on a boundary condition
      (Note~1), the two differentiation
      routes (Note~2), the big-$M$ pressure-reducing-valve penalty and its
      Woodbury correction (Note~3), the leak perturbation operator and the
      signature dictionary (Note~4), the batched status machine (Note~5),
      sparsity-pattern invariance and the stateful sparse solver (Note~6),
      the cost decomposition (Note~7), the calibration comparison protocol
      (Note~8), the verification protocol in full (Note~9), the sensor
      augmentation protocol with calibration and leak search under the
      augmented sets, control experiments and the coherence-driven placement test
      (Note~10), cluster-level leak diagnosability and its trade-off curve
      (Note~11), the layout-independent metric that scores sensor designs
      (Note~12) and the optimiser enhancements, the baseline tuning budget
      and the wall-clock reading (Note~13).
\item \textbf{Supplementary Discussion}: reference solutions that must not
      be used as training labels; the relation to learned surrogates; and the
      complete list of limitations.
\item \textbf{Figure~\ref{fig:networks}}: the drawn topology of nine of the
      verified networks, ordered by node count, three orders of magnitude
      of size in one plate.
\item \textbf{Tables~\ref{tab:full}--\ref{tab:optimiser}}: the per-network fidelity suite
      (Table~\ref{tab:full}), truncated-unrolling health (Table~\ref{tab:health}), the external
      per-coordinate gradient checks (Table~\ref{tab:external}), the 22-item degenerate-case audit
      (Table~\ref{tab:audit}), the 54-item regression suite (Table~\ref{tab:regression}), L-TOWN cost and
      memory (Table~\ref{tab:ltowncost}), the 16-network size sweep (Table~\ref{tab:sweep}), the
      calibration noise ladders (Table~\ref{tab:calib}), the sensor-augmentation account
      (Table~\ref{tab:augment}), calibration with the augmented sets
      (Table~\ref{tab:augcalib}), leak search with the augmented sets and
      its controls (Table~\ref{tab:leakrerun}), the coherence-driven placement test
      against random controls (Table~\ref{tab:coherence}), the exact-input control and its
      negative control (Table~\ref{tab:exactinput}), the cluster-level trade-off curve
      (Table~\ref{tab:cluster}), the layout-independent placement metric across every
      budget-by-noise cell (Table~\ref{tab:identmetric}) and the optimiser arms with the
      wall-clock reading (Table~\ref{tab:optimiser}).
\end{itemize}

\vspace{2mm}
\noindent\textbf{Provenance and anonymisation.} Every table and every
measured value quoted in the prose is generated by a single script from the
recorded measurement files it names, and the two operational
distribution models are carried at the full 52-network calibre of the main
text: \WrNets{} networks, \WrFrames{} frames. Throughout, the operational
models City~D and City~H are reported by counts and positional labels only;
no model node, link or sensor identifier of an operating network appears in
any table or sentence, and rows summarised by sensitivity band rather than
listed per coordinate (Table~\ref{tab:external}b, c) are summarised for exactly
that reason. Each model is released with the permission of the operating
utility.

\clearpage
\section*{Supplementary Notes}

\snote{1: reproducing the reference arithmetic}

\paragraph{Two forward paths, one parser.} The implementation offers an
\emph{EPANET-replica} path, which follows the reference control flow statement
by statement and carries every bit-level claim in this work, and a \emph{dense
batched} path, which assembles the Schur complement as a dense tensor by an index-scatter accumulation and solves many scenarios at once. They share one parser
and one set of coefficient formulas. Transcribing the head-loss laws, the pump
curves and the valve coefficients is mechanical; what separates $10^{-6}$~ft
agreement from $10^{-14}$~ft agreement is five things that are invisible in
the mathematics.

\paragraph{(i) Operation ordering in the linear solve.} EPANET performs one
symbolic factorisation at load time. It builds a de-duplicated adjacency list
(parallel pipes share a single off-diagonal slot and their conductances are
added), reorders the junction subgraph by multiple minimum
degree~\citep{liu1985modification}, simulates the elimination graph to
allocate fill-in slots, and stores the strictly lower triangle in compressed
form. The numerical phase is a left-looking Cholesky with $\sqrt{\cdot}$
adapted from \citet{george1981computer}. Both the elimination order and the
loop structure of the column algorithm are ported verbatim, because any other
ordering, and any other loop structure over the same sparsity pattern,
produces a different sequence of floating-point additions, and at the
conditioning quantified below that is not a difference in the last bits of the
head.

\paragraph{(ii) The transcendental library.} The reference engine's Windows build bundled with WNTR~\citep{klise2017wntr} is a MinGW-family build whose statically linked power routine agrees bit for bit with the one exported by the Windows C runtime and differs from other implementations, on a small
fraction of arguments, by one unit in the last place. Because the
Hazen--Williams resistance $r = 4.727\,L\,C^{-1.852}\,d^{-4.871}$ is evaluated
once per pipe and then multiplies every downstream quantity, a one-unit
difference there is amplified by the conditioning of the assembled system.
Every call to the power and logarithm routines is therefore routed through the same C runtime as the reference build; a vectorised power function from an
array library is not sufficient, and the residual it leaves grows with the
iteration count. This is the reason the bit-level claims are stated for one
platform and one binary, and for nothing else.

\paragraph{(iii) Literal constants, including the inexact ones.} The reference build was compiled without the C runtime's $\pi$ constant defined and uses the fallback literal
$\pi \approx 3.141592654$ from its own header rather than the nearest double.
The initial flow assignment $\pi d^{2}/4$ then differs in its last bits, and
the difference is amplified through the first iterations. The literal is
reproduced.

\paragraph{(iv) Accumulation fold order.} The stopping statistic is the
relative flow change, $\sum_k|\Delta Q_k|$ divided by $\sum_k|Q_k|$.
When its value sits near the convergence tolerance (the engine's ACCURACY option), a different summation order flips the iteration count, and a different iteration count means the two
implementations are no longer answering the same question. The left-fold order
of the reference C loop is reproduced exactly.

\paragraph{(v) Parsing semantics, including the clamps.} EPANET does not
necessarily solve the problem the input file describes. The ACCURACY option read from an input file is clamped to $[10^{-5},10^{-1}]$ during
parsing, while the same option set through the programmatic interface is not,
so a file requesting $10^{-8}$ is silently given $10^{-5}$. A
re-implementation that honours the file literally converges to a different
stopping point with every formula correct. The clamp is reproduced.

\paragraph{Why care with the formulas cannot substitute.} Let
$\hat{\vect{H}}$ be the head produced by a backward-stable double-precision
solve. Standard backward error analysis gives
\begin{equation}
  \frac{\|\hat{\vect{H}} - \vect{H}\|}{\|\vect{H}\|}
  \;\lesssim\; c(N)\,\varepsilon\,\kappa_2(\mat{A}),
  \qquad \varepsilon = 2^{-53},
  \label{eq:backward}
\end{equation}
with $c(N)$ a modest polynomial in the dimension. The conditioning is set by
the guard constants of the main text's Section~2.2: a closed link contributes
a conductance of $10^{-8}$, a fully open valve $10^{6}$, and an active pressure-reducing valve (PRV) a penalty of $10^{8}$ on the diagonal, the engine's large-coefficient constant (CBIG), so a single assembled matrix spans sixteen orders of magnitude. On
L-TOWN with all three PRVs active the converged Schur complement has
$\kappa_2(\mat{A}) = \SnKappaLtown$, and \eqref{eq:backward} with $c(N)=1$
then bounds the relative disagreement between two implementations differing
only in the order of their arithmetic by
$\varepsilon\kappa_2 \approx \SnBackwardBound$ (arithmetic on the two
constants just quoted, not a measurement). The directly measured figure sits
inside that bound: the same batched solve on CPU and on GPU, where nothing
differs but the reduction order of the underlying kernels, disagrees by
$1.023\times10^{-5}$~ft.

Two consequences follow. Agreement
at $10^{-14}$~ft is not obtainable by being careful with formulas; it is
obtainable only by reproducing the arithmetic. And a re-implementation that
agrees at $10^{-5}$--$10^{-6}$~ft is not defective (it is at the noise floor
implied by \eqref{eq:backward}), but it cannot verify anything finer, and in
particular cannot distinguish a modelling error from reduction-order noise.
The bit-level target is the only one at which a failed comparison is
unambiguously a modelling discrepancy rather than reduction-order noise.

\paragraph{(vi) The unit round-trip on the boundary conditions.} The five
items above concern the arithmetic. A sixth concerns the inputs. Every field
class the solver reads can be rebuilt bit for bit from the input file's own
text (diameters, lengths, resistances, minor-loss coefficients, demands,
emitter coefficients); reservoir heads and valve settings are the two a
parser is most likely to take through an internal metre representation and
back, and on a model written in US customary units that round trip costs one
unit in the last place. Five networks of the \WrNets{} are sensitive to it,
four of them above the acceptance threshold and a fifth above $10^{-12}$~ft.
The sensitivity is a property neither of the reference engine nor of those
models; it is a property of the input path.

The amplification is the conditioning of \eqref{eq:backward} applied to a
boundary condition. On the suite's largest network (12,527 nodes),
perturbing a single roughness value inside the reference library by one unit
in the last place moves the reference library's own answer by
$1.3\times10^{-6}$ to 6.9~ft, and a one-unit error in a fixed head enters the
same amplification. Reading those two field classes from the input text,
through a non-default parser entry point so that the default numerical path
is unchanged to the last bit, drives all five networks to bit-level
agreement over every frame (Table~\ref{tab:exactinput}a):
$\WrCtlBigWas$, $\WrCtlNetSixWas$, $6.086\times10^{-6}$,
$2.146\times10^{-6}$ and $\WrCtlAnytownWas$~ft all to zero, with
$\max|\Delta Q|\le\WrCtlMaxdQFixed$~cfs and with statuses, settings, frame
times and Newton iteration counts equal frame by frame. The
\WrCtlNetSixFrames-frame extended-period model among them is covered
across all its frames.

The negative control establishes the cause. On \WrCtlNullNets{} further
networks the same input path, reading the same two field classes from the
input text and, in a second arm, the elevations as well, changes no field,
and every measured maximum is identical (Table~\ref{tab:exactinput}b). The effect reaches exactly the models a
unit round trip can damage, the ones written in US customary units. The
$\WrCtlWorstdH$~ft that remains as the suite's worst deviation is on two
metric models the correction does not touch, and its cause is the
reduction-order noise of \eqref{eq:backward}, not the inputs. Nothing in this
note licenses the stronger claim that all \WrNets{} networks agree bit for
bit: \WrCtlExactZero{} of them do, against \WrCtlExactZeroWas{} under the
default input path, and the rest sit between $10^{-14}$ and $\WrCtlWorstdH$~ft.

\snote{2: the two differentiation routes}

\paragraph{Route A: truncated unrolling.} The dense forward path executes a
fixed number $K$ of global-gradient-algorithm (GGA) iterations with
out-of-place tensor operations, and ordinary reverse-mode autodifferentiation
is applied to the whole graph. What is returned is the exact derivative of the
$K$-step map, at $O(K)$ memory. Because $K$ is finite it is not the derivative
of the fixed point, and the gap is measured rather than assumed: this is why
the route is checked against a $10^{-4}$ threshold where the adjoint is
checked at $10^{-6}$, and why the implementation carries a health check for the truncated gradient. Over
\SnHealthNets{} public networks the truncated demand gradient converges in $K$
on four, is borderline on \SnHealthCond{} and does not converge on
\SnHealthBadN{} (\SnHealthBad; Table~\ref{tab:health}). This is
a list of measured networks, not a conditioning criterion, and the health
check should be run before the route is trusted on a new one. The route
additionally refuses every PRV network, L-TOWN included, at solver
construction.

\paragraph{Route B: the implicit-function adjoint.} Collect the unknowns as
$\vect{z} = (\vect{Q},\vect{q}_E,\vect{H})$ and write the converged state as
the root of $\vect{F}(\vect{z};\theta)=0$, whose three blocks are the
fixed-point form of EPANET's own update rules,
\begin{align}
  \text{link energy:}\quad & r_k = \varphi_k(Q_k) - (H_{n1(k)} - H_{n2(k)}) = 0,\\
  \text{emitter:}\quad     & r_{E,i} = K_{e,i}\,q_{E,i}^{\,1/\gamma} - (H_i - z_i) = 0,\\
  \text{continuity:}\quad  & m_i = \textstyle\sum_{k:\,n2(k)=i} Q_k
                             - \sum_{k:\,n1(k)=i} Q_k - q_{E,i} - d_i = 0,
\end{align}
where $\varphi_k$ is exactly the head-loss branch the solver used (including the branch clamped at the engine's low-flow resistance floor, RQtol) and $z_i$ is the node elevation. The
Jacobian has the block form
\begin{equation}
  \mat{J} =
  \begin{bmatrix}
    \mat{D} & \mat{0} & -\mat{A}_{12} \\
    \mat{0} & \mat{E}' & \mat{S} \\
    \mat{A}_{21} & \mat{P}_e & \mat{0}
  \end{bmatrix},
  \label{eq:jacobian}
\end{equation}
with $\mat{D} = \mathrm{diag}(\varphi_k')$ equal to the head-loss derivative the solver's own Newton update uses (on a clamped branch, the derivative of the clamped
expression, so that the Jacobian is consistent with the residual actually
driven to zero), $\mat{E}'$ the diagonal of emitter gradients and
$\mat{S} = \mat{P}_e = -\mat{I}$. Eliminating $\Delta\vect{Q}$ and
$\Delta\vect{q}_E$ from \eqref{eq:jacobian} returns exactly the Schur
complement $\mat{A}$ of the forward iteration, augmented by the emitter
diagonal: the adjoint system has the same structure, the same sparsity
pattern and the same conditioning as the forward solve. For a scalar loss
$L(\vect{z})$ the backward pass solves
$\mat{J}^{\!\top}\lambda = \partial L/\partial\vect{z}$ once and contracts
$\partial L/\partial\theta = -\lambda^{\!\top}\partial\vect{F}/\partial\theta$
at a memory cost independent of the iteration count.

\paragraph{The four closed forms.} Each supported parameter class has
$\partial\vect{F}/\partial\theta$ available analytically:
\begin{itemize}[leftmargin=1.4em,itemsep=1pt,topsep=2pt]
\item \textbf{nodal demand} $d_i$: continuity rows,
      $\partial m_i/\partial d_i = -1$;
\item \textbf{emitter coefficient} $K_{e,i}$: emitter rows,
      $\partial r_{E,i}/\partial K_{e,i} = |q_{E,i}|^{1/\gamma-1}q_{E,i}$, and
      exactly zero on the clamped branch;
\item \textbf{fixed head} $H_{0,f}$: link rows through $\mat{A}_{10}$,
      contributing $-1$ when the fixed-head node is the link's upstream end
      and $+1$ when it is the downstream end;
\item \textbf{pipe resistance} $r_k$: link rows,
      $\partial r_k/\partial r = \mathrm{sgn}(Q_k)|Q_k|^{n}$, and exactly zero
      on the RQtol-clamped branch, on valves and on closed links.
\end{itemize}
Pump speed $\omega$ and the pump curve coefficients $h_0$ and $R$ are
supported the same way, branch by branch of the pump coefficient routine. For
a constant-power pump on the clamped branch, EPANET's iteration uses $+g$
where the derivative of the residual is $-g$; that asymmetry is a half-step
device serving the Newton iteration, not a Jacobian entry, and the implicit
differentiation uses the derivative of the residual.

\paragraph{Newton polish, and why it is not optional.} In route B the state is
first converged with the GGA and then polished with a few full Newton steps on
$\vect{F}(\vect{z};\theta)=0$ using a sparse LU factorisation of the assembled
$\mat{J}$. The requirement is quantitative: on an ill-conditioned
network the GGA's Schur-complement half-iteration stalls at a relative error
near $10^{-7}$, whereas the assembled Newton step drives
$\|\vect{F}\|_\infty$ to $\sim\!10^{-14}$~ft (on the emitter-augmented Hanoi configuration, $\SnExtResid$~ft). Without the polish, finite-difference verification measures
the residual of the forward solve rather than the error of the gradient.

\paragraph{Three sources of non-smoothness, which are not equivalent.}
\emph{Clamped branches} (the RQtol floor on $g_k$, the CBIG/RQtol bracketing of constant-power pumps, and the emitter's
one-sided behaviour at $K_e=0$) give a piecewise-defined but continuous map:
exact inside a branch, one-sided at a boundary, with EPANET's own guard used as
the definition. In particular $\partial\!\cdot\!/\partial K_{e,i}$ at
$K_{e,i}=0$ is exactly zero, matching the engine's short circuit.
\emph{Pressure-driven demand} would add the three-segment Wagner
function~\citep{wagner1988water} with barrier penalties; it is not
implemented, so the class does not arise, at the cost of not supporting the
model. \emph{Status switching} is a genuine combinatorial event: a switch
changes the effective topology or replaces an equation. The backward pass
freezes the configuration the forward pass converged to, so the reported
derivative is the derivative of the smooth branch the solution sits on, and
with respect to a parameter whose perturbation would flip a status it is
one-sided at best. Nothing is smoothed; the \WrAuditItems-item
degenerate-case audit (Table~\ref{tab:audit}), \WrAuditOper{}
of whose cases run on the operating network's closed-link and throttle-valve
neighbourhoods, characterises exactly this regime.

\snote{3: the big-$M$ PRV penalty matrix, and the Woodbury correction}

This part of the construction is given in full, because a direct reuse of
the forward factorisation at an active PRV returns a wrong gradient with no
visible failure.

\paragraph{The reduction.} Eliminating the link and emitter rows of
$\mat{J}^{\!\top}\lambda = (\vect{g}_Q,\vect{g}_E,\vect{g}_H)$ leaves a
node-level system whose coefficient matrix is exactly $\bar{\mat{A}}$, the
converged GGA matrix the forward pass assembled last. Writing $\mat{P} =
\bar{\mat{D}}^{-1}$ for the per-branch conductance, $\mat{B}$ for the
junction incidence and $h'_{E}$ for the emitter derivative, the backward pass
is
\begin{equation}
  \bar{\mat{A}}\,\lambda_m
   = \mat{B}^{\!\top}\mat{P}\,\vect{g}_Q
     - \vect{e}_m\odot\vect{g}_E/h'_E - \vect{g}_H,
  \qquad
  \lambda_l = \mat{P}\bigl(\vect{g}_Q - (\lambda_m[j_2]-\lambda_m[j_1])\bigr),
  \qquad
  \lambda_e = (\vect{g}_E + \lambda_m)/h'_E ,
  \label{eq:reduction}
\end{equation}
so the whole backward pass is one triangular solve against the forward pass's
terminal factorisation plus sparse matrix--vector products. No new
factorisation is required, and independently instrumented counters confirm that none is performed.

\paragraph{Where the reuse breaks.} At an \emph{active} PRV the true Jacobian
row is a constraint row, $H_{n2} = h_{\mathrm{set}}$ with $D_{kk}=0$; the
valve's flow column then gives the constraint
$\lambda_m[j_2]-\lambda_m[j_1] = \vect{g}_Q[\mathrm{prv}]$, and the
corresponding $\lambda_l$ is a free quantity that enters no parameter gradient
(the resistance derivative is zero, there is no fixed-head coupling, and
$\mat{P}=0$, so the scatter vanishes identically). But the matrix the forward
pass actually factorised is the big-$M$ penalty matrix
\begin{equation}
  \hat{\mat{A}} \;=\; \bar{\mat{A}} + \mathrm{CBIG}\sum_{a=1}^{p}
    \vect{e}_{j_{2a}}\vect{e}_{j_{2a}}^{\!\top},
  \qquad \mathrm{CBIG} = 10^{8},
\end{equation}
for $p$ active PRVs. Solving the adjoint system naively with
$\hat{\mat{A}}^{-1}$ drives the adjoint variable at the valve's downstream row
to $\approx r_{j_2}/\mathrm{CBIG}$, which is numerically zero. On L-TOWN this
puts the demand gradient at that node at $-8.9\times10^{-8}$ where the true
value is $2.38\times10^{2}$: wrong by order one, with a small residual for the
wrong system and nothing visibly failing.

\paragraph{The correction.} $\hat{\mat{A}}$ and the true node-level matrix
$\mat{M}$ differ by a replacement of the $p$ downstream rows,
\begin{equation}
  \mat{M} \;=\; \hat{\mat{A}} + \sum_{a=1}^{p}
    \vect{e}_{j_{2a}}\bigl(\vect{m}_a - \hat{\vect{a}}_a\bigr)^{\!\top},
\end{equation}
where $\vect{m}_a^{\!\top}$ is the constraint row and
$\hat{\vect{a}}_a^{\!\top}$ the penalty row it replaces: a rank-$p$ update,
so the Woodbury identity applies. Writing $\vect{x}_0 = \hat{\mat{A}}^{-1}
\vect{r}$ for the naive solve and $\mat{S} = \hat{\mat{A}}^{-1}
[\vect{e}_{j_{21}},\dots,\vect{e}_{j_{2p}}]$ for the $p$ correction
directions, which cost $p$ triangular solves against the factorisation already
in hand, the corrected solve is
\begin{equation}
  \mat{M}^{-1}\vect{r} \;=\; \vect{x}_0 \;-\;
    \mat{S}\,\mat{C}^{-1}\bigl(\vect{x}_0[j_2]-\vect{x}_0[j_1]-\vect{r}[j_2]\bigr),
  \qquad
  C_{ab} = (\vect{s}_b)_{j_{2a}} - (\vect{s}_b)_{j_{1a}},
  \label{eq:woodbury}
\end{equation}
with $\mat{C}\in\mathbb{R}^{p\times p}$ inverted directly. The bracket is
exactly the residual of the $p$ valve constraints at the naive solution, so
\eqref{eq:woodbury} reads as: solve once with the penalty matrix, measure how
far the constraints are violated, and remove that violation along the $p$
directions the factorisation already supplies. One outer step of iterative
refinement against the true $\mat{M}$ then follows, which at
$\kappa_2 \approx \SnKappaLtown$ is a guard rather than a nicety.

\paragraph{Accuracy of the corrected adjoint.} With the correction, all four parameter
classes agree with a reference computed by a sparse LU factorisation of
$\mat{J}^{\!\top}$ to a relative error of $1.4\times10^{-10}$; the GPU adjoint
agrees with the CPU adjoint to $7.3\times10^{-11}$ over mixed batches whose
scenarios hold different valve states (active, open and closed PRVs in one
batch); and it agrees with central finite differences (each perturbed point
solved through the full status machine, switch-crossing coordinates discarded)
to $7.0\times10^{-7}$ at worst. A terminal refactorisation is therefore
required: at $\kappa \sim 10^{10}$--$10^{12}$, with a factorisation from a
stale status configuration the contraction factor of the iterative
refinement exceeds one and the refinement diverges (BWSN Network~1). Finally,
what the GPU adjoint cannot do it
refuses loudly rather than degrading: pump-parameter gradients,
Darcy--Weisbach, the clamped branch of constant-power pumps, cascaded PRVs
sharing a downstream node, float32 and second derivatives all raise at
construction or call time.

\snote{4: the leak perturbation operator and the signature dictionary}

\paragraph{Three leak paths, one sparsity pattern.} EPANET~2.2 represents a
leak in three superposable ways. \emph{(a)} An \emph{emitter}, a
pressure-dependent orifice discharge $q_{E,i} = C_i p_i^{\gamma}$ with
$p_i = H_i - z_i$ and $\gamma$ a global exponent, internally inverted into a
virtual link to a virtual reservoir at the node elevation with head loss
$h - z_i = K_{e,i}q_E^{1/\gamma}$ and $K_{e,i} = u\,C_i^{-1/\gamma}$ for a
unit-conversion factor $u$; assembly adds $1/g_{E,i}$ to $A_{ii}$ and
$(h_{L,E}+z_i)/g_{E,i}$ to $F_i$. \emph{(b)} \emph{Pressure-driven demand},
the Wagner function inverted into a virtual link to a virtual reservoir at
$z_i+p_{\min}$~\citep{germanopoulos1985technical,wagner1988water}, structurally
identical to (a). \emph{(c)} A \emph{known discharge} added to nodal demand,
$d_i \leftarrow d_i + Q_{\mathrm{leak},i}$, which changes only the right-hand
side. None of the three touches an off-diagonal entry of $\mat{A}$. The
exponent $\gamma$ is held fixed throughout this work and only the coefficient
$C_i$ is estimated: $\gamma$ enters the emitter residual through
$q_{E,i}^{1/\gamma}$, so treating it as a second unknown per candidate would
double the parameter count while the data of one operating window constrain
mainly the product's magnitude.

\paragraph{A leak is a diagonal rank-one update.} Inserting a leak at junction
$i$ therefore acts as
\begin{equation}
  \mat{A} \longmapsto \mat{A} + \beta_i\vect{e}_i\vect{e}_i^{\!\top},
  \qquad
  \vect{F} \longmapsto \vect{F} + \phi_i\vect{e}_i,
  \label{eq:leakop}
\end{equation}
with $\beta_i = 1/g_{E,i}\ge0$ and $\phi_i = (h_{L,E,i}+z_i)/g_{E,i}$ for path
(a), the analogous quantities for (b), and $\beta_i = 0$,
$\phi_i = -Q_{\mathrm{leak},i}$ for (c): a rank-one, diagonal, sign-definite
update of a symmetric positive definite matrix, plus a single-component update
of its right-hand side. Three consequences follow.

For a known discharge the matrix is unchanged, so
$\partial\vect{H}/\partial Q_{\mathrm{leak},i} = -\mat{A}^{-1}\vect{e}_i$:
the $i$-th column of $-\mat{A}^{-1}$, obtained by one solve against an
existing factorisation, and by symmetry also the sensitivity of node $i$ to a
unit demand anywhere. Because the update in \eqref{eq:leakop} is rank one,
Sherman--Morrison applies to the emitter case~\citep{sherman1950adjustment},
\begin{equation}
  (\mat{A} + \beta_i\vect{e}_i\vect{e}_i^{\!\top})^{-1}
   = \mat{A}^{-1}
     - \frac{\beta_i\,\mat{A}^{-1}\vect{e}_i\vect{e}_i^{\!\top}\mat{A}^{-1}}
            {1 + \beta_i(\mat{A}^{-1})_{ii}},
\end{equation}
so a candidate sweep over $M$ nodes replaces $M$ factorisations by $M$ solves,
and a batch of candidates is a low-rank rather than a rank-one update with the
same conclusion. And at fixed pressure
$\partial q_{E,i}/\partial C_i = p_i^{\gamma}$, so through the coupled system
the adjoint of Note~2 returns $\partial L/\partial C_i$ for all candidates
simultaneously with one solve rather than one forward simulation per
candidate. That is the operational difference between derivative-free leak
search and gradient-based leak inversion.

\paragraph{The signature dictionary.} Fix a sensor set $\mathcal{S}$ and a
candidate set $\mathcal{C}$, and define the \emph{signature} of candidate $i$
as the normalised column
\begin{equation}
  \vect{d}_i = \frac{\partial\vect{h}_{\mathcal{S}}}{\partial C_i}
    \Bigm/\Bigl\|\frac{\partial\vect{h}_{\mathcal{S}}}{\partial C_i}\Bigr\|_2
    \in \mathbb{R}^{|\mathcal{S}|\cdot T},
\end{equation}
stacked over $T$ operating frames. The matrix
$\mat{\Phi} = [\vect{d}_1,\dots,\vect{d}_{|\mathcal{C}|}]$ is the dictionary
that any sparse localisation method ($\ell_1$ minimisation, matching
pursuit~\citep{tropp2007signal}, or Bayesian selection in a surrogate's latent
space~\citep{mucke2023probabilistic}) implicitly works with, and its mutual
coherence $\mu = \max_{i\neq j}|\langle\vect{d}_i,\vect{d}_j\rangle|$ controls
whether sparse recovery can succeed at all. Without differentiability
$\mat{\Phi}$ must be built by finite differences, one simulation per
candidate per frame; here it is a by-product of the adjoint, and the
\SnInvCandidates-candidate dictionary of the main text costs \SnCohSeconds~s of
single-process CPU time.

\paragraph{Two medians, two denominators.} The coherence statistics of the
main text's Section~3.5 use two denominators, and neither may be quoted
against the other's population. The median over \emph{all} \SnCohPairs{}
candidate pairs is \SnCohMedianAll; that population includes \SnCohCross{}
cross-zone pairs which are exactly orthogonal, because \SnCohZoneCands{}
candidates sit behind a PRV and a pressure signal does not cross the closed
boundary. The median over the 1,495 \emph{within-zone} pairs is
\SnCohWithinMedian. Both are true; they answer different questions, and the
\SnCohGtThresh{} pairs above $0.999$ are \SnCohGtThresh{} of \SnCohPairs{}
overall and \SnCohGtThresh{} of 1,495 within-zone: all of them within-zone.

\paragraph{The dictionary is not an artefact of the stopping rule.} The
L-TOWN input file ships with an ACCURACY setting of $10^{-2}$, and both the base and
the probe solves behind the reported dictionary are tightened well past it. As
a control, the whole dictionary is built twice, at the shipped setting and
tightened. The coherence statistics are insensitive to
the choice: the maximum is identical to nine significant figures, the count of
pairs above $0.999$ is identical at \SnCohGtThresh, the median over all
\SnCohPairs{} pairs moves by $\SnCohMedianShift$, and each true leak's
closest-rival coherence agrees to five significant figures. One thing does
change, and it is the reason for tightening: at the shipped setting the solve
leaves spurious non-zero sensor responses across the valve boundary, so the
exact orthogonality of the cross-zone pairs is destroyed and the pair count is
inflated. The coherence conclusion is therefore robust to the stopping rule;
the zone decomposition quoted above is not, and requires the tightened solve.

\snote{5: the batched status machine}

EPANET's hydraulic loop interleaves two cadences of discrete logic: valve
statuses are re-examined on every iteration, whereas general link status checks run every CHECKFREQ iterations (the engine's status-check frequency option) and once more at convergence. The
batched port reproduces both cadences literally, and does \emph{not} bucket
scenarios by status. For each controlled link the coefficient contributions of
\emph{all} status branches are evaluated for the whole batch and combined by
masks; the transition rules are applied as one-hot selections over the same
four cases as the reference routine; and the loop terminates only when every
scenario in the batch is simultaneously converged and status-consistent.

An active PRV replaces its downstream row by the penalty construction of
Note~3, and the port writes the penalty on the diagonal only, so that
$\mat{A}$ remains bitwise symmetric. That is not an assumption but a checked
invariant: the regression suite (Table~\ref{tab:regression})
tests $\max|\mat{A}-\mat{A}^{\!\top}|$ on every Newton round of 86
network--path configurations (the operating network included) and finds it
exactly zero, and four seeded one-sided-assembly variants of the port, among
them ``write the penalty row one-sided'', are each detected.
The invariant the guard checks is a property of the assembly code, not of any
network; the two large public benchmarks on which parallel links in mixed
directions perturb $\mat{A}-\mat{A}^{\!\top}$ at the last bit
(main text, Section~2.5) lie outside the batched path by capability, not by
exemption.

Agreement with the serial replica is behavioural, not merely terminal. On 134
valve-forcing scenarios of L-TOWN, chosen to drive each PRV through all three
of its states, the batched and serial paths reach element-identical terminal
states in 134 of 134, and their non-empty status-transition subsequences agree
item by item in 131 of 134; the three divergences are stopping-time-only, one
path declaring convergence one iteration earlier, and are not transition
disagreements. Across five public PRV networks every divergence is attributable
either to a status decision sitting within rounding distance of its threshold
or to accumulated drift on penalty-degenerate branches, and the count of cases
in which the two paths saw the same decision inputs and transitioned
differently is zero. With the status machine the batched path covers 18 of
the 21-network public suite (11 without it). The remaining
three need valve types or head-loss laws the batched path does not implement:
Net6 and BWSN Network~2 are outside it because of a capability gate and not because of a memory limit: Net6 combines check-valve pipes (the engine's CVPIPE link type) with a PRV, and BWSN Network~2 combines them with pressure-sustaining and flow-control valves (PSV and FCV).

\snote{6: sparsity-pattern invariance and the stateful sparse solver}

\paragraph{The invariance, and why it holds.} The Schur complement is
$\mat{A} = \mat{A}_{21}\mat{D}^{-1}\mat{A}_{12}$ with
$\mat{A}_{12} = \mat{A}_{21}^{\!\top}$ the junction incidence and
$\mat{D} = \mathrm{diag}(g_k)$. Entrywise, for $i\neq j$,
\begin{equation}
  A_{ij} = -\!\!\sum_{k:\,\{n1(k),n2(k)\}=\{i,j\}}\!\! g_k^{-1},
  \qquad
  A_{ii} = \sum_{k \ni i} g_k^{-1} \;+\; \text{(emitter term)} ,
\end{equation}
so the off-diagonal pattern of $\mat{A}$ is exactly the adjacency of the
junction subgraph and does not depend on the values $g_k$. Every device the
status machine can apply changes only those values: a closed link is given
$g_k^{-1} = \mathrm{CBIG}^{-1} = 10^{-8}$ rather than being deleted, a fully
open valve $10^{6}$, an active PRV a CBIG term written on the diagonal.
By \eqref{eq:leakop} the same is true of a leak on any of its three paths.
Hence the pattern, the elimination ordering, the fill-in pattern and the
symbolic factorisation are invariant under the insertion, removal or
modification of any leak and under any status change, and the invariance is
structural rather than approximate. The one thing it requires of the
implementation is that closed links keep their slots, which is why the
reference engine's choice not to delete them matters here.

\paragraph{What the sparse route does with it.} The sparse route (compressed-row assembly handed to the vendor sparse direct solver; deliberately not the default) builds the compressed-row structure once at construction, asserting
at that point that every slot a PRV can write already exists, and thereafter
updates values alone. The assembled values are not approximately but exactly
those of the dense path: on L-TOWN all 17 iterations of the nominal frame give
$\max|\Delta\mat{A}| = 0$ against the dense assembly, and the terminal heads,
flows, statuses and iteration counts are bit-identical when both are handed to
the same linear solver.

\paragraph{The plan cache and its capacity rule.}
The vendor sparse direct solver is used statefully: the symbolic analysis is planned once per
batch shape and cached under a least-recently-used cache keyed by batch size,
floating-point precision, device and gradient slot. The rule that matters in training is
\begin{equation}
  (\text{distinct batch sizes passing through the solver within one optimiser
  step}) \times (\text{gradient slots}) \;\le\; \text{cache cap} .
\end{equation}
Below that, the cache evicts a plan that is about to be needed and silently
re-plans every step. Measured over six training shapes on two public networks,
the rule is broken by more than the gradient-accumulation case usually
described: an ordinary bucketed loader that runs the backward pass immediately
after each forward pass still triggers it as soon as the number of buckets
exceeds the cap, costing \SnPlanKyFour$\times$ the step time on ky4 and \SnPlanModena$\times$ on Modena, with the counter signature unambiguous
(\SnPlanKyFourPlans{} plans rebuilt per epoch at the default cap, zero at the
correct one). Raising the gradient-slot count is not the remedy: it multiplies
the number of cache keys, and in every measured cell its best outcome was
inside node-to-node noise while its worst was \SnPlanSlotsPen$\times$ slower
and \SnPlanSlotsCost~MiB more resident. The sparse route refuses rather than
degrades on float32, CPU execution, second derivatives and a missing vendor solver library.

\snote{7: the cost crossover, and why the end-to-end factor must be split}

\paragraph{The crossover is real and it is not one number.} The
sparse-versus-dense ratio was measured on \SnCrossNets{} public networks at
six batch sizes, forward-only and forward-plus-backward, on three
independently measured RTX~5090 nodes (the node-to-node spread of any cell is
at most \SnCrossSpread\%). Three readings follow, and the third is the one an
implementer needs. First, for the forward solve alone the crossover at
$B\ge64$ lies between $N_j = 268$ (Modena, \SnModenaFwdSixtyFour$\times$, sparse losing) and $N_j = 959$ (ky4, \SnKyFourFwd$\times$ at $B=1{,}024$, sparse winning); at $B=8$ it has already moved below 268 (Modena \SnModenaFwdEight$\times$). Second, for
forward-plus-backward the crossover is \emph{lower}, between $N_j = 92$ and
$N_j = 268$: Net3 is level at $B=8$ (\SnNetThreeFbEight$\times$) and loses above it, Modena wins at every batch size. No public network in
the dense-capable set has $92 < N_j < 268$, so an interval rather than a point
is reported. (On the scale axis of pure single-frame forward cost, a different
calibre, the main text's Section~3.2 places the crossover near 300
junctions, with the 541-junction operating network the first measured point
above it.) Third, on small networks the sparse route loses badly and the loss
grows with $B$: down to \SnNetOneWorst$\times$ at $B=1{,}024$ on the
nine-junction Net1. The mechanism is visible in the raw timings (the dense per-scenario cost keeps amortising with $B$ while the sparse direct solve is a floor that \emph{rises} with $B$ and is nearly independent of $N_j$), so
this is a structural property of the route, not a tuning artefact.

\paragraph{The end-to-end factor is a product of two independent things.}
Table~\ref{tab:ltowncost}a reports the L-TOWN wall time per
scenario, forward and forward-plus-backward, with the decomposition
$F_1F_2$: $F_1$ is the gain from moving the serial CPU adjoint onto the GPU at
the same dense linear algebra, $F_2$ the further gain from sparse-versus-dense
linear algebra. To confirm that the split is a property of the linear algebra
rather than of the pipeline, one round of linear algebra was additionally
timed three ways on the same assembled system: (A) dense with generic reverse-mode autodifferentiation through a dense Cholesky factorisation, which is what the shipped dense path does; (B) dense with a hand-written adjoint reusing the
same factor, which exists only in the measurement script and deliberately not
in the package; and (C) the sparse route through the vendor direct solver. All three compute the same
gradient, agreeing to between $4.3\times10^{-16}$ and $7.1\times10^{-9}$, so
the comparison is meaningful. The decomposition is unambiguous. On ky4 the sparse linear algebra genuinely wins, \SnKyFourBC$\times$ at $B=64$. On Modena it does \emph{not}: \SnModenaBC$\times$ at $B=1{,}024$, that is,
the sparse route loses by a factor of nearly five. The end-to-end numbers
those two networks show (\SnKyFourAC$\times$ on ky4 at $B=64$ and about $2\times$ on Modena) therefore come from the second factor
almost entirely: the debt the dense path's generic autodifferentiation
carries, \SnModenaAB$\times$ on Modena and \SnKyFourAB$\times$ on ky4. That debt grows monotonically with $N_j$ and with $B$, from a
fixed overhead of about \SnNetOneAB$\times$ on small networks, because the
vector--Jacobian product of a Cholesky factorisation costs $O(B\,N_j^{3})$
triangular solves and matrix products where a hand-written adjoint is one
back-substitution. The correct statement for Modena is therefore ``the
dense path's backward pass is expensive'', not ``sparse is faster''. The
product is accordingly never quoted as a single number, and the same applies
to the $F_1F_2$ split of Table~\ref{tab:ltowncost}.

\paragraph{Memory and admission.} Table~\ref{tab:ltowncost}b
gives peak device memory, measured as the tensor library's reserved peak plus device residency outside the tensor library, with the CUDA context subtracted, in one
fresh process per cell; the two measuring nodes agree bitwise on every cell.
The admission consequence is the one that changes what can be run: on L-TOWN
at $B=1{,}024$ the sparse forward-plus-backward path holds
\SnLtownMemSparse~MiB against the dense path's \SnLtownMemDense~MiB, and where
on a 32-GiB card the dense forward first fails at $B=1{,}280$ in a fresh
process, the sparse forward-plus-backward path was scanned to $B=16{,}384$
without reaching a boundary. Two cautions accompany this. Peak memory is a
fresh-process figure: within a long-running process, allocator fragmentation
at large $B$ can exhaust the device below the structural bound the memory
table reports. And a ratio of peak memories depends on which layer is being
compared: on ky4 at $B=64$ the sparsity-pattern activation ratio $N_j^2/\mathrm{nnz}$ is 286, the tensor library's peak-memory ratio is $57.9\times$ and
the ratio that actually decides admission is $33.6\times$.

\paragraph{Absolute cost against the reference engine.} Table~\ref{tab:sweep} times one steady-state frame on \WrSweepN{} networks:
14 public models spanning two and a half orders of magnitude in size, plus the
two operating networks. On the public rows the replica path costs
\WrSweepPubMin$\times$ (\SnSweepMinNet) to \WrSweepPubMax$\times$
(\SnSweepMaxNet) the reference library's, with fitted exponents
\SnSweepSlopeOurs{} for our solver and \SnSweepSlopeDll{} for the library, so
the gap is a constant factor rather than an asymptotic one; the smallest
factor anywhere in the table is the operating network City~D at
\WrSweepCityD$\times$, with City~H at \WrSweepCityH$\times$. What that factor
buys is a derivative, which the reference library does not supply at any
price. These are single-machine timings and are not a portable performance
claim.

\snote{8: the calibration comparison protocol}

\paragraph{Problems.} Calibration estimates Hazen--Williams roughness
coefficients from noisy junction heads by minimising a mean-squared head
misfit with box constraints $C \in [40, 160]$ and prior regularisation.
On Hanoi, 34 free pipes are estimated from 25 synthetic-diurnal frames
(20 training, 5 validation) at 25 training sensors, with
$\sigma = 0.1$~ft Gaussian head noise regenerated per seed; on City~D, 432
free pipes (of 475; 41 dead branches and 2 all-frame-clamped pipes are frozen
as structurally unidentifiable) from the utility's own 24-hour diurnal
pattern, 32 training sensors of the 40 designated, and the same noise model.
Table~\ref{tab:calib} reports the full noise ladders. The
gradient calibrator is Adam, then L-BFGS, then a Levenberg--Marquardt polish
whose Jacobian comes from the adjoint of Note~2.

\paragraph{The Modena rows.} The public Modena
rows of Table~\ref{tab:calib} are the two $\sigma = 0$ runs
(per-pipe and grouped truth) and nothing else: the truth was generated by the
same model the inversion searches, so both are inverse crimes by construction
and are labelled so in the table. They document that the machinery runs on a
third network; they are not evidence of calibration under uncertainty, which
is why the main text's calibration evidence is Hanoi and City~D, whose
ladders carry measured noise arms ($\sigma \in \{0.03, 0.1, 0.3\}$~ft). The
$\sigma = 0$ rows of Hanoi and City~D are inverse crimes for the same reason
and are labelled identically.

\paragraph{Baselines and tuning.} Five derivative-free baselines,
differential evolution (DE), particle swarm (PSO), CMA-ES, simulated
annealing (SA) and a DE$\to$Levenberg--Marquardt hybrid, were each tuned
over four configurations and three dedicated seeds before evaluation:
$\approx$240,000 model calls of tuning each on Hanoi, except SA, which is
serial, cannot use the batched forward, and received 60,576 (reduced and
declared); on City~D the tuning budgets were 3,168--3,600 calls. Evaluation
then used 30 fresh seeds (Hanoi) or 5 (City~D) at matched model-call budgets
$\{200, 600, 1{,}000, 2{,}000, 5{,}000\}$, extended to
$\{10{,}000, 20{,}000\}$ for DE, PSO, CMA-ES and the hybrid; backward passes
are costed at their measured $\approx$1\,\% of a forward call. Baselines
received batched forward evaluation, a more converged forward solve,
ground-truth-prior initialisation and generous snapshots. On City~D the
baselines are additionally tuned at 2,000 calls, equal to their evaluation
budget (Note~13); the one configuration where a baseline wins (City~D, hybrid
at 2,000 calls) is reported in the main text as a baseline win.

\paragraph{Censoring, stated per method.} SA's run-length ceiling is its own
declared budget: its hit-rate row is right-censored at 5,000 calls, and the
statement ``never reaches the endpoint within 20,000'' is measured only for
DE and PSO. Hit rates are metric-specific: on the training loss DE, PSO
and SA are at 0\,\% where on the identifiable-subspace parameter-error RMSE
(the roughness-error projection onto the well-conditioned subspace of the
sensitivity matrix, a parameter-space quantity distinct from any pressure
RMSE) the same table reads
hybrid 73.3\,\%, CMA-ES 26.7\,\%, DE 10.0\,\%, PSO 6.7\,\%, SA 6.7\,\%,
so every hit-rate quoted in the main text names its metric.

\paragraph{Statistics.} Effect sizes are Vargha--Delaney $A_{12}$ with
two-sided Wilcoxon rank-sum tests. Both implementations were cross-checked
against brute-force re-implementations (direct pairwise counting for
$A_{12}$; exact rank-sum enumeration at these sample sizes) and agree
exactly; the attainable two-sided floor at $n = 5$ per arm is
$p = 0.0625$, which is why the City~D baseline win is quoted with exactly
that $p$-value. Robustness arms on City~D (demand model error, sensor
bias, and eight-start multi-start) are in Table~\ref{tab:calib}b, including the one-in-four demand-uncertainty repeat
that degrades to $1.62\times10^{-2}$.

\snote{9: the verification protocol in full}

\begin{figure}[!ht]
\centering
\includegraphics[width=\linewidth]{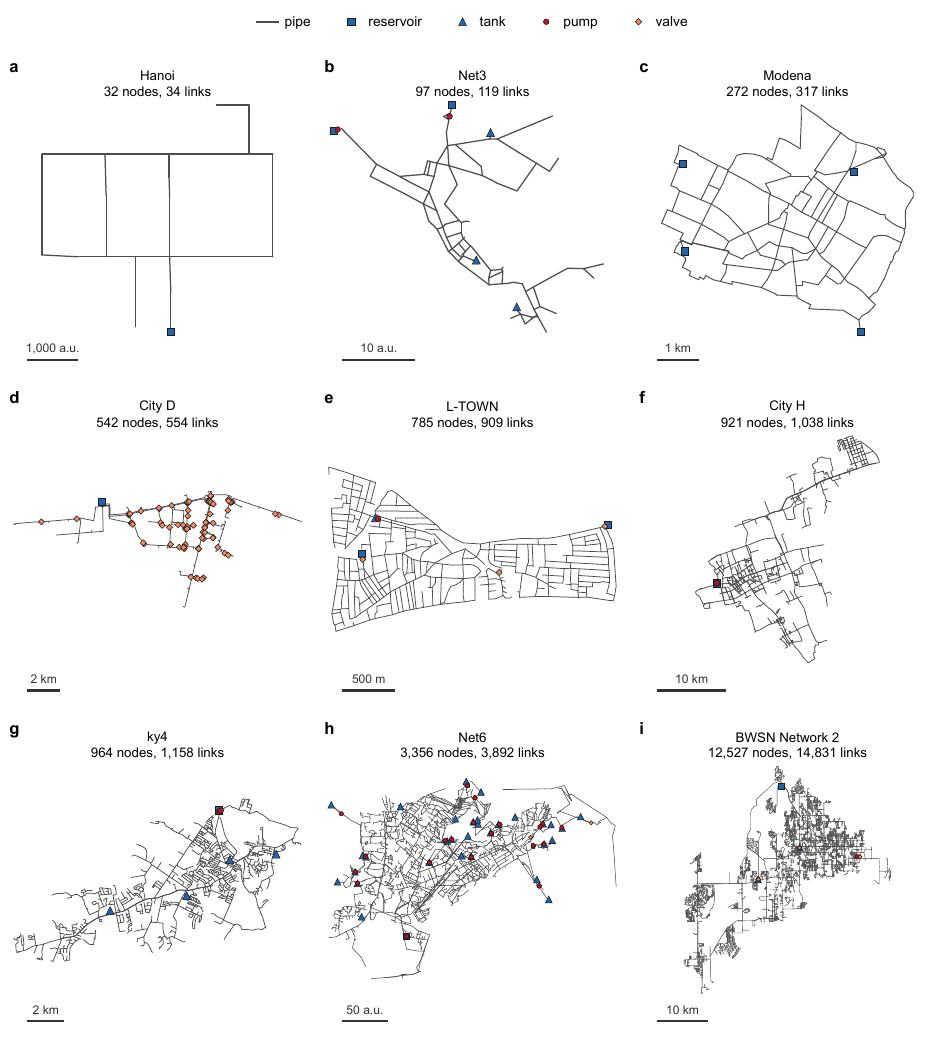}
\caption{Nine of the 52 verified networks. \textbf{a--i}, Pipes, reservoirs,
tanks, pumps and valves; junctions are not drawn. Panels are ordered by node
count and independently scaled, each with its own bar: metric where
coordinates carry the model's length unit, arbitrary units (a.u.) otherwise.
City~D and City~H are drawn from the released models; their coordinates are
local, rigidly transformed.}
\label{fig:networks}
\end{figure}

\paragraph{Forward fidelity.} The sweep comprises \WrNets{} networks:
\WrPublicNets{} public benchmark models, \WrExampleNets{} EPANET distribution
examples, \WrSyntheticNets{} randomly generated networks and
\WrOperationalNets{} operational rows (City~D, its emitter variant and
City~H), all compared against the double-precision EPANET~2.2 dynamic library
obtained through WNTR, frame by frame over the full extended-period
simulation, \WrFrames{} frames in all. Acceptance requires
$\max|\Delta H| < 10^{-6}$~ft, $\max|\Delta Q| < 10^{-6}$~cfs and per-frame
equality of the integer time sequence, the link statuses, the valve and pump
settings and the Newton iteration counts. Table~\ref{tab:full}
lists every network and every measured maximum, and
Figure~\ref{fig:networks} draws nine of them, three orders of
magnitude of node count in one plate: the public panels are the verified
files themselves, resolved through the same index the alignment harness
reads, and the two operating panels are drawn from the released model files,
whose coordinates are metres under a rigid-body transform carrying no
georeference, so the plate can be redrawn from the published data alone.
With inputs read exactly
(Note~1), all \WrNets{} networks meet the criterion: every one agrees to
$\le10^{-12}$~ft and \WrCtlExactZero{} agree exactly, against
\WrCtlExactZeroWas{} under the default input path; the largest deviation is
$\WrCtlWorstdH$~ft in head and $\WrCtlWorstdQ$~cfs in flow, and the three
operational rows sit at $1.421\times10^{-14}$~ft. Under the default input
path the corresponding figures are \WrLeTwelve{} networks at
$\le10^{-12}$~ft, \WrExactZero{} exact, and a worst head deviation of 6.836~ft on
the largest network (Table~\ref{tab:exactinput}). The suite spans both implemented head-loss laws
(\WrHW{} Hazen--Williams, \WrDW{} Darcy--Weisbach) and three flow-unit
systems (\WrUnitsLPS{} LPS, \WrUnitsGPM{} GPM, \WrUnitsCMH{} CMH); the other
seven unit systems EPANET supports are unit-tested only. Every value quoted
in this document, Table~\ref{tab:full} included, is the maximum
over \emph{all} frames.

\begin{sloppypar}
\paragraph{Gradient correctness.} Four independent checks are used, and they
differ in what they would catch. Against Richardson-extrapolated central
finite differences the implicit adjoint reaches a worst relative error of
$\WrThreeWayWorst$ over \WrThreeWayCoords{} coordinates and four parameter
classes per network, the operating network included, at the same
threshold as the public one. The Darcy--Weisbach path is differentiated
analytically through all three flow regimes and checked separately over
\SnDwCoords{} coordinates: $\SnDwTurb$ in the fully turbulent regime and
$\SnDwLam$ on the laminar and transitional branch, where the gradient
magnitudes are themselves of order $10^{-8}$ and the absolute agreement is
$\sim\!10^{-13}$. The external checks (Table~\ref{tab:external})
are independent of the replica's own arithmetic: a loss is computed by EPANET's
own compiled library, with demands and settings written through the
programmatic interface and heads read back, and differenced against our
analytic gradient. On the public emitter-augmented Hanoi configuration, eight coordinates spanning emitter, demand, reservoir-head and roughness classes agree to a
worst $\SnExtWorst$ against a $\SnExtTol$ threshold
(Table~\ref{tab:external}a); on City~D the roughness class is checked in
breadth (24 non-clamped pipes spanning 1.7 decades of sensitivity to a
worst $\WrCityDExtWorst$, plus six clamped pipes proven signal-free, their
finite-difference residuals of at most $\WrCityDClampMax$ matching the
independently predicted noise floor; Table~\ref{tab:external}b), and the
City~D emitter variant covers four parameter classes to a worst
$\WrEmitVarWorst$ (Table~\ref{tab:external}c). The finite-difference scheme
takes symmetric steps at several fractions of a per-coordinate base and fits
the odd model $\mathrm{d}L(\delta) = 2g\delta + b\delta^{3}$ by least
squares, with denominators taken from measured applied values rather than
nominal ones. Finally, \WrAuditItems{} hand-constructed degenerate cases
(Table~\ref{tab:audit}), \WrAuditOper{} of them on the
operating network's closed-link neighbourhoods, throttle-valve endpoints and
mixed batches, are each checked against their own threshold; all pass, the
worst reaching \WrAuditWorstFrac{} of its own threshold, and \WrAuditZero{}
are exactly zero because the correct gradient is structurally absent: the
low-flow linearisation, closed links and valves make the derivative of head
loss with respect to resistance analytically zero, and naive
autodifferentiation of a re-implementation reproduces those zeros without
warning.
\end{sloppypar}

\paragraph{The regression suite.} Table~\ref{tab:regression}
lists the suite item by item: \WrRegTotal{} items, 53 numerical checks and
one packaging guard. The 53 numerical checks all pass (391~s for the full
suite on the reference machine). \WrRegOper{} of
the numerical items exercise the two operational models (alignment, replay,
extended-period simulation, physical cross-validation, batch consistency and
the three-way gradient cross-check) and are marked in the table. The
packaging guard scans file contents, paths and version-control metadata
for thirteen name variants of the two utilities and requires zero hits, the
mechanical form of the anonymisation rule stated at the head of this
document; over every file, path and compiled PDF of this submission package
it reports zero.

\snote{10: sensor augmentation, calibration and leak search with the
augmented sets, and the control experiments}

\paragraph{Every design here is virtual.} New sensors are simulated on the
calibrated model; none has been installed, and field installation is a
later stage. The augmentation mode of the main text (Section~2.7) holds an
installed set $S_0$ fixed and adds $k$ junctions one at a time from the
candidate pool, either for the Bayesian D-optimal objective of the
placement link (the same $\sigma_{\rm prior}=15$ and
$\sigma_{\rm noise}=0.1$~ft) or for coverage, the number of target pipes
whose maximum sensitivity over sensors and frames crosses the census
threshold; threshold, frames and structural mask are those of the census,
so ``restored'' means exactly the census's transition from an all-zero to a
non-zero sensitivity column. At every step an assertion checks that the
identifiable set has not shrunk, that no per-pipe posterior variance has
risen and that no pipe visible to $S_0$ has been lost; the curves
carry zero violations. On City~D, $S_0$ is the 40 designated sensors, the
target set is the census's 149 unobservable pipes, and the candidate pool is
all 541 junctions, which includes the three work-order leak junctions (a
fact that matters in the leak search below). On L-TOWN, $S_0$ is the 33
sensors of the main-text inversion, the target set is the 62 pipes they
cannot inform, and the pool excludes the three injected leak junctions, as
the original sensor draw did. On Hanoi, $S_0$ is ten random sensors over the
25 synthetic frames of the calibration ladder, with one unobservable pipe.

\paragraph{Curves (Table~\ref{tab:augment}a, b).} On City~D coverage
restores 37/62/95/133/149 of the 149 at $k=5/10/20/40/80$, 80\,\% at
$+\WrAugCoverKEighty$ and all at $+\WrAugCoverKAll$ (the coverage sequence
saturates at step \WrAugCoverSat{} and reverts to the D-optimal objective
thereafter); D-optimal restores 21/23/35/52/76, and its well-conditioned
subspace CRLB trace falls from \WrAugSZeroCrlbSub{} to
\WrAugDoptTwentyCrlbSub{} at $+20$ while that subspace widens from 20 to
\WrAugDoptTwentyKSub{} dimensions. Reselecting $40+k$ sensors from scratch
recovers more of the 149 at small $k$ but loses 38/38/24/21/15 pipes the
designated sensors could see; augmentation loses none. The eps-rank CRLB
of the main text's Section~3.6 is dominated on City~D by directions at the
rank threshold and moves by up to $10^{-6}$ relative between platforms, so
only counts and the well-conditioned trace are quoted for the augmented
sets. On L-TOWN coverage restores all 62 at $+\WrAugLtownKAll$ (80\,\% at
$+\WrAugLtownKEighty$) where reselecting 33 from scratch loses
\WrAugLtownReselLost; on Hanoi the single pipe is restored at
$+\WrAugHanoiCoverK$ (coverage) or $+\WrAugHanoiDoptK$ (D-optimal), and the
two objectives select identical sets at $k=5$, 10 and 20. An independent
re-computation of the City~D curve, with a separate Fisher-matrix and census
implementation on two hosts, reproduces every count, the well-conditioned
traces agreeing to $10^{-9}$ relative.

\paragraph{Calibration with the augmented sets
(Table~\ref{tab:augcalib}).} The calibrator, truth, noise realisation and
20\,\% sensor hold-out of Note~8 are kept and only the sensor set changes.
``Restored'' pipes are the target pipes that cross the census threshold
under the augmented set; their error is $|\hat C - C_{\rm true}|$ and the
prior's error is $|130 - C_{\rm true}|$. At $\sigma=0.1$~ft the restored
pipes move off the prior (median \WrAugDoptTwentyMed{} against
\WrAugDoptTwentyPrior{} under D-optimal $+20$; \WrAugCoverTwentyMed{}
against \WrAugCoverTwentyPrior{} under coverage $+20$); the RMSE over all
informative pipes moves from \WrAugSZeroInfoRmse{} to
\WrAugInfoRmseMin--\WrAugInfoRmseMax; at $\sigma=0.3$~ft the $+20$ sets are
worse than $S_0$ alone (\WrAugTwentyInfoRmseThreeMin--%
\WrAugTwentyInfoRmseThreeMax{} against \WrAugSZeroInfoRmseThree) and
coverage's restored pipes sit above their prior. One noise seed per cell on
City~D; the Hanoi ladder (three seeds) moves from \WrAugHanoiSZeroRmse{} to
\WrAugHanoiTwentyRmse{} at $+20$ and $\sigma=0.1$~ft.

\paragraph{Leak search with the augmented sets (Table~\ref{tab:leakrerun}).}
The work-order inversion of the main text (49 candidates, three injected
leaks, $\sigma=0.1$~ft, the same regularisation and step budget, the same
inverse-crime setting) is repeated with $S_0\cup S_k$. Two facts about the
baseline fix how the comparison is read. First, the 40-sensor set of the
main-text demonstration and the census set $S_0$ are different draws sharing
22 junctions; both return none of the three under this experiment's noise
realisation. Second, that noise field is generated over all junctions and
read at the sensors, so every set sees one realisation, distinct from the
realisation of the main-text demonstration; the ``none of three to one of
three'' comparison is within this experiment. Each of the six augmented sets
returns $L_2$ with a discharge error of 1--4\,\% and nothing else. The
control experiments attribute that hit: the coverage sequence places a sensor
on $L_2$'s own junction at its third step and the D-optimal sequence at its
fifth; removing that one sensor from coverage $+20$ returns none of three;
$S_0$ plus that one sensor alone returns the same one of three (3\,\%
error); twenty random additions from the same pool recover $L_2$ in none of
five draws, and forty in none of one. $L_1$ (3.0~L\,s$^{-1}$) changes no
junction head by more than \WrAugLeakTOneFootprint~ft under any set, below
the 0.1~ft noise, so no sensor set can recover it. $L_3$ is recovered by
\WrAugLeakRandTThree{} of the five random $+20$ sets and by D-optimal $+20$
once the $L_2$-junction sensor is removed, but by none of the six designed
sets, nor by $S_0$ plus a sensor on its own junction; its nearest rival's
coherence is \WrAugLeakRivalMin--\WrAugLeakRivalMax{} in every set, and the
designed sequences never place a sensor within three hops of it in 80
steps. Coherence medians are over all candidate pairs; on City~D no pair is
exactly orthogonal, so the two denominators of Note~4 coincide.
Augmentation lowers the median from \WrAugLeakSZeroMedian{} to
\WrAugLeakMedMin--\WrAugLeakMedMax{} and the pairs above 0.999 from
\WrAugLeakSZeroGt{} to \WrAugLeakGtMin--\WrAugLeakGtMax, without lowering the
rivals that matter. On L-TOWN the ten augmented sets, passed through the
main-text inversion in a noiseless and a 0.1~ft group, all return the same
one of three as $S_0$, with the two lost leaks at rank
\WrAugLtownLostRankA{} and between ranks 27 and \WrAugLtownLostRankB{} of
60 and their rivals above 0.9985 in every set
(Table~\ref{tab:leakrerun}b).

\paragraph{Coherence-driven placement against random controls
(Table~\ref{tab:coherence}).} A second placement objective, greedily
adding sensors to minimise the same signature dictionary's coherence
(the diagnostic of Note~4, not the census's Fisher information), was run
on City~D at $k=20$ and $k=40$ and compared against budget-matched random
draws over the same candidate pool and rejection rule as Note~10's
augmentation curves: 20 fair seeds at each $k$ (seeds that draw a leak
junction are discarded and redrawn, so every random set is a valid
candidate set), plus eight sensor sets chosen purely to minimise coherence (the greedy coherence sequence at 5, 10, 20, 40 and 80 added sensors; its worst-pair variant, which minimises the largest pairwise coherence, at 20; and its full-pool variant, drawn without the leak-junction rejection, at 20 and 40), for 40 random draws in total. The pooled significance tests
below go further and ask a stricter question: is $L_2$'s recovery a
property of \emph{any} purposefully chosen sensor set, or specifically of
a coherence-minimising one? They pool those eight coherence-minimising
sets with the six identifiability-driven augmented sets
above (the coverage augmentations at 20, 40 and 80 added sensors and the D-optimal augmentations at 20, 40 and 80), 14 designed sets in total, against the 40 random draws -- a design pool that already
recovers $L_2$ in every one of its six identifiability-driven members
(Table~\ref{tab:leakrerun}a), so a coherence effect would have to survive
being tested \emph{alongside} that fact, not instead of it. The result
reads differently leak by leak (Table~\ref{tab:coherence}a). $L_2$ is
recovered by 12 of the 14 pooled designed sets against 3 of the 40
random draws, a one-sided Fisher exact test of \WrCohLTwoFisherP; of the
eight coherence-minimising sets alone, six recover $L_2$
(the greedy sequence's 5- and 80-sensor sets do not). What actually sorts recovery is
not which objective chose the set but each set's own rival-coherence
gap: across all 55 reconstructible configurations (every pooled designed
set, every random draw, and every other City~D sensor set this note or
the main text records) the gap of every configuration that recovers
$L_2$ exceeds the gap of every configuration that does not, with no
overlap -- true of the identifiability-driven recoveries as much as the
coherence-minimising ones, which is why the gap, not the label, is the
mechanism reported in the main text. At the matched $k=20$ budget alone
(the worst-pair variant with 20 added sensors against 20 random $k=20$ draws) the same conclusion holds by an exact permutation test
(\WrCohKTwentyP, none of 20 random draws recovering $L_2$); at $k=40$
(the greedy coherence sequence at 40 added sensors against 20 random $k=40$ draws) the matched-budget test
alone does not reach significance (\WrCohKFortyP, 3 of 20 random draws
recovering $L_2$), and the shortfall is not one of sample size: the random
group's 95\,\% Clopper--Pearson interval on its own hit rate is
[\WrCohKFortyCILo, \WrCohKFortyCIHi] (pooled over both budgets,
[\WrCohPooledCILo, \WrCohPooledCIHi], Table~\ref{tab:coherence}b), wide
enough that additional random draws at $k=40$ alone would not settle it
either way. $L_3$'s recovery is statistically indistinguishable from random
placement (3 of 14 pooled designed sets against 7 of 40 random draws,
$p=0.51$) and its rival-coherence gap ranges overlap
(Table~\ref{tab:coherence}a); it is not a coherence effect, and no claim
in the main text credits it as one.
$L_1$ is recovered by neither objective at any budget: its dictionary
column norm (0.15--0.64) sits an order of magnitude below $L_2$'s
(5.1--17.1), so the 0.1~ft noise floor swamps its signature however far
the rival coherence is pushed down (to 0.121 at best on this pool); it is
an amplitude limit, not a coherence one.

\paragraph{The two inversion stages.} The
work-order inversion above adds a discrete support-refinement stage
(non-linear orthogonal matching pursuit and a support-swap polish)
after its Adam-plus-$\ell_1$ coarse stage; the L-TOWN inversion routine of the main text's Section~3.4 (Adam with an $\ell_1$ penalty) has only the coarse stage, ranking candidates by coefficient magnitude with no
discrete search. Re-scoring every one of the 55 City~D configurations on
the coarse stage alone -- whether the truth is among the top three by
magnitude, the criterion the L-TOWN inversion routine itself reports -- and
repeating the Fisher and gap tests on that coarser judgement leaves both
conclusions unchanged: $L_2$ stays separated (\WrCohLTwoStageOneP, 13 of 14
coherence-selected sets against 8 of 40 random draws, rival-coherence gap
still non-overlapping) and $L_3$ stays statistically indistinguishable
from random (\WrCohLThreeStageOneP, 5 of 14 against 20 of 40). The discrete
support-refinement step changes zero $L_2$ configurations from lost to
recovered across the 55 and six from recovered to lost, so it does not
produce the coherence result; it only removes coarse-stage hits the flow
error then fails to confirm. One further asymmetry appears only at the
coarse stage: coherence-selected sets rank $L_1$ in
the top three by magnitude significantly more often than random draws do
(\WrCohLOneStageOneP, 6 of 14 against 2 of 40), yet not one of those coarse
hits survives the discrete refinement step under the final flow-error
criterion, where both groups still recover $L_1$ zero times. Coherence
can move $L_1$ up the coarse ranking without ever producing an accurate
recovery, which is consistent with, not contrary to, its being
amplitude-limited rather than coherence-limited.

\paragraph{Why no placement closes the L-TOWN loop
(Table~\ref{tab:coherence}c).} The same candidate-pool geometry that
explains City~D's graded outcome bounds what any placement on L-TOWN
could achieve for the two leaks Section~3.4's inversion never recovers
(truth ranks 19 and 32 of 60). Of the full 746-position fair candidate
pool, only 1 and 14 positions, respectively, can push that leak's
coherence with its nearest rival below the 0.999 threshold Note~4 uses;
an oracle greedy search dedicated to just that one pair, run for 80
steps and free to ignore every other leak and every budget constraint,
drives the pair's own coherence down to 0.996775 and 0.990741 but leaves
the next-closest rival at 0.999400 and 0.997689, both still inside the
inseparable regime. The one leak L-TOWN's inversion always recovers has
746 of 746 fair positions that separate it from its rival, and an oracle
search on that pair alone reaches 0.903043. No placement objective, on
this pool, buys the other two leaks anywhere close to that margin.

The conclusion carried to the main text is therefore:
placement for roughness identifiability does not close the leak-search
loop, but a second objective, placement against coherence itself,
closes it for one of the three leaks for the reason coherence predicts,
is indistinguishable from random placement for a second, and cannot
close it for the third because that leak's limit is amplitude, not
coherence; the asymmetry between the two inversions used in this note
does not drive the coherence result, and the candidate-pool geometry
that bounds City~D's coherence effect also bounds what placement can do
on L-TOWN.

\snote{11: from single-point localisation to cluster-level diagnosability}

\paragraph{The construction.} Note~4's dictionary makes the failure of
single-point localisation measurable; it also says what \emph{is} resolvable,
because two candidates whose signatures are coherent above a threshold are
exactly the pair the data cannot separate. Group them, and the inversion
answers the question the dictionary can answer. Candidates are clustered by
constrained agglomerative complete linkage: two clusters may merge only if
they are adjacent in the network's own Voronoi partition (multi-source
Dijkstra over pipe lengths, no tuning parameter), and the merge is accepted
only while the minimum cross-coherence stays at or above $\tau$, so at
termination every pair inside a cluster has $\mu_{ij}\ge\tau$. The inversion
then replaces the non-negative Lasso by its group form,
$\min_{x\ge0}\tfrac12\|\mat{A}x-y\|^2 + \lambda\sum_g\sqrt{|g|}\,\|x_g\|_2$,
solved by FISTA with a non-negative group proximal step; with every group a
singleton it reduces exactly to the single-point inversion, which is the
control reported beside it. $\lambda$ is chosen by a discrepancy rule that
never sees the truth: sweeping $\lambda$ downwards with warm starts, the
first value whose residual sum of squares falls below
$\max\{\sigma^2(n+2\sqrt{2n}),\ \min\mathrm{RSS}+2\sigma^2\sqrt{2n}\}$, the
second term a floor because a linearised dictionary does not in general reach
the pure-noise level. Clusters are ranked by $\|x_g\|_2$ and reported with
the radius, diameter and connected pipe length of the district they name.
On the operating network (542 nodes, 554 links, 49 candidates, 1,000
observations) the whole pipeline costs 0.06~s for the pairwise pipe
distances, 0.002~s for the clustering and 1.94~s for a 14-point warm-started
$\lambda$ path on one CPU core.

\paragraph{What it buys, and what it costs.} Table~\ref{tab:cluster} is the trade-off curve, and the curve is the result:
a hit rate quoted without the radius that produced it is not one. At
$\tau=0$ the operating network is a single cluster of radius 9,043~m covering
all 83.9~km of main, and the top-3 hit rate is 3 of 3, the hit rate of
saying ``somewhere in the network''. At the other end, at $\tau>1$ every
cluster is a singleton and the reading is the single-point inversion, 0 of 3.
Between them, at $\tau=\WrClusTauCityD$, 25 clusters of mean radius 179~m put
two of the three injected leaks inside the top three; those three clusters
hold 7 of the 49 candidates and $\WrClusTopThreeKm$~km of main,
$\WrClusTopThreeShare$\,\% of the network, the largest of them
$\WrClusTopThreeRad$~m in radius. Panel~(c) gives that inspection burden at
five thresholds, because the mean radius over all clusters is dominated by
singletons and understates what a crew would walk. On L-TOWN the
corresponding point is $\tau=\WrClusTauLtown$: 22 clusters, mean radius 98~m,
1.96~km of district, all three leaks inside the top three.

\paragraph{Four things this result is not.} First, the thresholds quoted are
chosen after seeing the whole sweep; both sweeps are printed in full so that
the choice can be discounted. Second, the design does not beat chance on hit
rate. Against 20 size-matched random groupings that preserve the cluster-size
distribution while destroying topology and coherence, the one-sided
exchangeability $p$ never falls below 0.09 on either network (panel~d); what
the design wins is the radius, 179~m against 1,700~m on the operating network
and 40~m against 298~m on L-TOWN at the same hit rate. Third, the equal-budget
single-point control is not a formality: given the pooled candidate count of
the three ranked clusters, the single-point ranking recovers two of three at
$\tau\le0.9$ and none at $\tau\ge0.95$, so the cluster reading wins only in
the tight-radius regime and loses in the loose one. Fourth, the leak that
noise defeats is not rescued. Its own signal-to-noise ratio
$\|C\vect{a}\|^2/\sigma^2$ is 0.50 against 166 and 151 for the other two; it
is a singleton at every $\tau\ge0.5$ because its most coherent rival is not
Voronoi-adjacent to it; and forcing it into a group with its twelve most
coherent neighbours moves it only from 49th to 36th, the group's
non-centrality parameter and its degrees of freedom rising together so that
the standardised score falls from 177.9 to 159.0. Aggregation redistributes
the burden of saying \emph{which}; it does not create signal. On L-TOWN,
where the same leak label has signal-to-noise $5.6\times10^{4}$, the same
construction moves it from 23rd to 3rd.

\snote{12: scoring a sensor layout on a ruler that does not move with it}

\paragraph{Why a layout-independent ruler is needed.} The natural score for an augmented
layout, roughness error over the pipes that layout makes informative or
over its own identifiable subspace, is not comparable across layouts,
because the set being averaged over is itself a function of the design. Over
the designs compared here the informative set runs from \WrDriftInfoLo{} to
\WrDriftInfoHi{} pipes and the design-specific subspace rank from
\WrDriftRankLo{} to \WrDriftRankHi{}; a design that adds sensors is scored on
a larger and differently conditioned set than the one it is being compared
with. Head-space misfit is not an alternative. At $\sigma=0.1$~ft it is
\WrHeadNoiseShare\,\% noise variance, and across eight optimiser restarts on
\emph{identical} data and an identical layout the training head misfit moves
\WrMsTrain\,\% and the held-out-frame misfit \WrMsVal\,\% while the roughness
error in the reference subspace moves \WrMsIdent\,\%: the head metrics cannot
resolve what is being compared.

\paragraph{The ruler.} A reference subspace is fixed once, before any design
is considered, from the adjoint sensitivity spectrum of the \emph{fully
instrumented} candidate pool (\WrIdentPool{} junctions, 25 frames): direction
$v_j$ is admitted when its posterior standard deviation would fall to
$\sigma_{\mathrm{prior}}/\gamma$, an absolute criterion involving no layout.
At $\gamma=2$ this admits $k_{\mathrm{ref}}=\WrIdentKRef{}$ of the 432
directions and the score is
$\mathrm{RMSE}_{\mathrm{ident}}=\|V_{\mathrm{ref}}^{\!\top}
(\hat{C}-C_{\mathrm{true}})\|_2/\sqrt{k_{\mathrm{ref}}}$, against a prior
reference of \WrIdentPriorRmse{} (every pipe left at its initial value). The
same subspace scores every layout, designed or random, and $\gamma=3$ and
$\gamma=10$ are reported beside it.

\paragraph{What the designs are worth on it.} Table~\ref{tab:identmetric}a is the full grid: two budgets, three noise
levels, two objectives, each against random additions drawn from the same
pool at the same budget. In all twelve combinations no random layout matched
the design, so each cell's one-sided exact randomisation $p$ sits at its
attainable floor, $1/(R+1)$: $0.0099$ in the cell run to $R=100$ and $0.0476$
in the five run to $R=20$. That the five coarser cells stop at $0.0476$ is a
sample-size limit and nothing else; reaching $p<0.01$ there needs $R\ge100$
and no change of statistic would substitute. $R$ was fixed per cell in
advance of looking at any $p$, and the whole grid rests on a single noise
realisation, so the $p$ values are conditional on it.

\paragraph{The costs.} Panel~(b) reads the
main cell on five rulers. The conclusion survives $\gamma=3$, $\gamma=10$ and
the unprojected 432-pipe RMSE, but in the \WrIdentNull{} reference directions
the projection discards, 60 of 100 random layouts are at least as good as the
design ($p=0.60$): concentrating a finite observation budget on identifiable
directions is a choice to give up the rest, and the number says so. Panel~(c)
prices the alternative practice: selecting the best of 100 random additions by
training head MSE returns a layout ranked 25th of 100 on the identifiability
ruler, by held-out-frame RMSE one ranked 56th, a coin flip. Across the
whole grid there are \WrDisagreeN{} of \WrDisagreeOutOf{} (cell, design,
head-metric) combinations in which the design is ahead of \emph{every} random
layout on parameters and behind most of them on heads. Panel~(d) relates the
design-time prediction to the achieved error: the linear-Gaussian prediction
correlates with the achieved error at Spearman $\rho=0.38$ within a budget
($n=100$) and $0.55$ when budgets are pooled, so the prediction separates
budget levels reliably and orders layouts within a budget only weakly.
Because the D-optimal objective and this evaluation subspace both descend
from the same sensitivity matrix, the coverage row, whose objective is
criterion-independent, is the one that carries the general claim.

\snote{13: optimiser enhancements, the baseline tuning budget and the
wall clock}

\paragraph{Device memory.} The experiment driver releases the terminal
Cholesky factor after each backward pass; device memory holds at 67.5~MiB
across the arms.

\paragraph{The noise floor, and which metric can resolve anything.} At the
City~D configuration the training loss of the \emph{true} parameter vector is
$1.010\times10^{-2}$, and the plain first-order arm already reaches
$9.49\times10^{-3}$: below the floor, fitting noise. Near the floor the loss
and the parameter error are anti-correlated across arms
($\rho=-0.96$), so arms are compared on parameter error, and on two of them:
the design's own identifiable subspace (the placement objective's own ruler,
of rank 41 and prior 30.41 for all 77 runs, so internally consistent) and the
layout-independent ruler of Note~12.

\paragraph{What the enhancements are worth.} Table~\ref{tab:optimiser}a lists every arm. On the internal ruler the
Schur-complement diagonal preconditioner is a clear win (17.71 to 12.40,
better on 10 of 10 seeds, sign test $p=0.0020$) while making the training
loss \emph{worse} on 10 of 10, which is what a noise floor implies. On the
layout-independent ruler the same paired comparison is 7 of 10 with
$p=0.3438$ and the median gap shrinks from 4.40 to 0.41: the gain is real in
the subspace the design itself defines and not established outside it, and
both readings are reported because neither alone is the answer. The exact
Gauss--Newton diagonal is \emph{worse} than the cheap Jacobi approximation
(1 of 10, $p=0.0215$). Batched multi-start buys at most $2.8\times$ per model
call at $B=32$, not $B\times$ (panel~b), because 20 frames already saturate
the device; on the parameter error, added to the preconditioner, it is worse
on 10 of 10 ($p=0.0020$). No arm beats the 200-call gradient
configuration, which leads on both rulers.

\paragraph{Baseline tuning budget.} The City~D baselines are tuned at 2,000
model calls, equal to their evaluation budget and matching the Hanoi
protocol, over four configurations and three dedicated seeds; each of the
four algorithms selects the same configuration as under the
3,168--3,600-call tuning of Note~8. Differential evolution at ten times the
budget (20,000 calls) reaches a parameter error of 16.66, still worse than
the gradient configuration's 200-call result.

\paragraph{The wall clock.} Panel~(c) gives the
other axis. The gradient configuration takes \WrGdWall~s for its 200 calls;
the DE$\to$LM hybrid takes \WrHybridWall~s for 1,997, on the same node, and
that row rests on \WrHybridSeeds{} seed, not five. Per model call the
baselines are the cheaper side (0.081~s against 0.113~s), so the accounting
gives them no timing advantage; but dividing the hybrid's wall clock by
the eight-way parallelism it is promised in the fairness protocol gives
\WrHybridWallEight~s against \WrGdWall~s, a tie. The equal-budget claim
therefore stands on the model-call axis, 200 against 1,997, and not on the
wall clock.

\clearpage
\section*{Supplementary Discussion}

\subsection*{Some EPANET reference solutions must not be used as labels}

A bit-level re-implementation is an unusually sensitive instrument for
auditing the reference engine, because every disagreement has to be explained
rather than absorbed. Two of the candidate explanations are properties of the
reference solution rather than of the replica, and the finding is useful
independently of anything else in this paper.

On ky5, controls close two pumps and isolate a dead-end branch, which
then retains a stale through-flow at the input file's own ACCURACY setting; tightening the setting reduces the residual by more than two orders
of magnitude. On BWSN Network~2, three junctions form dead-end
pockets that reach the main network only through pumps and flow-control valves
closed for 31 or 32 of the 32 frames. They carry a stale through-flow whose
head is close to arbitrary: perturbing a single roughness value inside the
reference library by one unit in the last place moves the library's own answer
at those junctions by 6.857~ft. Both BWSN Network~1 and BWSN Network~2 continue to oscillate without converging when the
accuracy is tightened to $10^{-8}$ over 1,000 trials. The replica reproduces
these states faithfully, as a replica should, but they are simulator
artefacts.

The implication is general and cheap to act on. Any work that generates
supervision by running EPANET over benchmark
networks~\citep{hajgato2021reconstructing,kerimov2024towards,ashraf2024physics,
ashraf2025scalable}, and any use of large simulated
corpora~\citep{truong2025ditec}, will train on those artefacts unless it
compares each frame's residual against the input file's own stopping criterion
and flags the frames that stopped rather than converged. The same applies to
the mainline benchmark used here: L-TOWN ships with an ACCURACY setting of $10^{-2}$, whose stopping solution sits
$1.7\times10^{-4}$~ft from the converged one, and every reference solution
used in this work is tightened first.

\subsection*{Relation to the surrogate literature}

Learned surrogates and a differentiable replica answer different questions and
are complementary rather than competing. A surrogate is asked to be fast, can
be orders of magnitude faster than EPANET, and can generalise across
topologies. A replica is asked to be correct; per solve it is slower than the
reference engine (Note~7), and it generalises exactly as far as its feature
coverage does: to any network built from the elements the package documents
as supported, a strict subset of what EPANET reads, which the sweep exercises
on \WrNets{} networks rather than on all of them.

Two consequences are worth stating. First, a differentiable replica is a
supervision source that supplies not only $H(\theta)$ but
$\partial H/\partial\theta$, so a surrogate could be trained to match the
sensitivity field and not only the pressure field. No such surrogate is
trained here, and nothing is claimed about what it would buy. Second, it supplies the
reference against which a surrogate's \emph{gradients}, and not merely its
predictions, can be audited. At present that literature reports prediction
error and is silent on gradient error, even where the gradient is what the
surrogate is used for~\citep{strotherm2025goflow}. Extending EPANET without
modifying it is a recurring strategy (language
bindings~\citep{eck2016reading,arandia2018simulations}, plugin
architectures~\citep{sela2019plugin}, domain
decomposition~\citep{diao2014speedup} and digital twins built on
it~\citep{park2026digitaltwin}), and treating the engine as a callable black
box is the right decision when the quantity of interest is the output, and the
wrong one when it is a derivative of the output.

\subsection*{Limitations, in full}

The main text states the binding limitations where they arise; this is the
complete set, in the order in which a reader is likely to be constrained by
them.

\begin{enumerate}[label=(\arabic*),leftmargin=*,itemsep=0.3em]
\item \textbf{Unimplemented features, and coverage is per path.}
Pressure-breaker and general-purpose valves, pressure-driven demand analysis,
tanks with volume curves and the entire water-quality module are not
implemented, and are rejected at parse or construction time with an explicit
error rather than approximated. A positive damping limit (the engine's DAMPLIMIT option, DAMPLIMIT${}>0$) and non-zero head-error or flow-change limits (its HEADERROR and FLOWCHANGE options) are likewise rejected.
Chezy--Manning is implemented but unvalidated, because no network in the suite
uses it. The batched differentiable path is narrower than the replica path:
Darcy--Weisbach networks and PSV, FCV, PBV and GPV valves are
construction-time rejections there, which is why Net6 and BWSN Network~2 lie outside it: a capability gate, not a memory
limit. With the status machine of Note~5 the batched path reaches 18 of the 21
public benchmark networks (11 without it).

\item \textbf{Bit-level agreement is platform-dependent and is a claim about
one reference binary.} The $10^{-14}$~ft figures were obtained on Windows x64
calling the power and logarithm routines of the Windows C runtime, against the reference engine's Windows build shipped inside WNTR. On another platform, against a
differently compiled EPANET, or against a WNTR release carrying a different
binary, a single one-unit difference in the power routine is enough to move the
answer and agreement degrades to roughly $10^{-6}$~ft, far inside
engineering tolerance but no longer bit-level. The $10^{-14}$ figures must not
be quoted for another build without re-measuring.

\item \textbf{The exact-input control is what makes the whole suite pass,
and it is not the default path.} Reading reservoir heads and valve settings
from the input text is a non-default parser entry point (Note~1). The default
path, on which every other result in this package is computed, is unchanged
to the last bit, and on a model written in US customary units it
still loses one unit in the last place on those two field classes. Five rows
of Table~\ref{tab:full} carry that loss.

\item \textbf{``All 52 pass'' is not ``all 52 agree bit for bit.''}
\WrCtlExactZero{} networks agree exactly; the other \WrCtlNotExact{} sit
between $10^{-14}$ and $\WrCtlWorstdH$~ft, and the exact-input control does
not move them, because their residual comes from reduction order rather than
from the inputs.

\item \textbf{Two reference solutions are unconverged transients.} ky5 and BWSN Network~2 stop rather than converge at their input files' own ACCURACY setting, and neither may be used as a training label; the replica
reproduces them faithfully, which is not the same as their being right. Net6
is verified over all 609 of its frames, with frame times, statuses and Newton
iteration counts equal frame by frame (Table~\ref{tab:exactinput}a).

\item \textbf{Gradients are gradients of a frozen status configuration.}
Derivatives with respect to parameters that would flip a valve or pump status
are one-sided at best (Note~2).

\item \textbf{The unrolled route is not usable everywhere, and the refusals
are explicit.} The unrolled-iteration path refuses every PRV network including
L-TOWN, raising at construction or at the call rather than degrading; every
PRV gradient in this paper comes from the implicit adjoint. Its truncated
demand gradient does not converge in $K$ on \SnHealthBad{} and is borderline
on \SnHealthCond{} (Table~\ref{tab:health}).

\item \textbf{The GPU adjoint and the sparse route refuse rather than
degrade.} The GPU adjoint raises on pump-parameter gradients,
Darcy--Weisbach, the clamped branch of constant-power pumps, cascaded PRVs
sharing a downstream node, float32 and second derivatives; the sparse route raises on float32, CPU execution and a missing vendor solver library; float32 with a
PRV network is refused at construction on every path. None of these fall back
silently.

\item \textbf{The dense and unrolled paths do not have the accuracy of the
replica path.} The bit-level claims apply to the replica path only. GPU double
precision differs from CPU double precision by an amount set mainly by the
network rather than by the card, from $1.2\times10^{-10}$~ft on a pipes-only
model upward, while repeating one measurement eight times on a single card
already spreads it by $1.73\times$; single precision is unusable, drifting by
126.6~ft with the iteration count rising from 4--5 to 15--23. The batched
paths additionally carry run-to-run jitter on identical inputs of up to
$5.7\times10^{-6}$~ft in head, on both the dense and the sparse route, from
the non-deterministic reduction order of the index-scatter assembly; acceptance
criteria on these paths are therefore set relative to each path's own
run-to-run jitter and never as a fixed threshold.

\item \textbf{Coverage is \WrNets{} networks and a documented feature subset,
not ``all of EPANET''.} Three of the ten supported flow-unit systems are
exercised bit-exactly; the other seven are unit-tested only.

\item \textbf{The dense default path is slower than the reference engine}, by
a factor of \WrSweepPubMin{} to \WrSweepPubMax{} per single steady-state frame
across the public rows of the \WrSweepN-network size sweep, \WrSweepCityD{} to
\WrSweepCityH$\times$ on the operating rows (Table~\ref{tab:sweep}), and at $B=1$ the GPU loses to the CPU. The large
factors reported for L-TOWN are against our own prior serial CPU-adjoint
pipeline at 1.0-core occupancy, not against EPANET, which supplies no gradient
at any price; a perfectly eight-way parallel CPU adjoint would divide them by
up to eight, and that division is arithmetic rather than a measurement. Below
the crossover the sparse route is slower than the dense one, down to
\SnNetOneWorst$\times$ at $B=1{,}024$ on a nine-junction network.

\item \textbf{The leak demonstrations are one network and one injected
triple each, and the two inversions used differ in exactly one stage.}
The operating-network work-order inversion injects its three leaks
into the same model the inversion searches (an inverse crime by
construction, stated in the main text), and its noisy counterpart recovers
none of them; the L-TOWN inversion is one candidate set and one injected
triple, with a single-stage $\ell_1$ objective deliberately not augmented
with the support-refinement step City~D's inversion has. Repeating both
with the identifiability-driven augmented sensor sets of Note~10 recovers
no further leak that a control does not attribute to a sensor placed on
the leak's own junction. A second, coherence-driven placement objective
does recover one of the three leaks the noisy work-order inversion loses,
for a reason the signature coherence predicts, and is indistinguishable
from random placement for the other, and re-scoring that result on the
coarser stage the two inversions share leaves both conclusions unchanged
(Note~10): the stage the two inversions do not share is not what drives
them. What generalises is the diagnosis (identifiability census and
signature coherence), not the recovery, with one qualified exception
(Note~10).

\item \textbf{Every added sensor is virtual.} The augmentation
prescriptions of Note~10 are designed and verified in simulation on the
calibrated model; none has been installed, and installation,
re-calibration on field data and the field-side check of the restored
identifiability are later work.

\item \textbf{This is research code}, not validated for operational use.
\end{enumerate}

\clearpage
\section*{Supplementary Figures}

\clearpage
\section*{Supplementary Tables}

\begin{table}[p]
\centering
\caption{Complete forward verification suite: all 52 networks against the
EPANET~2.2 library, maxima over every frame. Newton iteration counts are
identical on all 52 (column omitted). $^{\rm p}$constant-power pump;
*custom curve; $^{\ddagger}$the deviation shown is the default input path's,
and this row reaches the acceptance criterion only under the exact-input
control of Note~1 (Table~\ref{tab:exactinput}).}
\label{tab:full}
\begingroup
\setlength{\tabcolsep}{2.6pt}\scriptsize
\begin{tabular}{@{}llrrllL{29mm}rrrl@{}}
\toprule
Network & Source & $N$ & $L$ & Loss & Units & Features & Frames &
max$|\Delta H|$ & max$|\Delta Q|$ & Outcome \\
 & & & & & & & & (ft) & (cfs) & \\
\midrule
Anytown & Public & 22 & 41 & H-W & GPM & \tiny pump1* & 9 & 6.082e-12 & 4.750e-12 & pass$^{\ddagger}$ \\
Anytown (WNTR) & Public & 25 & 46 & H-W & GPM & \tiny pump3* tank2 & 1441 & 0.000e+0 & 1.776e-15 & pass \\
Balerma & Public & 447 & 454 & D-W & LPS & \tiny pipes only & 1 & 2.842e-14 & 8.882e-16 & pass \\
BWSN Network 1 & Public & 129 & 178 & H-W & GPM & \tiny pump2 PRV8 tank2 ctrl1 rule4 & 207 & 6.086e-6 & 7.046e-6 & pass$^{\ddagger}$ \\
BWSN Network 2 & Public & 12527 & 14831 & H-W & GPM & \tiny pump4* PSV1 FCV4 tank2 ctrl1067 & 32 & 6.836e+0 & 1.406e-5 & pass$^{\ddagger}$ \\
C-Town (BATADAL) & Public & 396 & 444 & H-W & LPS & \tiny pump11 PRV3 FCV1 tank7 ctrl20 & 850 & 5.684e-14 & 8.882e-16 & pass \\
D-Town & Public & 407 & 459 & H-W & LPS & \tiny pump11 PRV4 TCV1 tank7 ctrl24 & 819 & 5.684e-14 & 1.776e-15 & pass \\
Fossolo (poly1) & Public & 37 & 58 & H-W & LPS & \tiny pipes only & 1 & 0.000e+0 & 6.939e-18 & pass \\
Hanoi & Public & 32 & 34 & H-W & LPS & \tiny pipes only & 1 & 1.421e-14 & 1.421e-14 & pass \\
ky10 & Public & 935 & 1061 & H-W & GPM & \tiny pump13$^{\rm p}$ PRV5 tank13 ctrl6 & 1 & 0.000e+0 & 2.220e-16 & pass \\
ky3 & Public & 279 & 354 & H-W & GPM & \tiny pump5$^{\rm p}$ tank3 ctrl2 & 25 & 0.000e+0 & 8.882e-16 & pass \\
ky4 & Public & 964 & 1158 & H-W & GPM & \tiny pump2$^{\rm p}$ tank4 ctrl2 & 1 & 0.000e+0 & 4.441e-16 & pass \\
ky5 & Public & 430 & 507 & H-W & GPM & \tiny pump11$^{\rm p}$ tank3 ctrl8 & 31 & 0.000e+0 & 3.553e-15 & pass \\
L-TOWN & Public & 785 & 909 & H-W & CMH & \tiny pump1 PRV3 tank1 ctrl2 & 2031 & 2.842e-14 & 1.110e-16 & pass \\
L-TOWN (Real) & Public & 785 & 909 & H-W & CMH & \tiny pump1 PRV3 tank1 ctrl2 & 2030 & 2.842e-14 & 1.110e-16 & pass \\
Modena & Public & 272 & 317 & H-W & LPS & \tiny pipes only & 1 & 2.842e-14 & 8.882e-16 & pass \\
Net1 & Public & 11 & 13 & H-W & GPM & \tiny pump1 tank1 ctrl2 & 27 & 0.000e+0 & 4.441e-16 & pass \\
Net2 & Public & 36 & 40 & H-W & GPM & \tiny tank1 & 56 & 0.000e+0 & 2.220e-16 & pass \\
Net3 & Public & 97 & 119 & H-W & GPM & \tiny pump2 tank3 ctrl18 & 183 & 2.146e-6 & 5.684e-7 & pass$^{\ddagger}$ \\
Net6 & Public & 3356 & 3892 & H-W & GPM & \tiny pump61$^{\rm p}$ PRV2 tank32 ctrl124 & 70/609 & 1.525e-5 & 4.739e-6 & pass$^{\ddagger}$ \\
Pescara & Public & 71 & 99 & H-W & LPS & \tiny pipes only & 1 & 1.421e-14 & 8.882e-16 & pass \\
Richmond (skeleton) & Public & 48 & 51 & H-W & LPS & \tiny pump7* tank6 ctrl14 & 91 & 1.137e-13 & 2.220e-16 & pass \\
Richmond (standard) & Public & 872 & 957 & H-W & LPS & \tiny pump7* PRV1 tank6 ctrl16 & 55 & 1.137e-13 & 2.220e-16 & pass \\
\addlinespace
EXA4 & Example & 396 & 444 & H-W & LPS & \tiny pump11 PRV3 TCV1 tank7 ctrl22 & 1 & 5.684e-14 & 8.882e-16 & pass \\
EXA5 & Example & 193 & 273 & H-W & LPS & \tiny pump2 PRV3 tank1 ctrl4 rule1 & 52 & 5.684e-14 & 8.882e-16 & pass \\
EXA6 & Example & 111 & 123 & H-W & GPM & \tiny pump1 tank1 & 25 & 0.000e+0 & 8.882e-16 & pass \\
\addlinespace
\textbf{City D} & Operating & 542 & 554 & H-W & LPS & \tiny TCV79 & 25 & 1.421e-14 & 1.776e-15 & pass \\
\textbf{City D (emitter)} & Operating & 542 & 554 & H-W & LPS & \tiny TCV79 emit5 & 25 & 1.421e-14 & 3.553e-15 & pass \\
\textbf{City H} & Operating & 921 & 1038 & H-W & LPS & \tiny pump6 & 25 & 1.421e-14 & 7.105e-15 & pass \\
\addlinespace
Synthetic-M0 & Synthetic & 104 & 176 & H-W & LPS & \tiny pipes only & 1 & 0.000e+0 & 5.551e-17 & pass \\
Synthetic-M1 & Synthetic & 46 & 67 & H-W & LPS & \tiny pipes only & 1 & 0.000e+0 & 1.110e-16 & pass \\
Synthetic-M10 & Synthetic & 39 & 63 & H-W & LPS & \tiny pipes only & 1 & 0.000e+0 & 5.551e-17 & pass \\
Synthetic-M11 & Synthetic & 72 & 133 & H-W & LPS & \tiny pipes only & 1 & 2.842e-14 & 2.776e-17 & pass \\
Synthetic-M12 & Synthetic & 89 & 162 & H-W & LPS & \tiny pipes only & 1 & 2.842e-14 & 2.220e-16 & pass \\
Synthetic-M13 & Synthetic & 72 & 105 & H-W & LPS & \tiny pipes only & 1 & 2.842e-14 & 2.220e-16 & pass \\
Synthetic-M14 & Synthetic & 106 & 220 & H-W & LPS & \tiny pipes only & 1 & 2.842e-14 & 1.110e-16 & pass \\
Synthetic-M15 & Synthetic & 47 & 79 & H-W & LPS & \tiny pipes only & 1 & 0.000e+0 & 1.388e-17 & pass \\
Synthetic-M16 & Synthetic & 98 & 142 & H-W & LPS & \tiny pipes only & 1 & 2.842e-14 & 2.220e-16 & pass \\
Synthetic-M17 & Synthetic & 89 & 178 & H-W & LPS & \tiny pipes only & 1 & 2.842e-14 & 5.551e-17 & pass \\
Synthetic-M18 & Synthetic & 117 & 235 & H-W & LPS & \tiny pipes only & 1 & 0.000e+0 & 5.551e-17 & pass \\
Synthetic-M19 & Synthetic & 85 & 168 & H-W & LPS & \tiny pipes only & 1 & 2.842e-14 & 5.551e-17 & pass \\
Synthetic-M2 & Synthetic & 101 & 135 & H-W & LPS & \tiny pipes only & 1 & 0.000e+0 & 5.551e-17 & pass \\
Synthetic-M3 & Synthetic & 36 & 52 & H-W & LPS & \tiny pipes only & 1 & 0.000e+0 & 1.388e-17 & pass \\
Synthetic-M4 & Synthetic & 81 & 143 & H-W & LPS & \tiny pipes only & 1 & 0.000e+0 & 1.388e-17 & pass \\
Synthetic-M5 & Synthetic & 87 & 141 & H-W & LPS & \tiny pipes only & 1 & 0.000e+0 & 4.441e-16 & pass \\
Synthetic-M6 & Synthetic & 37 & 61 & H-W & LPS & \tiny pipes only & 1 & 2.842e-14 & 5.551e-17 & pass \\
Synthetic-M7 & Synthetic & 65 & 105 & H-W & LPS & \tiny pipes only & 1 & 2.842e-14 & 5.551e-17 & pass \\
Synthetic-M8 & Synthetic & 74 & 148 & H-W & LPS & \tiny pipes only & 1 & 2.842e-14 & 2.220e-16 & pass \\
Synthetic-M9 & Synthetic & 39 & 50 & H-W & LPS & \tiny pipes only & 1 & 0.000e+0 & 6.939e-18 & pass \\
Synthetic-S0 & Synthetic & 43 & 68 & H-W & LPS & \tiny pipes only & 1 & 1.421e-14 & 1.110e-16 & pass \\
Synthetic-S1 & Synthetic & 117 & 181 & H-W & LPS & \tiny pipes only & 1 & 2.842e-14 & 1.110e-16 & pass \\
Synthetic-S2 & Synthetic & 73 & 122 & H-W & LPS & \tiny pipes only & 1 & 0.000e+0 & 5.551e-17 & pass \\
\bottomrule
\end{tabular}
\endgroup

\end{table}

\begin{table}[htbp]
\centering
\caption{Health of the truncated-unrolling demand gradient on the seven
public networks it was measured on: relative error of the $K$-step gradient
against the implicit adjoint. The route refuses PRV networks at
construction.}
\label{tab:health}
\begin{tabular}{@{}lrl@{}}
\toprule
Network & rel.\ error at $K$ & outcome \\
\midrule
Net1 & $1.25\times10^{-13}$ & usable \\
Hanoi & $1.52\times10^{-11}$ & usable \\
Net2 & $4.03\times10^{-10}$ & usable \\
Modena & $5.47\times10^{-11}$ & usable \\
Pescara & $1.10\times10^{-7}$ & conditional \\
Net3 & $3.98\times10^{-3}$ & not recommended \\
ky4 & $1.80\times10^{-4}$ & not recommended \\
\bottomrule
\end{tabular}

\end{table}

\begin{table}[p]
\centering
\caption{External gradient checks against the compiled EPANET~2.2 library.
(a)~The public emitter-augmented Hanoi configuration, per coordinate. (b)~City~D roughness class by
sensitivity band and (c)~City~D emitter variant by parameter class,
summarised without identifiers per the anonymisation rule.}
\label{tab:external}
\begingroup\setlength{\tabcolsep}{4pt}\small
\begin{tabular}{@{}llrrrr@{}}
\multicolumn{6}{@{}l}{(a) The public emitter-augmented Hanoi configuration (Hanoi + 5 emitters), per coordinate; threshold $10^{-4}$} \\[1pt]
\toprule
Class & Coordinate & $x_0$ & EPANET-FD & Adjoint & Rel.\ error \\
\midrule
emitter coefficient $C$ & \texttt{12} & 1.2 & -0.305558467 & -0.305558468 & $8.60\times10^{-10}$ \\
 & \texttt{17} & 1 & 0.221943309 & 0.22194331 & $6.35\times10^{-9}$ \\
 & \texttt{30} & 1.5 & 0.115063779 & 0.115063777 & $1.79\times10^{-8}$ \\
nodal demand $d$ & \texttt{21} & 9.1228 & 9.35450703 & 9.35448105 & $2.78\times10^{-6}$ \\
 & \texttt{22} & 4.7576 & 12.5026841 & 12.5026823 & $1.43\times10^{-7}$ \\
 & \texttt{13} & 9.221 & -4.47928896 & -4.47903978 & $5.56\times10^{-5}$ \\
reservoir head $H_0$ & \texttt{1} & 328.08 & -1.21584257 & -1.2158337 & $7.30\times10^{-6}$ \\
pipe roughness $C_{\rm HW}$ & \texttt{21} & 130 & -0.809110601 & -0.80911228 & $2.07\times10^{-6}$ \\
\bottomrule
\end{tabular}

\vspace{7pt}
\begin{tabular}{@{}L{78mm}rrl@{}}
\multicolumn{4}{@{}l}{(b) City D, roughness class in breadth; 30 pipes by sensitivity band, no identifiers} \\[1pt]
\toprule
Sensitivity band & Pipes & Worst rel.\ error & Outcome \\
\midrule
$|\mathrm{d}L/\mathrm{d}C| \ge 3\times10^{-2}$ (12 pipes sampled of 22) & 12 & $1.26\times10^{-5}$ & PASS \\
$[10^{-2}, 3\times10^{-2})$ (12 sampled of 29) & 12 & $4.78\times10^{-5}$ & PASS \\
clamped at the low-flow floor (all 6; analytic gradient exactly 0) & 6 & $|g_{\rm FD}| \le$ $6.45\times10^{-7}$ & PASS \\
\bottomrule
\end{tabular}

\vspace{7pt}
\begin{tabular}{@{}L{78mm}rrl@{}}
\multicolumn{4}{@{}l}{(c) City D emitter variant, four parameter classes over 8 coordinates} \\[1pt]
\toprule
Parameter class & Coords & Worst rel.\ error & Outcome \\
\midrule
emitter coefficient $C$ & 3 & $9.43\times10^{-6}$ & PASS \\
nodal demand $d$ & 3 & $2.39\times10^{-5}$ & PASS \\
reservoir head $H_0$ & 1 & $3.93\times10^{-7}$ & PASS \\
pipe roughness $C_{\rm HW}$ & 1 & $3.04\times10^{-6}$ & PASS \\
\bottomrule
\end{tabular}
\endgroup

\end{table}

\begin{table}[p]
\centering
\caption{Gradient audit on degenerate cases: 22 hand-constructed cases,
each against its own threshold; $^{\rm o}$marks the six on the operating
network. ``$0$ (structural)'' means the correct derivative is analytically
zero and the computed one is exactly zero.}
\label{tab:audit}
\begingroup\setlength{\tabcolsep}{4pt}\small
\begin{tabular}{@{}L{86mm}rrrl@{}}
\toprule
Degenerate case & Worst & Threshold & Fraction & Outcome \\
\midrule
near-clamp step, dead-end pipe, resistance gradient & $0$ (structural) & exactly $0$ & -- & PASS \\
near-clamp step, dead-end junction, demand gradient & $2.08\times10^{-11}$ & $1.00\times10^{-6}$ & $<0.001$ & PASS \\
near-clamp step, finite-difference check, 9 coordinates & $3.57\times10^{-11}$ & $1.00\times10^{-6}$ & $<0.001$ & PASS \\
true clamp, clamped low-flow link, resistance gradient must be exactly zero & $0$ (structural) & exactly $0$ & -- & PASS \\
true clamp, dead-end junction, demand gradient & $4.87\times10^{-8}$ & $1.00\times10^{-6}$ & 0.049 & PASS \\
true clamp, finite-difference check, 9 coordinates & $2.49\times10^{-10}$ & $1.00\times10^{-6}$ & $<0.001$ & PASS \\
operating network, closed-link neighbourhood: step inside the clamped branch, $|\Delta G|$ against its noise ceiling$^{\rm o}$ & $1.74\times10^{-1}$ & $1.00\times10^{0}$ & 0.174 & PASS \\
operating network, base load added: implicit vs unrolled cross-check$^{\rm o}$ & $2.73\times10^{-7}$ & $1.00\times10^{-4}$ & 0.003 & PASS \\
operating network, base load added: finite differences, micro-switch averaged$^{\rm o}$ & $1.22\times10^{-5}$ & $1.00\times10^{-3}$ & 0.012 & PASS \\
operating network, throttle-valve endpoints: demand gradient vs finite differences$^{\rm o}$ & $5.75\times10^{-7}$ & $1.00\times10^{-6}$ & 0.575 & PASS \\
operating network, throttle valves and closed links: resistance gradient must be exactly zero$^{\rm o}$ & $0$ (structural) & exactly $0$ & -- & PASS \\
implicit adjoint at a zero-emitter node: finite and exactly zero & $0$ (structural) & exactly $0$ & -- & PASS \\
truncated unrolling at a zero-emitter node: finite and exactly zero & $0$ (structural) & exactly $0$ & -- & PASS \\
control: finite differences at a live emitter coordinate & $8.02\times10^{-11}$ & $1.00\times10^{-6}$ & $<0.001$ & PASS \\
zero-demand junction, demand gradient & $1.56\times10^{-9}$ & $1.00\times10^{-6}$ & 0.002 & PASS \\
two-reservoir head gradient against finite differences & $4.03\times10^{-12}$ & $1.00\times10^{-6}$ & $<0.001$ & PASS \\
head-shift identity & $1.32\times10^{-15}$ & $1.00\times10^{-9}$ & $<0.001$ & PASS \\
direct reservoir-head term & $1.11\times10^{-16}$ & $1.00\times10^{-12}$ & $<0.001$ & PASS \\
GPU against CPU on the same batch, synthetic 0009 & $3.48\times10^{-16}$ & $1.00\times10^{-8}$ & $<0.001$ & PASS \\
operating network, GPU against CPU on the same batch$^{\rm o}$ & $2.76\times10^{-16}$ & $1.00\times10^{-8}$ & $<0.001$ & PASS \\
the tensor library's own finite-difference gradient check, synthetic 0015 & $0$ (structural) & exactly $0$ & -- & PASS \\
the tensor library's own finite-difference gradient check, synthetic 0006 & $0$ (structural) & exactly $0$ & -- & PASS \\
\bottomrule
\end{tabular}
\endgroup

\end{table}

\begin{table}[p]
\centering
\caption{The regression suite, item by item: 53 numerical checks, all
passing, plus one packaging guard, clean over this submission package
(Note~9). $^{\rm o}$operating-network
item; $^{\ddagger}$two networks, City~D and one public;
$^{\S}$deliberately seeded mutants that must fail.}
\label{tab:regression}
\begingroup\setlength{\tabcolsep}{3pt}\footnotesize
\begin{tabular}{@{}L{56mm}L{78mm}@{}}
\toprule
Item & Measured (every threshold met; ft and cfs) \\
\midrule
\addlinespace\multicolumn{2}{@{}l}{\itshape Steady-state alignment against the reference library} \\
\quad 23 synthetic networks & worst $\max|\Delta H| = 2.84\times10^{-14}$ \\
\quad City D$^{\rm o}$ & $\Delta H\,1.42\times10^{-14}$, $\Delta Q\,1.78\times10^{-15}$, $\Delta Q_{\rm cl}\,4.60\times10^{-8}$ \\
\quad City D (emitter)$^{\rm o}$ & $\Delta H\,1.42\times10^{-14}$, $\Delta Q\,3.55\times10^{-15}$, $\Delta Q_{\rm cl}\,4.65\times10^{-8}$, $\Delta e\,2.78\times10^{-16}$ \\
\addlinespace\multicolumn{2}{@{}l}{\itshape Snapshot replay} \\
\quad replay EXA4 & $\Delta H\,5.68\times10^{-14}$, $\Delta Q\,8.88\times10^{-16}$ \\
\quad replay EXA6 & $\Delta H\,0$, $\Delta Q\,8.88\times10^{-16}$ \\
\quad replay City H$^{\rm o}$ & $\Delta H\,1.42\times10^{-14}$, $\Delta Q\,7.11\times10^{-15}$ \\
\quad replay ky3 & $\Delta H\,0$, $\Delta Q\,8.88\times10^{-16}$ \\
\quad replay ky5 & $\Delta H\,0$, $\Delta Q\,3.55\times10^{-15}$ \\
\addlinespace\multicolumn{2}{@{}l}{\itshape Autonomous extended-period simulation} \\
\quad EPS EXA4 & $\Delta H\,5.68\times10^{-14}$, $\Delta Q\,8.88\times10^{-16}$ \\
\quad EPS EXA5 & $\Delta H\,5.68\times10^{-14}$, $\Delta Q\,8.88\times10^{-16}$ \\
\quad EPS EXA6 & $\Delta H\,0$, $\Delta Q\,8.88\times10^{-16}$ \\
\quad EPS City H$^{\rm o}$ & $\Delta H\,1.42\times10^{-14}$, $\Delta Q\,7.11\times10^{-15}$ \\
\quad EPS ky3 & $\Delta H\,0$, $\Delta Q\,8.88\times10^{-16}$ \\
\quad EPS ky5 & $\Delta H\,0$, $\Delta Q\,3.55\times10^{-15}$ \\
\quad EPS Anytown & $\Delta H\,6.08\times10^{-12}$, $\Delta Q\,4.75\times10^{-12}$ \\
\addlinespace\multicolumn{2}{@{}l}{\itshape Physical cross-validation} \\
\quad cross-validation EXA4 & $\text{mass}\,1.14\times10^{-6}$, $\text{dem}\,6.94\times10^{-18}$, $\text{H--W}\,1.34\times10^{-8}$ \\
\quad cross-validation EXA5 & $\text{mass}\,1.62\times10^{-6}$, $\text{dem}\,6.94\times10^{-18}$, $\text{H--W}\,6.43\times10^{-15}$ \\
\quad cross-validation EXA6 & $\text{mass}\,1.28\times10^{-6}$, $\text{dem}\,5.55\times10^{-17}$, $\text{H--W}\,3.73\times10^{-10}$ \\
\quad cross-validation City H$^{\rm o}$ & $\text{mass}\,6.17\times10^{-7}$, $\text{dem}\,2.22\times10^{-16}$, $\text{H--W}\,3.55\times10^{-15}$ \\
\quad cross-validation ky3 & $\text{mass}\,3.07\times10^{-9}$, $\text{dem}\,5.55\times10^{-17}$, $\text{H--W}\,5.86\times10^{-15}$ \\
\quad cross-validation ky5 & $\text{mass}\,5.99\times10^{-6}$, $\text{dem}\,1.39\times10^{-17}$, $\text{H--W}\,1.79\times10^{-13}$ \\
\quad cross-validation City D$^{\rm o}$ & $\text{mass}\,6.76\times10^{-7}$, $\text{dem}\,4.44\times10^{-16}$, $\text{H--W}\,1.43\times10^{-14}$ \\
\addlinespace\multicolumn{2}{@{}l}{\itshape Gradient cross-checks} \\
\quad 2 networks $\times$ 4 parameter classes & worst implicit $5.85\times10^{-8}$, worst unrolled $1.21\times10^{-6}$, City D included$^{\ddagger}$ \\
\quad the tensor library's own finite-difference gradient check, synthetic 0009 & pass \\
\quad City D, batch-of-8 vs single-scenario consistency$^{\rm o}$ & demand $0$, resistance $0$ (exact) \\
\addlinespace\multicolumn{2}{@{}l}{\itshape Degenerate-case gradient audit} \\
\quad 22 degenerate cases, own thresholds & 22/22 pass (6 on City D; Table~\ref{tab:audit}) \\
\addlinespace\multicolumn{2}{@{}l}{\itshape Capability parity} \\
\quad D--W, Chezy--Manning and FCV vs the library & $\Delta H \le 2.842\times10^{-14}$~ft on all three \\
\addlinespace\multicolumn{2}{@{}l}{\itshape Darcy--Weisbach gradient} \\
\quad adjoint vs central differences, Balerma & worst $6.142\times10^{-7}$ \\
\addlinespace\multicolumn{2}{@{}l}{\itshape Symmetry guard} \\
\quad $\max|\mathbf{A}-\mathbf{A}^{\!\top}|$ per Newton round & 86/86 bitwise symmetric; 4/4 mutants red$^{\S}$ \\
\addlinespace\multicolumn{2}{@{}l}{\itshape Status-schedule parity} \\
\quad status schedule and PRV subsequences & 0 disagreements; 3/3 mutants red \\
\addlinespace\multicolumn{2}{@{}l}{\itshape Bit-level reference comparison} \\
\quad L-TOWN bit for bit; float32 refused & 785/785 heads, 909/909 flows \\
\addlinespace\multicolumn{2}{@{}l}{\itshape Packaging guard} \\
\quad tracked contents, paths, commit message & 13 name variants; 0 hits over every file, path and compiled PDF of this submission package (Note~9) \\
\bottomrule
\end{tabular}
\endgroup

\end{table}

\begin{table}[htbp]
\centering
\caption{L-TOWN cost against batch size on two RTX~5090 nodes
(minimum--maximum), double precision. (a)~Wall time per scenario with the
two-factor decomposition $F_1$ (adjoint onto GPU) and $F_2$ (sparse versus
dense). (b)~Peak device memory; fwd, forward only; f+b, forward plus
backward.}
\label{tab:ltowncost}
\begingroup\setlength{\tabcolsep}{2.6pt}\scriptsize
\begin{tabular}{@{}rrrrrrrrrr@{}}
\multicolumn{10}{@{}l}{(a) Wall time per scenario (ms); forward and forward-plus-backward} \\[1pt]
\toprule
 & \multicolumn{3}{c}{forward} & \multicolumn{6}{c}{forward + backward} \\
\cmidrule(lr){2-4}\cmidrule(lr){5-10}
$B$ & dense & sparse & d$\div$s & CPU adj. & GPU dense & GPU sparse & $F_1$ & $F_2$ & $F_1F_2$ \\
\midrule
1 & 85.98--100.51 & 56.97--79.01 & 1.27--1.51 & 420.06--422.51 & 152.41--152.90 & 107.80--109.03 & 2.75--2.77 & 1.40--1.42 & 3.88--3.90 \\
8 & 20.57--22.98 & 7.71--10.76 & 2.13--2.67 & 349.51--350.87 & 47.07--47.13 & 15.18--15.22 & 7.42--7.44 & 3.09--3.11 & 22.96--23.12 \\
64 & 5.93--6.19 & 1.38--2.01 & 3.08--4.31 & 334.26--335.33 & 12.01--12.08 & 3.195--3.205 & 27.76--27.83 & 3.75--3.78 & 104.29--104.95 \\
256 & 3.78--3.85 & 0.69--1.03 & 3.75--5.48 & 332.90--334.39 & 6.642--6.670 & 1.895--1.909 & 50.12--50.13 & 3.49--3.50 & 175.19--175.66 \\
512 & 3.48--3.50 & 0.57--0.85 & 4.12--6.10 & 333.40--333.49 & 5.816--5.824 & 1.671--1.679 & 57.26--57.33 & 3.46--3.49 & 198.52--199.60 \\
1024 & 3.33--3.35 & 0.52--0.77 & 4.36--6.42 & OOM & OOM & 1.612--1.617 & -- & -- & 198.06--201.23 \\
\bottomrule
\end{tabular}

\vspace{7pt}
\begin{tabular}{@{}rrrrrr@{}}
\multicolumn{6}{@{}l}{(b) Peak device memory (MiB), one fresh process per cell} \\[1pt]
\toprule
$B$ & fwd dense & fwd sparse & f+b dense & f+b sparse & dense $\div$ sparse \\
\midrule
1 & 112 & 46 & 194 & 214 & 0.91 \\
8 & 292 & 48 & 300 & 86 & 3.49 \\
64 & 1\,626 & 124 & 1\,634 & 160 & 10.21 \\
256 & 7\,338 & 296 & 7\,346 & 332 & 22.13 \\
512 & 14\,564 & 590 & 14\,572 & 626 & 23.28 \\
1024 & 29\,068 & 1\,212 & 29\,096 & 1\,286 & 22.63 \\
\bottomrule
\end{tabular}
\endgroup

\end{table}

\begin{table}[htbp]
\centering
\caption{Cost per steady-state frame across the size sweep: 14 public
networks plus the two operating networks, replica path against the reference
library, single machine. The factor buys a derivative the reference engine
does not supply.}
\label{tab:sweep}
\begingroup\setlength{\tabcolsep}{4pt}\small
\begin{tabular}{@{}lrrrrr@{}}
\toprule
Network & $N$ & $L$ & replica (ms) & reference (ms) & replica $\div$ ref. \\
\midrule
Hanoi & 32 & 34 & 0.498 & 0.0184 & 27.0 \\
synthetic 0003 & 36 & 52 & 1.052 & 0.0294 & 35.8 \\
Fossolo (poly1) & 37 & 58 & 1.590 & 0.0219 & 72.6 \\
Pescara & 71 & 99 & 2.824 & 0.0594 & 47.5 \\
Net3 & 97 & 119 & 1.831 & 0.0715 & 25.6 \\
Modena & 272 & 317 & 6.489 & 0.1871 & 34.7 \\
EXA4 & 396 & 444 & 8.181 & 0.2140 & 38.2 \\
Balerma & 447 & 454 & 5.473 & 0.1181 & 46.3 \\
\textbf{City D} (operating) & 542 & 554 & 6.371 & 0.3038 & 21.0 \\
L-TOWN & 785 & 909 & 36.199 & 0.7390 & 49.0 \\
Richmond (standard) & 872 & 957 & 13.982 & 0.3449 & 40.5 \\
\textbf{City H} (operating) & 921 & 1038 & 15.007 & 0.3200 & 46.9 \\
ky10 & 935 & 1061 & 24.195 & 0.5110 & 47.3 \\
ky4 & 964 & 1158 & 23.036 & 0.5339 & 43.1 \\
Net6 & 3356 & 3892 & 79.354 & 2.6096 & 30.4 \\
BWSN Network 2 & 12527 & 14831 & 478.214 & 9.5961 & 49.8 \\
\bottomrule
\end{tabular}
\endgroup

\end{table}

\begin{table}[htbp]
\centering
\caption{Calibration noise ladders (a) and City~D robustness arms (b).
$^{\rm ic}$inverse crime: truth generated by the model being inverted;
Modena has only such rows, which is why it appears here and not in Results
(Supplementary Note~8).}
\label{tab:calib}
\begingroup\setlength{\tabcolsep}{4pt}\small
\begin{tabular}{@{}lrlrrl@{}}
\multicolumn{6}{@{}l}{(a) Noise ladders; median final training MSE over the seeds of each arm} \\[1pt]
\toprule
Network & Free pipes & $\sigma$ (ft) & Seeds & Median MSE & Note \\
\midrule
Hanoi (public) & 34 & 0 (truth per-pipe) & 1 & $3.00\times10^{-11}$ & inverse crime$^{\rm ic}$ \\
Hanoi (public) & 34 & 0 (truth grouped) & 1 & $9.73\times10^{-28}$ & inverse crime$^{\rm ic}$ \\
Hanoi (public) & 34 & 0.03 & 3 & $8.85\times10^{-4}$ &  \\
Hanoi (public) & 34 & 0.1 & 3 & $9.83\times10^{-3}$ &  \\
Hanoi (public) & 34 & 0.3 & 3 & $8.85\times10^{-2}$ &  \\
\addlinespace
Modena (public) & 317 & 0 (truth per-pipe) & 1 & $3.25\times10^{-6}$ & inverse crime$^{\rm ic}$ \\
Modena (public) & 317 & 0 (truth grouped) & 1 & $1.96\times10^{-5}$ & inverse crime$^{\rm ic}$ \\
\addlinespace
\textbf{City D} (operating) & 432 & 0 (truth per-pipe) & 1 & $2.98\times10^{-6}$ & inverse crime$^{\rm ic}$ \\
\textbf{City D} (operating) & 432 & 0 (truth grouped) & 1 & $3.39\times10^{-6}$ & inverse crime$^{\rm ic}$ \\
\textbf{City D} (operating) & 432 & 0.03 & 5 & $8.43\times10^{-4}$ &  \\
\textbf{City D} (operating) & 432 & 0.1 & 5 & $9.40\times10^{-3}$ &  \\
\textbf{City D} (operating) & 432 & 0.3 & 5 & $8.48\times10^{-2}$ &  \\
\bottomrule
\end{tabular}

\vspace{7pt}
\begin{tabular}{@{}L{86mm}rr@{}}
\multicolumn{3}{@{}l}{(b) City D robustness arms at $\sigma = 0.1$~ft (median final training MSE)} \\[1pt]
\toprule
Arm & Runs & MSE \\
\midrule
demand model error: truth $\times\,U[0.85, 1.15]$, inverted on nominal demand & 4 & $9.51\times10^{-3}$ (worst $1.62\times10^{-2}$) \\
paired control, demand known & 4 & $9.11\times10^{-3}$ \\
sensor bias 5\,\% ($\sigma_b = 0.05$) & 5 & $9.86\times10^{-3}$ \\
multi-start: 8 Latin-hypercube starts, $C \in [80, 145]$ & 8 & $9.34\times10^{-3}$--$9.42\times10^{-3}$ ($1.0087\times$ spread) \\
\bottomrule
\end{tabular}
\endgroup

\end{table}

\begin{table}[htbp]
\centering
\caption{Sensor augmentation, all virtual (designed and verified in
simulation; none installed). (a)~City~D: recovery of the 149 unobservable
pipes when $k$ sensors are added to the 40 designated, both objectives,
against reselection from scratch. (b)~The same prescription on L-TOWN and
Hanoi (Supplementary Note~10).}
\label{tab:augment}
\begingroup\setlength{\tabcolsep}{3pt}\footnotesize
\begin{tabular}{@{}lrrrrrrrr@{}}
\multicolumn{9}{@{}l}{(a) City D: 149 pipes unobservable under the 40 designated sensors; 541 candidates, 25 frames, census threshold, $\sigma_{\rm noise}=0.1$~ft} \\[1pt]
\toprule
Design & $k$ & Sensors & Recovered & Still unobs. & Lost & Rank & $k_{\rm sub}$ & CRLB$_{\rm sub}$ \\
\midrule
S$_0$ (designated) & 0 & 40 & 0 & 149 & 0 & 110 & 20 & 742 \\
cover $+k$ & 5 & 45 & 37 & 112 & 0 & 122 & 22 & 489 \\
cover $+k$ & 10 & 50 & 62 & 87 & 0 & 132 & 28 & 686 \\
cover $+k$ & 20 & 60 & 95 & 54 & 0 & 144 & 33 & 656 \\
cover $+k$ & 40 & 80 & 133 & 16 & 0 & 163 & 36 & 347 \\
cover $+k$ & 80 & 120 & 149 & 0 & 0 & 202 & 59 & 486 \\
\addlinespace
D-opt $+k$ & 5 & 45 & 21 & 128 & 0 & 118 & 24 & 457 \\
D-opt $+k$ & 10 & 50 & 23 & 126 & 0 & 123 & 28 & 359 \\
D-opt $+k$ & 20 & 60 & 35 & 114 & 0 & 133 & 36 & 222 \\
D-opt $+k$ & 40 & 80 & 52 & 97 & 0 & 157 & 56 & 512 \\
D-opt $+k$ & 80 & 120 & 76 & 73 & 0 & 196 & 79 & 1430 \\
\addlinespace
reselect $40{+}k$ & 5 & 45 & 59 & 128 & 38 & 128 & -- & -- \\
reselect $40{+}k$ & 10 & 50 & 59 & 128 & 38 & 134 & -- & -- \\
reselect $40{+}k$ & 20 & 60 & 68 & 105 & 24 & 149 & -- & -- \\
reselect $40{+}k$ & 40 & 80 & 73 & 97 & 21 & 166 & -- & -- \\
reselect $40{+}k$ & 80 & 120 & 94 & 70 & 15 & 201 & -- & -- \\
reselect 40 & 0 & 40 & 52 & 135 & 38 & 122 & -- & -- \\
\bottomrule
\end{tabular}

\vspace{6pt}
\begin{tabular}{@{}llrrrrr@{}}
\multicolumn{7}{@{}l}{(b) Public networks: recovered of the unobservable pipes at $k$ added (nothing lost at any $k$ under either objective)} \\[1pt]
\toprule
Network & Design & $+5$ & $+10$ & $+20$ & $+40$ & $+80$ \\
\midrule
L-TOWN, 33 fixed, 62 unobservable & add, coverage & 21 & 33 & 46 & 62 & 62 \\
 & add, D-optimal & 2 & 2 & 3 & 7 & 10 \\
 & reselect $33+k$: recovered / lost & 6/99 & 6/99 & 8/99 & 13/98 & 15/2 \\
\addlinespace
Hanoi, 10 fixed, 1 unobservable & add, coverage & 1 & 1 & 1 & -- & -- \\
 & add, D-optimal & 1 & 1 & 1 & -- & -- \\
\bottomrule
\end{tabular}
\endgroup

\end{table}

\begin{table}[htbp]
\centering
\caption{Calibration with the augmented sensor sets, the calibrator,
truth, noise and hold-out of Supplementary Note~8 unchanged. (a)~City~D:
error of the restored pipes against their prior, and the RMSE over all
informative pipes. (b)~Hanoi (Supplementary Note~10).}
\label{tab:augcalib}
\begingroup\setlength{\tabcolsep}{3pt}\footnotesize
\begin{tabular}{@{}lrlrrrrrr@{}}
\multicolumn{9}{@{}l}{(a) City D, 432 free pipes, one noise seed per cell, 20\,\% of sensors held out} \\[1pt]
\toprule
Set & Sensors & $\sigma$ (ft) & Inform. & Restored & Median $|\Delta C|$ & Prior & Better & RMSE, inform. \\
\midrule
S$_0$ & 40 & 0.03 & 279 & 0 & -- & -- & -- & 24.4 \\
S$_0$ & 40 & 0.1 & 279 & 0 & -- & -- & -- & 25.6 \\
S$_0$ & 40 & 0.3 & 279 & 0 & -- & -- & -- & 26.3 \\
\addlinespace
D-opt $+20$ & 60 & 0.03 & 314 & 35 & 11.5 & 21.9 & 21/35 & 22.9 \\
D-opt $+20$ & 60 & 0.1 & 314 & 35 & 11.5 & 21.9 & 24/35 & 23.2 \\
D-opt $+20$ & 60 & 0.3 & 314 & 35 & 12.2 & 21.9 & 20/35 & 30.3 \\
\addlinespace
D-opt $+40$ & 80 & 0.03 & 331 & 52 & 12.1 & 18.3 & 26/52 & 22.4 \\
D-opt $+40$ & 80 & 0.1 & 331 & 52 & 13.2 & 18.3 & 26/52 & 23.6 \\
D-opt $+40$ & 80 & 0.3 & 331 & 52 & 11.9 & 18.3 & 23/52 & 23.9 \\
\addlinespace
cover $+20$ & 60 & 0.03 & 374 & 95 & 14.8 & 19.1 & 45/95 & 23.5 \\
cover $+20$ & 60 & 0.1 & 374 & 95 & 15.5 & 19.1 & 44/95 & 24.3 \\
cover $+20$ & 60 & 0.3 & 374 & 95 & 19.7 & 19.1 & 40/95 & 29.5 \\
\addlinespace
cover $+40$ & 80 & 0.03 & 412 & 133 & 16.9 & 19.5 & 58/133 & 23.0 \\
cover $+40$ & 80 & 0.1 & 412 & 133 & 18.4 & 19.5 & 56/133 & 24.2 \\
cover $+40$ & 80 & 0.3 & 412 & 133 & 22.0 & 19.5 & 49/133 & 30.0 \\
\bottomrule
\end{tabular}

\vspace{6pt}
\begin{tabular}{@{}lrrrrr@{}}
\multicolumn{6}{@{}l}{(b) Hanoi, 34 free pipes, median RMSE over the informative pipes across three noise seeds} \\[1pt]
\toprule
Set & Sensors & Unobservable & $\sigma=0.03$ & $\sigma=0.1$ & $\sigma=0.3$ \\
\midrule
S$_0$ (10 random) & 10 & 1 & 18.3 & 18.7 & 19.4 \\
D-opt $+5$ & 15 & 0 & 9.9 & 11.0 & 14.4 \\
D-opt $+10$ & 20 & 0 & 7.8 & 10.2 & 17.0 \\
D-opt $+20$ & 30 & 0 & 6.2 & 7.0 & 9.8 \\
\bottomrule
\end{tabular}
\endgroup

\end{table}

\begin{table}[htbp]
\centering
\caption{Leak search with the augmented sensor sets.
(a)~The work-order search on City~D with the designated set, the six
augmented sets and the control experiments; $^{\dagger}$an added sensor sits
on $L_2$'s junction, $^{\ddagger}$the main-text 40-sensor set, sharing 22
junctions with $S_0$. (b)~The L-TOWN inversion with its ten augmented sets
(Supplementary Note~10).}
\label{tab:leakrerun}
\begingroup\setlength{\tabcolsep}{3pt}\footnotesize
\begin{tabular}{@{}lrrllrrrr@{}}
\multicolumn{9}{@{}l}{(a) Work-order search with the augmented sets, City D: 49 candidates, $\sigma=0.1$~ft, one noise realisation for every set} \\[1pt]
\toprule
Sensor set & Sensors & Hits & Which & Error & Coh. median & $>0.999$ & Rival L$_2$ & Rival L$_3$ \\
\midrule
40 of main-text Section~3.4$^{\ddagger}$ & 40 & 0/3 & none & -- & 0.934 & 51 & 1.00000 & 0.9993 \\
S$_0$ (designated) & 40 & 0/3 & none & -- & 0.928 & 48 & 1.00000 & 0.9994 \\
cover $+20$$^{\dagger}$ & 60 & 1/3 & L$_2$ & 2.3\,\% & 0.796 & 22 & 0.99898 & 0.9995 \\
D-opt $+20$$^{\dagger}$ & 60 & 1/3 & L$_2$ & 3.7\,\% & 0.897 & 34 & 0.99893 & 0.9995 \\
cover $+40$$^{\dagger}$ & 80 & 1/3 & L$_2$ & 2.3\,\% & 0.789 & 18 & 0.99893 & 0.9996 \\
D-opt $+40$$^{\dagger}$ & 80 & 1/3 & L$_2$ & 1.5\,\% & 0.846 & 27 & 0.99900 & 0.9996 \\
cover $+80$$^{\dagger}$ & 120 & 1/3 & L$_2$ & 2.1\,\% & 0.806 & 20 & 0.99909 & 0.9997 \\
D-opt $+80$$^{\dagger}$ & 120 & 1/3 & L$_2$ & 1.1\,\% & 0.791 & 23 & 0.99934 & 0.9997 \\
\addlinespace
cover $+20$ minus the L$_2$-junction sensor & 59 & 0/3 & none & -- & 0.796 & 27 & 1.00000 & 0.9995 \\
D-opt $+20$ minus that sensor & 59 & 1/3 & L$_3$ & 7.0\,\% & 0.902 & 39 & 1.00000 & 0.9995 \\
S$_0$ plus that sensor alone & 41 & 1/3 & L$_2$ & 3.0\,\% & 0.928 & 44 & 0.99905 & 0.9994 \\
S$_0$ plus an L$_3$-junction sensor & 41 & 0/3 & none & -- & 0.920 & 47 & 1.00000 & 0.9995 \\
random $+20$, draw 1 & 60 & 1/3 & L$_3$ & 3.9\,\% & 0.933 & 38 & 1.00000 & 0.9995 \\
random $+20$, draw 2 & 60 & 1/3 & L$_3$ & 4.2\,\% & 0.919 & 37 & 1.00000 & 0.9995 \\
random $+20$, draw 3 & 60 & 0/3 & none & -- & 0.930 & 37 & 1.00000 & 0.9995 \\
random $+20$, draw 4 & 60 & 1/3 & L$_3$ & 4.2\,\% & 0.913 & 35 & 1.00000 & 0.9995 \\
random $+20$, draw 5 & 60 & 0/3 & none & -- & 0.940 & 46 & 1.00000 & 0.9995 \\
random $+40$ & 80 & 0/3 & none & -- & 0.913 & 33 & 1.00000 & 0.9997 \\
\bottomrule
\end{tabular}

\vspace{6pt}
\begin{tabular}{@{}lrrrrrrrr@{}}
\multicolumn{9}{@{}l}{(b) L-TOWN inversion with the augmented sets: 60 candidates, 1,770 pairs (275 orthogonal); hits of 3 in the noiseless and the $\sigma=0.1$~ft group} \\[1pt]
\toprule
Sensor set & Sensors & Clean & Noisy & Med. all & Med. zone & $>0.999$ & Rival A & Rival B \\
\midrule
S$_0$ (33 of main-text Section~3.4) & 33 & 1/3 & 1/3 & 0.8258 & 0.8713 & 42 & 0.999231 & 0.999965 \\
D-opt $+5$ & 38 & 1/3 & 1/3 & 0.8248 & 0.8678 & 40 & 0.998821 & 0.999965 \\
D-opt $+10$ & 43 & 1/3 & 1/3 & 0.8260 & 0.8716 & 40 & 0.998797 & 0.999966 \\
D-opt $+20$ & 53 & 1/3 & 1/3 & 0.8262 & 0.8704 & 41 & 0.998912 & 0.999967 \\
D-opt $+40$ & 73 & 1/3 & 1/3 & 0.8281 & 0.8682 & 43 & 0.998878 & 0.999966 \\
D-opt $+80$ & 113 & 1/3 & 1/3 & 0.8243 & 0.8662 & 44 & 0.998560 & 0.999917 \\
cover $+5$ & 38 & 1/3 & 1/3 & 0.8252 & 0.8737 & 41 & 0.999204 & 0.999966 \\
cover $+10$ & 43 & 1/3 & 1/3 & 0.8304 & 0.8807 & 37 & 0.999289 & 0.999966 \\
cover $+20$ & 53 & 1/3 & 1/3 & 0.8200 & 0.8627 & 38 & 0.999362 & 0.999860 \\
cover $+40$ & 73 & 1/3 & 1/3 & 0.8132 & 0.8567 & 31 & 0.999175 & 0.999849 \\
cover $+80$ & 113 & 1/3 & 1/3 & 0.8161 & 0.8618 & 30 & 0.999057 & 0.999859 \\
\bottomrule
\end{tabular}
\endgroup

\end{table}

\begin{table}[htbp]
\centering
\caption{Coherence-driven placement on City~D against random controls, the
same test re-scored on the coarse stage L-TOWN's inversion shares, and the
L-TOWN candidate-pool bound. (a)~Fisher test and rival-coherence gap by
leak and inversion stage. (b)~$L_2$'s matched-budget exact test and the
random group's 95\,\% Clopper--Pearson interval. (c)~Separable positions
and the oracle-pair-greedy floor for L-TOWN's two never-recovered leaks
(Supplementary Note~10).}
\label{tab:coherence}
\begingroup\setlength{\tabcolsep}{4pt}\footnotesize
\begin{tabular}{@{}llrrrrrl@{}}
\multicolumn{8}{@{}l}{(a) City D: 14 designed sets (8 coherence-min., 6 identifiability-driven) vs 40 random draws, by leak / stage} \\[1pt]
\toprule
Leak & Stage & Design & Random & Fisher $p$ & Gap, recovered & Gap, lost & Separated \\
\midrule
L$_1$ & final & 0/14 & 0/40 & $1.00\times10^{0}$ & -- & 0.667 & no \\
L$_1$ & stage1 & 6/14 & 2/40 & $2.39\times10^{-3}$ & 0.55 & 0.624 & no \\
L$_2$ & final & 12/14 & 3/40 & $1.05\times10^{-7}$ & 3.12e-05 & 2.8e-05 & yes \\
L$_2$ & stage1 & 13/14 & 8/40 & $2.10\times10^{-6}$ & 3.01e-07 & 3.01e-07 & yes \\
L$_3$ & final & 3/14 & 7/40 & $5.12\times10^{-1}$ & 0.000327 & 0.000606 & no \\
L$_3$ & stage1 & 5/14 & 20/40 & $8.92\times10^{-1}$ & 0.000327 & 0.00059 & no \\
\bottomrule
\end{tabular}

\vspace{6pt}
\begin{tabular}{@{}lrrrl@{}}
\multicolumn{5}{@{}l}{(b) L$_2$, same-budget exact permutation test and the random group's 95\% Clopper--Pearson CI} \\[1pt]
\toprule
Budget & Random draws & Random hits & $p$ (exact) & Random rate, 95\% CI \\
\midrule
$k=20$ (worst-pair variant, $+20$) & 20 & 0/20 & $4.76\times10^{-2}$ & 0.000--0.168 \\
$k=40$ (greedy coherence sequence, $+40$) & 20 & 3/20 & $1.90\times10^{-1}$ & 0.032--0.379 \\
\addlinespace
pooled random rate, L$_1$ & 40 & 0/40 & -- & 0.000--0.088 \\
pooled random rate, L$_2$ & 40 & 3/40 & -- & 0.016--0.204 \\
pooled random rate, L$_3$ & 40 & 7/40 & -- & 0.073--0.328 \\
\bottomrule
\end{tabular}

\vspace{6pt}
\begin{tabular}{@{}lrrrrr@{}}
\multicolumn{6}{@{}l}{(c) L-TOWN: candidate-pool bound for the two leaks no placement ever recovers} \\[1pt]
\toprule
Truth rank & Rival coh. at $S_0$ & Separable of & Pool & Oracle floor & Rival after \\
\midrule
rank 19 of 60 & 0.999965 & 1 & 746 & 0.996775 & 0.999400 \\
rank 32 of 60 & 0.999231 & 14 & 746 & 0.990741 & 0.997689 \\
\bottomrule
\end{tabular}
\endgroup

\end{table}

\begin{table}[htbp]
\centering
\caption{The exact-input control: reservoir heads and valve settings read
from the input text instead of through a unit round-trip (Supplementary
Note~1). (a)~Every network the control changes: the metre round-trip path against
the input-text path, over all frames. (b)~The negative control: the same code path on eight networks
whose fields it leaves untouched.}
\label{tab:exactinput}
\begingroup
\setlength{\tabcolsep}{3.4pt}\scriptsize
\begin{tabular}{@{}lrlrrcrr@{}}
\toprule
 & & & \multicolumn{2}{c}{default input path}
 & fields & \multicolumn{2}{c}{exact-input control} \\
\cmidrule(lr){4-5}\cmidrule(lr){7-8}
Network & Frames & Outcome & max$|\Delta H|$ & max$|\Delta Q|$
 & res./valve & max$|\Delta H|$ & max$|\Delta Q|$ \\
 & & & (ft) & (cfs) & changed & (ft) & (cfs) \\
\midrule
\multicolumn{8}{@{}l}{\emph{(a) every network the control changes}}\\
Net3 & 183 & fail & $2.15\times10^{-6}$ & $5.68\times10^{-7}$ & 1 / 0 & $0$ & $3.55\times10^{-15}$ \\
BWSN Network 1 & 207 & fail & $6.09\times10^{-6}$ & $7.05\times10^{-6}$ & 1 / 3 & $0$ & $8.88\times10^{-16}$ \\
BWSN Network 2 & 32 & fail & $6.84\times10^{0}$ & $1.41\times10^{-5}$ & 2 / 1 & $0$ & $7.11\times10^{-15}$ \\
Net6 & 609 & fail & $2.30\times10^{-5}$ & $4.74\times10^{-6}$ & 1 / 0 & $0$ & $7.11\times10^{-15}$ \\
Anytown & 9 & pass & $6.08\times10^{-12}$ & $4.75\times10^{-12}$ & 2 / 0 & $0$ & $1.78\times10^{-15}$ \\
\addlinespace
\multicolumn{8}{@{}l}{\emph{(b) controls: the same code path on eight
networks it leaves alone}}\\
EXA5 & 52 & pass & $5.68\times10^{-14}$ & $8.88\times10^{-16}$ & 0 / 0 & $5.68\times10^{-14}$ & $8.88\times10^{-16}$ \\
\textbf{City D} & 25 & pass & $1.42\times10^{-14}$ & $1.78\times10^{-15}$ & 0 / 0 & $1.42\times10^{-14}$ & $1.78\times10^{-15}$ \\
\textbf{City H} & 25 & pass & $1.42\times10^{-14}$ & $7.11\times10^{-15}$ & 0 / 0 & $1.42\times10^{-14}$ & $7.11\times10^{-15}$ \\
C-Town (BATADAL) & 850 & pass & $5.68\times10^{-14}$ & $8.88\times10^{-16}$ & 0 / 0 & $5.68\times10^{-14}$ & $8.88\times10^{-16}$ \\
D-Town & 819 & pass & $5.68\times10^{-14}$ & $1.78\times10^{-15}$ & 0 / 0 & $5.68\times10^{-14}$ & $1.78\times10^{-15}$ \\
L-TOWN & 2,031 & pass & $2.84\times10^{-14}$ & $1.11\times10^{-16}$ & 0 / 0 & $2.84\times10^{-14}$ & $1.11\times10^{-16}$ \\
Richmond (skeleton) & 91 & pass & $1.14\times10^{-13}$ & $2.22\times10^{-16}$ & 0 / 0 & $1.14\times10^{-13}$ & $2.22\times10^{-16}$ \\
Richmond (standard) & 55 & pass & $1.14\times10^{-13}$ & $2.22\times10^{-16}$ & 0 / 0 & $1.14\times10^{-13}$ & $2.22\times10^{-16}$ \\
\bottomrule
\end{tabular}
\endgroup

\end{table}

\begin{table}[htbp]
\centering
\caption{Cluster-level leak diagnosability: the hit-rate against
inspection-radius trade-off (Supplementary Note~11). (a),~(b)~Full
threshold sweeps with the geometry-only, coherence-only and equal-budget
single-point controls. (c)~The inspection burden of the three
highest-ranked clusters. (d)~Size-matched random grouping.}
\label{tab:cluster}
\begingroup
\setlength{\tabcolsep}{3.0pt}\scriptsize
\begin{tabular}{@{}lrrrrrrrr@{}}
\toprule
$\tau$ & clusters & mean & mean & \multicolumn{2}{c}{cluster level}
 & geometry & coherence & equal-budget \\
 & & radius & district & top-1 & top-3 & only & only & single point \\
 & & (m) & (km) & & & top-3 & top-3 & top-3 \\
\midrule
\multicolumn{9}{@{}l}{\emph{(a) operating network, 49 work-order
candidates, 40 sensors, $\sigma=0.1$~ft}}\\
0 & 1 & 9043 & 83.92 & 3/3 & 3/3 & 3/3 & 3/3 & 3/3 \\
0.5 & 4 & 1902 & 20.98 & 1/3 & 2/3 & 2/3 & 3/3 & 2/3 \\
0.8 & 5 & 1506 & 16.78 & 1/3 & 2/3 & 1/3 & 3/3 & 2/3 \\
0.9 & 6 & 1427 & 13.99 & 1/3 & 1/3 & 2/3 & 1/3 & 2/3 \\
0.95 & 8 & 1086 & 10.49 & 1/3 & 2/3 & 2/3 & 2/3 & 0/3 \\
0.98 & 12 & 788 & 6.99 & 1/3 & 2/3 & 1/3 & 2/3 & 0/3 \\
0.99 & 13 & 729 & 6.46 & 1/3 & 2/3 & 1/3 & 2/3 & 0/3 \\
0.995 & 17 & 379 & 4.94 & 1/3 & 2/3 & 1/3 & 2/3 & 0/3 \\
0.999 & 25 & 179 & 3.36 & 1/3 & 2/3 & 1/3 & 2/3 & 0/3 \\
0.9995 & 28 & 140 & 3.00 & 1/3 & 1/3 & 1/3 & 1/3 & 0/3 \\
0.9999 & 30 & 114 & 2.80 & 1/3 & 1/3 & 1/3 & 1/3 & 0/3 \\
$>1$ & 49 & 0 & 1.71 & 0/3 & 0/3 & 0/3 & 0/3 & 0/3 \\
\addlinespace
\multicolumn{9}{@{}l}{\emph{(b) L-TOWN, 60 candidates, 33 sensors,
256 scenarios, $\sigma=0.1$~ft}}\\
0 & 1 & 1575 & 43.16 & 3/3 & 3/3 & 3/3 & 3/3 & -- \\
0.5 & 2 & 934 & 21.58 & 3/3 & 3/3 & 3/3 & 3/3 & -- \\
0.8 & 4 & 542 & 10.79 & 2/3 & 3/3 & 3/3 & 3/3 & -- \\
0.9 & 6 & 399 & 7.19 & 2/3 & 3/3 & 3/3 & 3/3 & -- \\
0.95 & 9 & 295 & 4.80 & 1/3 & 2/3 & 2/3 & 3/3 & -- \\
0.98 & 15 & 185 & 2.88 & 1/3 & 2/3 & 3/3 & 2/3 & -- \\
0.99 & 17 & 160 & 2.54 & 1/3 & 3/3 & 3/3 & 3/3 & -- \\
0.995 & 22 & 98 & 1.96 & 1/3 & 3/3 & 2/3 & 3/3 & -- \\
0.999 & 36 & 40 & 1.20 & 1/3 & 2/3 & 1/3 & 2/3 & -- \\
0.9995 & 41 & 30 & 1.05 & 0/3 & 1/3 & 1/3 & 1/3 & -- \\
0.9999 & 53 & 7 & 0.81 & 0/3 & 1/3 & 1/3 & 1/3 & -- \\
$>1$ & 60 & 0 & 0.72 & 0/3 & 1/3 & 1/3 & 1/3 & -- \\
\bottomrule
\end{tabular}

\vspace{2mm}
\begin{tabular}{@{}rrrrrrrr@{}}
\multicolumn{8}{@{}l}{\emph{(c) what a crew would actually walk: the three
highest-ranked clusters, operating network}}\\
\toprule
$\tau$ & clusters & candidates & pipe in & share of & largest of the
 & mean radius & mean district \\
 & & in top 3 & top 3 (km) & network & three radii (m) & all clusters (m)
 & all clusters (km) \\
\midrule
0.9 & 6 & 23 & 34.70 & 41.3\,\% & 2344 & 1427 & 13.99 \\
0.95 & 8 & 14 & 18.53 & 22.1\,\% & 2344 & 1086 & 10.49 \\
0.99 & 13 & 10 & 10.43 & 12.4\,\% & 1041 & 729 & 6.46 \\
0.995 & 17 & 8 & 8.66 & 10.3\,\% & 750 & 379 & 4.94 \\
0.999 & 25 & 7 & 7.42 & 8.8\,\% & 750 & 179 & 3.36 \\
\bottomrule
\end{tabular}

\vspace{2mm}
\begin{tabular}{@{}llrrrrrr@{}}
\multicolumn{8}{@{}l}{\emph{(d) size-matched random grouping: same cluster
size distribution, topology and coherence destroyed}}\\
\toprule
Network & $\tau$ & design & radius & random & radius & $\ge$ design
 & $p$ \\
 & & top-3 & (m) & top-3 & (m) & & \\
\midrule
operating network & 0.9 & 1/3 & 1427 & 2.10/3 & 4415 & 20/20 & 1.000 \\
 & 0.99 & 2/3 & 729 & 0.65/3 & 3544 & 1/20 & 0.095 \\
 & 0.999 & 2/3 & 179 & 0.75/3 & 1700 & 2/20 & 0.143 \\
L-TOWN & 0.9 & 3/3 & 399 & 2.30/3 & 1223 & 8/20 & 0.429 \\
 & 0.99 & 3/3 & 160 & 1.70/3 & 698 & 3/20 & 0.190 \\
 & 0.999 & 2/3 & 40 & 1.50/3 & 298 & 8/20 & 0.429 \\
\bottomrule
\end{tabular}
\endgroup

\end{table}

\begin{table}[htbp]
\centering
\caption{Sensor layouts scored on a reference subspace fixed before any
design is chosen (Supplementary Note~12). (a)~All twelve budget-by-noise
cells against same-budget random additions. (b)~The main cell on five
rulers, including the discarded directions. (c)~The cost of selecting a
layout by head misfit. (d)~Design-time prediction against achieved error.}
\label{tab:identmetric}
\begingroup
\setlength{\tabcolsep}{3.2pt}\scriptsize
\begin{tabular}{@{}llrrrrrrrr@{}}
\multicolumn{10}{@{}l}{\emph{(a) every cell: the designed additions against
random additions from the same pool at the same budget}}\\
\toprule
Budget & $\sigma$ (ft) & objective & $R$
 & \multicolumn{4}{c}{identifiable-subspace roughness RMSE}
 & random & $p$ \\
\cmidrule(lr){5-8}
 & & & & design & best & median & worst & $\ge$ design & \\
\midrule
$+20$ & 0.03 & coverage & 20 & 15.87 & 18.58 & 20.48 & 21.18 & 0/20 & 0.0476 \\
 &  & D-optimal & 20 & 17.50 & 18.58 & 20.48 & 21.18 & 0/20 & 0.0476 \\
$+20$ & 0.1 & coverage & 100 & 17.45 & 19.70 & 21.73 & 23.81 & 0/100 & 0.0099 \\
 &  & D-optimal & 100 & 18.04 & 19.70 & 21.73 & 23.81 & 0/100 & 0.0099 \\
$+20$ & 0.3 & coverage & 20 & 18.13 & 21.87 & 24.15 & 26.56 & 0/20 & 0.0476 \\
 &  & D-optimal & 20 & 21.85 & 21.87 & 24.15 & 26.56 & 0/20 & 0.0476 \\
$+40$ & 0.03 & coverage & 20 & 14.47 & 16.60 & 19.22 & 21.13 & 0/20 & 0.0476 \\
 &  & D-optimal & 20 & 15.64 & 16.60 & 19.22 & 21.13 & 0/20 & 0.0476 \\
$+40$ & 0.1 & coverage & 20 & 16.24 & 18.04 & 20.11 & 22.01 & 0/20 & 0.0476 \\
 &  & D-optimal & 20 & 18.02 & 18.04 & 20.11 & 22.01 & 0/20 & 0.0476 \\
$+40$ & 0.3 & coverage & 20 & 19.50 & 20.06 & 22.48 & 24.45 & 0/20 & 0.0476 \\
 &  & D-optimal & 20 & 19.99 & 20.06 & 22.48 & 24.45 & 0/20 & 0.0476 \\
\bottomrule
\end{tabular}

\vspace{2mm}
\begin{tabular}{@{}lrrrr@{}}
\multicolumn{5}{@{}l}{\emph{(b) the same cell ($+20$, $\sigma=0.1$~ft,
$R=100$, coverage objective) read on five different rulers}}\\
\toprule
Ruler & design & random median & random $\ge$ design & $p$ \\
\midrule
identifiable subspace, $\gamma=2$ ($k=98$) & 17.45 & 21.73 & 0/100 & 0.0099 \\
identifiable subspace, $\gamma=3$ ($k=70$) & 15.83 & 21.83 & 0/100 & 0.0099 \\
identifiable subspace, $\gamma=10$ ($k=29$) & 9.86 & 20.05 & 0/100 & 0.0099 \\
all 432 pipes, no projection & 25.10 & 25.79 & 0/100 & 0.0099 \\
reference null space ($432-98$ directions) & 26.93 & 26.86 & 60/100 & 0.6040 \\
\bottomrule
\end{tabular}

\vspace{2mm}
\begin{tabular}{@{}lrrrrr@{}}
\multicolumn{6}{@{}l}{\emph{(c) the price of choosing a layout by head
misfit: pick the best of 100 random additions by each head metric}}\\
\toprule
Selection criterion & its ident. & rank & percentile & best in
 & regret \\
 & RMSE & & & pool & \\
\midrule
training head MSE & 21.05 & 25/100 & 24\,\% & 19.70 & +1.36 \\
held-out-frame head RMSE & 21.80 & 56/100 & 55\,\% & 19.70 & +2.10 \\
held-out-sensor head RMSE & 20.65 & 10/100 & 9\,\% & 19.70 & +0.95 \\
\bottomrule
\end{tabular}

\vspace{2mm}
\begin{tabular}{@{}lrrrr@{}}
\multicolumn{5}{@{}l}{\emph{(d) how well the design-time linear-Gaussian
prediction orders the achieved error}}\\
\toprule
Pool & Spearman $\rho$ & $p$ & predicted spread & achieved spread \\
\midrule
$+20$ only ($n=100$) & 0.38 & 1.3e$-$04 & 5.6\,\% & 19.0\,\% \\
$+40$ only ($n=20$) & 0.69 & 6.0e$-$04 & 4.7\,\% & 19.6\,\% \\
both budgets pooled ($n=120$) & 0.55 & 6.3e$-$11 & 8.2\,\% & 26.9\,\% \\
\bottomrule
\end{tabular}
\endgroup

\end{table}

\begin{table}[htbp]
\centering
\caption{Optimiser enhancements and the wall-clock reading on City~D
(Supplementary Note~13). (a)~Every arm at a 2,000-call evaluation budget.
(b)~What batching multiple starts buys per model call. (c)~The
derivative-free baselines on the same node, with the eight-way-parallel
boundary the fairness protocol promises them.}
\label{tab:optimiser}
\begingroup
\setlength{\tabcolsep}{3.6pt}\scriptsize
\begin{tabular}{@{}lrrrrr@{}}
\multicolumn{6}{@{}l}{\emph{(a) City D, 432 free pipes, evaluation budget
2,000 model calls, medians over noise seeds}}\\
\toprule
Arm & seeds & calls & training loss & roughness RMSE
 & wall (s) \\
\midrule
reference gradient configuration & 4 & 200 & 9.109e-03 & 11.32 & 631 \\
first order, no preconditioner & 10 & 1976 & 9.492e-03 & 17.71 & 223 \\
first order, Schur-complement diagonal & 10 & 1976 & 9.541e-03 & 12.40 & 221 \\
first order, exact Gauss--Newton diagonal & 10 & 1976 & 9.459e-03 & 16.04 & 232 \\
first order, 8 starts batched & 10 & 1976 & 9.575e-03 & 15.00 & 139 \\
first order, 8 starts $+$ Schur & 10 & 1976 & 9.542e-03 & 13.63 & 141 \\
first order, 16 starts $+$ Schur & 10 & 1976 & 9.580e-03 & 14.09 & 127 \\
Levenberg--Marquardt tail & 5 & 1997 & 9.459e-03 & 15.17 & 2331 \\
LM tail $+$ Schur & 4 & 1998 & 9.565e-03 & 11.37 & 2326 \\
LM tail $+$ 8 starts $+$ Schur & 4 & 1997 & 9.052e-03 & 13.69 & 2383 \\
\bottomrule
\end{tabular}

\vspace{2mm}
\begin{tabular}{@{}rrrr@{}}
\multicolumn{4}{@{}l}{\emph{(b) what batching many starts buys: it is not a
factor of $B$}}\\
\toprule
starts $B$ & s per step & s per model call & speed-up per call \\
\midrule
1 & 0.1123 & 0.1123 & 1.00$\times$ \\
2 & 0.1507 & 0.0754 & 1.49$\times$ \\
4 & 0.2326 & 0.0581 & 1.93$\times$ \\
8 & 0.3773 & 0.0472 & 2.38$\times$ \\
16 & 0.6861 & 0.0429 & 2.62$\times$ \\
32 & 1.2770 & 0.0399 & 2.81$\times$ \\
\bottomrule
\end{tabular}

\vspace{2mm}
\begin{tabular}{@{}lrrrrr@{}}
\multicolumn{6}{@{}l}{\emph{(c) the derivative-free baselines on the same
node and the same budget: the wall-clock reading}}\\
\toprule
Algorithm & seeds & calls & training loss & wall (s)
 & wall\,/\,8 (s) \\
\midrule
differential evolution & 5 & 1984 & 1.449e-01 & 156 & 20 \\
particle swarm & 5 & 1984 & 8.000e-02 & 156 & 20 \\
CMA-ES & 5 & 1980 & 2.488e-01 & 156 & 20 \\
DE $\to$ LM hybrid & 1 & 1997 & 9.400e-03 & 4756 & 595 \\
differential evolution, $10\times$ budget & 3 & 20000 & 1.433e-02 & 1691 & 211 \\
\bottomrule
\end{tabular}
\endgroup

\end{table}

\clearpage
\bibliography{refs}

\end{document}